\documentclass[11pt,preprintnumbers,pra,aps,
nofootinbib,tightenlines,floatfix, twocolumn,superscriptaddress,longbibliography]{revtex4-2}

\usepackage[colorlinks=true]{hyperref}
\hypersetup{
  linkcolor  = TealBlue!90!Black,
  citecolor  = YellowGreen!90!Black,
  urlcolor   = Purple!90!Black,
  colorlinks = true,
}
\usepackage[dvipsnames]{xcolor}
\usepackage[all]{hypcap}
\usepackage{amsmath,amssymb}
\usepackage{graphicx}
\usepackage{physics}
\usepackage{mathtools}
\usepackage{balance}
\usepackage{cancel}
\usepackage[utf8]{inputenc}
\usepackage{csquotes}
\usepackage{empheq}
\usepackage{longtable}
\usepackage{multirow}
\usepackage[section]{placeins}
\usepackage{comment}
\usepackage[customcolors]{hf-tikz}
\usepackage[top=1in,bottom=1in,left=1.8cm,right=1.6cm]{geometry}
\usepackage{tikz}
\usepackage{tcolorbox}
\usepackage{algpseudocode}
\usetikzlibrary{positioning}

\allowdisplaybreaks
\makeatletter
\renewcommand\onecolumngrid{
\do@columngrid{one}{\@ne}
\def\set@footnotewidth{\onecolumngrid}
\def\footnoterule{\kern-6pt\hrule width 1.5in\kern6pt}%
}
\makeatother
\DeclareMathOperator{\sign}{sgn}

\newcommand\scalemath[2]{\scalebox{#1}{\mbox{\ensuremath{\displaystyle #2}}}}

\catcode`,\active

\catcode`\,12

\newcount\hour \newcount\hourminute \newcount\minute
\hour=\time \divide \hour by 60
\hourminute=\hour \multiply \hourminute by 60
\minute=\time \advance \minute by -\hourminute

\DeclareMathOperator{\nullspace}{null}
\newcommand{\vct}[1]{\operatorname{vec}\!\left(#1\right)}

\begin{document}

\title{Real-time simulation of scattering in \texorpdfstring{$\phi^4$}{phi4} theory: localized excitation, scalable wave packet preparation, and peak-generating artifacts}

\author{Boyu Gao}
\email{boyu.gao@duke.edu}
\affiliation{{Duke Quantum Center and Department of Physics, Duke University, Durham, NC 27708, USA}}
\author{Natalie Klco}
\email{natalie.klco@duke.edu}
\affiliation{{Duke Quantum Center and Department of Physics, Duke University, Durham, NC 27708, USA}}

\begin{abstract}

Probing real-time dynamics is essential for understanding non-perturbative scattering and interactions in quantum field theory. A common challenge in real-time simulation is preparing accurate asymptotic incoming states. Through efficiently solvable eigenvalue problems, we jointly optimize a superposition of single particle momentum eigenstates and a single site interpolating operator to maximize preparation fidelity from the interacting vacuum. This leads to a localized excitation in the interacting theory analogous to a position eigenstate. Weighted sums of the resulting interpolating operator over lattice sites prepare wave packets of arbitrary spatial extent. To ensure that this construction produces the correct continuum limit, relative phases among single particle momentum eigenstates in this superposition are calibrated using a relation derived from Lorentz symmetry. Analyzing single particle wave packet propagation through the position space profile reveals oscillatory Airy distortions, which originate from interference between saddle point contributions at different momenta that share the same lattice group velocity. These oscillations systematically weaken when approaching the continuum limit, and reliable regimes are identified as a function of simulation time and wave packet momentum profile. Two-particle scattering simulations in interacting scalar field theory are also implemented, revealing artificial inelasticity arising from coarse local field digitization or deviation from the Nyquist-Shannon sampling point. With these artifacts under control, present simulations provide emerging evidence for physical inelastic processes in $\phi^4$ theory. The constructions introduced in this work may provide new understanding of state preparation for the quantum simulation of scattering.

\end{abstract}

\date{\today}
\maketitle

{
\footnotesize
\tableofcontents
}

\section{Introduction}
\label{sec:I}

Scattering processes reveal the interactions and particle content of quantum field theories. Their amplitudes are encoded in the $S$ matrix and can be calculated at weak coupling through a perturbative expansion using Feynman diagrams~\cite{Feynman:1949zx,Peskin:1995ev,weinberg1995quantum}. Understanding scattering in the strongly coupled regime where perturbation theory breaks down is essential for modeling phenomena such as hadronization~\cite{Andersson:1983ia,Webber:1983if}. One way to study scattering at strong coupling is through the Hamiltonian formalism, by preparing well-separated incoming wave packets and evolving them in real-time under the interacting Hamiltonian. Such simulations have been carried out in $(1+1)$ dimensions for spin chains~\cite{PhysRevResearch.3.013078,Karpov:2020pqe,Milsted:2020jmf,Vovrosh:2022bpj,Jha:2024jan,farrell2026digital,Zemlevskiy:2026kpc}, $\phi^4$ scalar field theory~\cite{qr72-51v1,Ingoldby:2025bdb,Sayegh:2025rlp}, and lattice gauge theories~\cite{Rigobello:2021fxw,Belyansky:2023rgh,Papaefstathiou:2024zsu,Barata:2025rjb,Morgavi:2026elf,Davoudi:2025rdv}.

\begin{figure*}[!t]
  \centering
  \includegraphics[width=\textwidth]{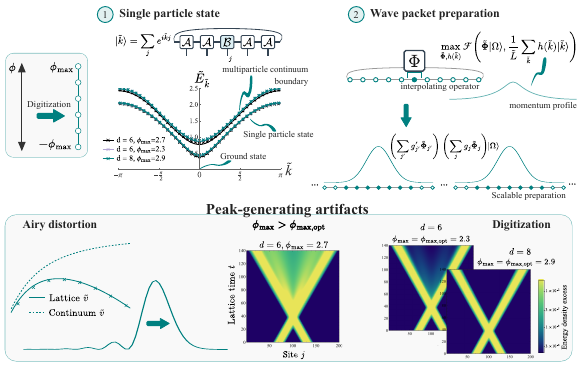}
  \caption{At each lattice site, the field is represented by $d$ discrete values, evenly distributed within the interval $[-\phi_{\max}, \phi_{\max}]$. (1) With total sites $\tilde{L} = 41$ and periodic boundary condition (PBC), the quasiparticle ansatz is used to compute the two lowest energy eigenstates orthogonal to the ground state at each momentum sector (crosses). The solid curves show predictions from the free lattice dispersion relation governed by the lattice physical mass, and a lower bound on the multiparticle continuum~\cite{PhysRevB.67.024412}. (2) Wave packet preparation follows from jointly optimizing the localized excitation and the interpolating operator. Incoming wave packets are prepared from superposing the weighted interpolating operators. (Bottom) Peak-generating artifacts. The left panel shows that the non-monotonic lattice group velocity can produce Airy oscillations in the envelope of a single propagating wavepacket. The middle and right panels show excess energy density during scattering in lattice $\phi^4$ theory with $\tilde{m}_0=0.1$, $\tilde{\lambda}_0=0.8$ and $\tilde{L}=200$, highlighting artifacts caused by over-claiming $\phi_{\text{max}}$ and digitization.}
  \label{fig:prepandscatterdiagram}
\end{figure*}

To prepare well-separated incoming wave packets, one approach starts from analytically known free theory wave packets and turns on the interaction adiabatically, with backward time evolution steps inserted to compensate dynamical phases~\cite{Jordan:2011ci,Jordan:2012xnu}. Alternatively, Haag--Ruelle scattering theory~\cite{PhysRev.112.669,ruelle1962asymptotic,Duncan:2012aja,PRXQuantum.5.020311} provides a framework for constructing asymptotic incoming states directly from the interacting vacuum. Together, these approaches guide the state preparation for both classical and quantum simulations of scattering. The present work follows the Haag--Ruelle scattering framework to construct approximate asymptotic incoming states, as illustrated in step 2 of Fig.~\ref{fig:prepandscatterdiagram}.

The matrix product state (MPS)~\cite{White:1992zz,White:1993zza,PhysRevLett.75.3537,PhysRevLett.93.040502,PhysRevLett.93.207204,schollwock2011density,Haegeman:2011zz,Haegeman:2011lcd,Pirvu:2011ijm,PhysRevLett.111.080401,Haegeman:2015ezw,Vanderstraeten:2019voi} provides one practical framework to carry out the classical real-time simulation of scattering in $(1+1)$ dimensions. With this framework, the quasiparticle ansatz~\cite{Haegeman:2011lcd,Pirvu:2011ijm,PhysRevLett.111.080401,Vanderstraeten:2019voi} offers a variational description of single particle momentum eigenstates on top of an MPS ground state. As shown in step 1 of Fig.~\ref{fig:prepandscatterdiagram}, it replaces a ground state tensor with a variational excitation tensor, $\mathcal{B}$, and coherently superposes these insertions to construct states of definite momentum. Single particle momentum eigenstates are thus obtained as the lowest energy solutions in this subspace orthogonal to the ground state, with exponential convergence to the exact eigenstates as the support of the excitation tensor increases~\cite{PhysRevLett.111.080401}. Constructing a wave packet by superposing these states with weights $h(\tilde{k})$ generally involves momentum dependent excitation tensors $\mathcal{B}(\tilde{k})$.

Some implementations simplify this construction by using a fixed excitation tensor, evaluated at the central momentum throughout $h(\tilde{k})$~\cite{Milsted:2020jmf,Jha:2024jan}. Others retain the momentum dependence of the excitation tensors to improve the accuracy of wave packet preparation~\cite{PhysRevResearch.3.013078,Belyansky:2023rgh,Morgavi:2026elf}. In particular, Reference~\cite{PhysRevResearch.3.013078} optimizes over $h(\tilde{k})$ to localize single particle excitations, yielding position dependent excitation tensors, $\mathcal{B}(j)$. Reference~\cite{Morgavi:2026elf} optimizes the relative phases of an equal weighted momentum space profile to localize single particle excitations, then constructs a unitary interpolating operator that prepares this state from the vacuum. In both works, these building blocks are then used to construct wave packets with arbitrary spatial extent.

Using single particle momentum eigenstates obtained from the quasiparticle ansatz, this work \emph{jointly} optimizes $h(\tilde{k})$ and a single site interpolating operator to maximize the fidelity of preparing this state from the vacuum, as illustrated in step 2 of Fig.~\ref{fig:prepandscatterdiagram}. The procedure considered in this work gives localization an operational meaning tailored to single particle wave packet preparation with interpolating operators. It differs from localization defined with respect to norm of excitation tensors~\cite{PhysRevResearch.3.013078} or excess energy density~\cite{Morgavi:2026elf}. 

Moreover, superposing single particle momentum eigenstates requires aligning the arbitrary phases of independently computed $\mathcal{B}(k)$. In Refs.~\cite{PhysRevResearch.3.013078,Morgavi:2026elf}, the relative phases are embedded as variational parameters in the localization optimization. Alternatively, Ref.~\cite{Belyansky:2023rgh} fixes the phase of single particle momentum eigenstates by requiring a selected element to be real and positive (see also Ref.~\cite{farrell2026digital}). In the present work, this phase convention is fixed by a relation derived from Lorentz symmetry in the continuum. This provides a common reference for momentum space profiles that smoothly connects sequential calculations toward the continuum.

To prepare incoming wave packets on quantum computers, some approaches optimize circuit parameters on small systems and extrapolate them to larger lattices~\cite{PhysRevD.109.114510,qr72-51v1}, while others use variational energy minimization to determine the parameters of preparation circuits or interpolating operator ansatz~\cite{Davoudi:2024wyv,Davoudi:2025rdv,farrell2026digital,Zemlevskiy:2026kpc}. As shown in step 2 of Fig.~\ref{fig:prepandscatterdiagram}, the present construction does not require an interpolating operator ansatz to construct arbitrary-sized wave packets. Furthermore, the single site support imposed on the interpolating operator makes the optimization tractable, avoiding the difficulties associated with barren plateaus~\cite{McClean:2018jps,Cerezo:2020mtn}.

Lattice scalar field theory with $\phi^4$ interaction provides a simple interacting setting for studying the localization, propagation, and scattering of wave packets, with a controlled approach to the continuum limit. In the free theory, applying the local field operator to the vacuum,  $\phi|0\rangle$, creates a localized single particle excitation, which provides an analogue of a position eigenstate in quantum field theory~\cite{weinberg1995quantum,Peskin:1995ev,Jordan:2011ci,Jordan:2012xnu}. In this work, the localized excitation identified in the interacting theory, $\Phi|\Omega\rangle$, is found to allow a similar interpretation.

With local field space digitization, as shown in Fig.~\ref{fig:prepandscatterdiagram}, the dispersion relations are explicitly calculated, with no evidence of an isolated bound state below the multiparticle continuum~\footnote{Since the quasiparticle ansatz defines a variational subspace at fixed momentum, its higher energy solutions can also approximate other energy eigenstates in the same momentum sector.}. The digitized lattice scalar field therefore provides an opportunity to study inelastic processes beyond $2\to2$ scattering, with $2\to4$ expected to be the lowest particle production channel, since odd particle final states are forbidden by the $\mathbb{Z}_2$ symmetry. Earlier studies of scattering in $\phi^4$ theory report particle production using Hamiltonian truncation~\cite{Ingoldby:2025bdb}, and substantial inelasticity using MPS simulations with wave packet created via local field operators~\cite{Sayegh:2025rlp}. In the simulations presented here, coarse field sampling (relative to Nyquist-Shannon optimal) enhances the excess energy density in the outgoing state, as shown in the lower panel of Fig.~\ref{fig:prepandscatterdiagram}~\footnote{These results, together with further numerical simulations, indicate that choosing $\phi_{\max}$ above its optimal value enhances the inelastic signal, whereas choosing it below its optimal value suppresses the inelastic signal.}. This warrants further examination of the inelastic signals to distinguish physical particle production from digitization artifacts.

In addition to digitization specifications, the lattice dispersion relation also produces an artifact that may impact inelastic analyses. The free theory lattice dispersion relation approximately describes the interacting single particle spectrum when the bare mass is replaced by the lattice physical mass, as shown in step 1 of Fig.~\ref{fig:prepandscatterdiagram} and Appendix Fig.~\ref{fig:Adistortion}. Similar description has been observed in lattice Monte Carlo studies of interacting scalar field theories~\cite{Rummukainen:1995vs,Guo:2018xbv}. With this dispersion relation, components with distinct lattice momenta can share the same group velocity and interfere while propagating, producing oscillations as in the lower left panel of Fig.~\ref{fig:prepandscatterdiagram} that distort the wave packet. Because these oscillations are characterized by Airy functions at long evolution times (see Appendix~\ref{app:latartifact}), we refer to their presence as Airy distortion. A related pattern has been observed in the vacuum subtracted energy density of scattering in Ising field theory~\cite{farrell2026digital}, suggesting a common origin.

This article is organized as follows. After presenting an overview of lattice scalar field theory and its local field space digitization in Section~\ref{sec:II}, Section~\ref{sec:III1} examines digitization for the ground state and single particle momentum eigenstates in $(1+1)$ dimensions, with details elaborated in Appendices~\ref{app:mpsintro1}--\ref{app:mpsintro3}. Section~\ref{sec:III2} then leverages the calculated low lying eigenstates to construct the localized excitation and the interpolating operator. The details for the optimization of the interpolating operator, its application to the interacting vacuum, and scattering simulations with MPS are discussed in Appendices~\ref{app:mpsintro4}--\ref{app:mpsintro6}. Airy distortions in single particle propagation is discussed in Section~\ref{sec:IV1} and Appendix~\ref{app:latartifact}, while digitization artifacts for scattering are analyzed in Section~\ref{sec:IV2}. Appendix~\ref{app:scatteringcountlim} presents more scattering simulations, with one step toward the thermodynamic and continuum limit of wave packet scattering. Appendices~\ref{app:freefield1} and~\ref{app:freefield2} present a polynomially efficient continuous variable (CV) formalism for counterpropagating wave packets in the free theory and derive an analytic expression for the Renyi-2 entropy. The free scalar field calculations provide benchmarks for the MPS calculations and help assess digitization artifacts.

\section{Lattice scalar field}
\label{sec:II}

The Hamiltonian for 1+1 dimensional real scalar field defined on a lattice of $\hat{L}$ sites is,
\begin{equation}
\scalemath{0.94}{\hat{H} = \sum_{j} \left[ \frac{1}{2}\hat{\pi}_j^2 + \frac{1}{2} \hat{m}_0^2 \hat{\phi}_j^2 + \frac{1}{2}(\hat{\phi}_{j+1}-\hat{\phi}_j)^2 + \frac{\hat{\lambda}_0}{4!}\hat{\phi}_j^4 \right]} ,
\label{eq:Hamiltonianmain}
\end{equation}
where the canonical commutation relations (CCR) reads $[\hat{\phi}_i, \hat{\pi}_j] = i \delta_{ij}$. The connection between the dimensionless lattice scalar field theory defined in Eq.~\eqref{eq:Hamiltonianmain} and its continuum formulation expressed in physical units is discussed in Appendix~\ref{app:fieldintro}. 

In this work, the single particle momentum eigenstates of Eq.~\eqref{eq:Hamiltonianmain} are studied on a finite size lattice under periodic boundary conditions (PBC), where the sum spans $j = -\ell, \dots, \ell$ with $\ell \equiv (\hat{L}-1)/2$ and $\hat{L}$ odd. As shown in Sec.~\ref{sec:III} and detailed in Appendix~\ref{app:fieldintro1}, this indexing choice allows sign calibration of single particle momentum eigenstates, yielding a smooth continuum limit that respects the Lorentz symmetry. The allowed lattice momenta lie within the Brillouin zone, $\hat{k} = \frac{2\pi n}{\hat{L}}$, where integer $n \in (-\frac{\hat{L}}{2},\frac{\hat{L}}{2})$. 

The operators that create the single particle states are then implemented in open boundary conditions (OBC) within the MPS framework for scattering simulations. For sufficiently large systems, the scattering process remains localized far from the boundaries, rendering boundary effects negligible (see Appendix Fig.~\ref{fig:freescatterplot}), while avoiding the substantial computational overhead associated with the ring geometry required by PBC~\cite{Verstraete:2004zza}. Here, $(\phi_{\hat{L}+1},\pi_{\hat{L}+1}) = (\phi_0,\pi_0) = 0$ and the sum spans $j=(0, ..., \hat{L})$. 

The theory reduces to a free lattice scalar field when $\hat{\lambda}_0 = 0$. With PBC, the corresponding lattice dispersion relation is given by 
\begin{equation}  \hat{E}_{\hat{m}_0,\hat{k}}\equiv\hat{E}_{\hat{k}}= \sqrt{\hat{m}^2_0 + 4 \sin^2\!\left(\frac{\hat{k}}{2}\right)} \ \ \ .
\label{eq:freedispersion}
\end{equation}
Define the creation and annihilation operators,
\begin{equation}
   \hat{a}_{\hat{k}}^{\dagger} = \sum_j e^{i\hat{k}j} \left(\sqrt{\frac{\hat{E}_{\hat{k}}}{2}} \hat{\phi}_j - i \sqrt{\frac{1}{2\hat{E}_{\hat{k}}}} \hat{\pi}_j\right) \ \ \ ,
\label{eq:akdagger}
\end{equation}
where $\hat{a}_{\hat{k}}^{\dagger}$ satisfies the CCR, $[\hat{a}_{\hat{p}}, \hat{a}_{\hat{k}}^{\dagger}] = \hat{L} \delta_{\hat{p},\hat{k}}$ and $[\hat{a}_{\hat{p}}, \hat{a}_{\hat{k}}] = 0$. The free scalar field Hamiltonian can be diagonalized,
\begin{equation}
\hat{H}_{0} =  \sum_{\hat{k}} \hat{E}_{\hat{k}} \left( \frac{1}{\hat{L}}a^{\dagger}_{\hat{k}} \hat{a}_{\hat{k}} + \frac{1}{2}  \right) \ \ \ .
\label{eq:H0main}
\end{equation}
Field and conjugate momentum operators may be expanded using ladder operators as,
\begin{align}
\hat{\phi}_j &= \frac{1}{\hat{L}} \sum_{\hat{k}} \sqrt{\frac{1}{2 \hat{E}_{\hat{k}}}} \left( \hat{a}_{\hat{k}} e^{i \hat{k} j} + \hat{a}_{\hat{k}}^\dagger e^{-i \hat{k} j} \right) \ \ \ , 
\label{eq:phimain} \\
\hat{\pi}_j &= \frac{1}{\hat{L}} \sum_{\hat{k}} (-i) \sqrt{\frac{\hat{E}_{\hat{k}}}{2}} \left( \hat{a}_{\hat{k}} e^{i \hat{k} j} - \hat{a}_{\hat{k}}^\dagger e^{-i \hat{k} j} \right)  \ \ \ .
\label{eq:pimain}
\end{align}
With $|0\rangle$ denoting the free vacuum, $\hat{a}_{\hat{k}}|0\rangle = 0$. Note the ground state is dimensionless and normalized to unity. The single particle momentum eigenstates and their normalization read,
\begin{equation}
| \hat{k} \rangle_0 = \hat{a}_{\hat{k}}^\dagger | 0 \rangle, \quad _0 \langle \hat{p} |\hat{k} \rangle_0 = \hat{L} \delta_{\hat{p},\hat{k}} \ \ \ .
\label{eq:freesingleparticlenormmain}
\end{equation} 
Single particle wave packet is normalized to unity (c.f. Ref.~\cite{Peskin:1995ev}),
\begin{equation}
|\psi\rangle_{1P}
= \frac{1}{\hat{L}} \sum_{\hat{k}} f_{\hat{k}} |\hat{k}\rangle_{0}, \quad \frac{1}{\hat{L}} \sum_{\hat{k}} |f_{\hat{k}}|^2 = 1   \ \ \ ,  
\label{eq:wpnormalizationmain}
\end{equation}
such that $_{1P} \langle \psi | \psi\rangle_{1P}  = 1$, consistent with the normalization convention adopted for the vacuum. Single particle excitation localized at site $j$ is given by,
\begin{equation}
\hat{\phi}_j | 0 \rangle =  \frac{1}{\hat{L}} \sum_{\hat{k}} e^{-i \hat{k} j} \sqrt{\frac{1}{2 \hat{E}_{\hat{k}}}} |\hat{k}\rangle_0 \ \ \ .
\label{eq:freesingleparticlemain}
\end{equation}
In the nonrelativistic limit, this definition recovers the Fourier relation between position and momentum space wavefunctions. $| \hat{k} \rangle_0$ can equivalently be expressed in terms of the field operator as,
\begin{equation}
    | \hat{k} \rangle_0 = \sqrt{2\hat{E}_{\hat{k}}} \sum_j e^{i\hat{k}j}  \hat{\phi}_j  | 0 \rangle \ \ \ .
\label{eq:freex}
\end{equation}
Wave packets formed by superpositions of single particle momentum eigenstates with weight $f_{\hat{k}}$ can then be constructed using the field operator. With,
\begin{equation}
g_{0,j} \equiv \frac{1}{\hat{L}} \sum_{\hat{k}} e^{i\hat{k}j}  f_{\hat{k}}  \sqrt{2\hat{E}_{\hat{k}}} \ \ \ ,
\label{eq:gfconnectionfree}
\end{equation}
one obtains,
\begin{equation}
\frac{1}{\hat{L}}\sum_{\hat{k}} f_{\hat{k}}  |\hat{k}\rangle_0 = \sum_j g_{0,j} \hat{\phi}_j  | 0 \rangle \ \ \  .
\label{eq:fkgphi}
\end{equation}
The field operator therefore provides a framework for constructing wave packets, particularly useful in position space quantum simulations. In this work, this construction is extended to the interacting theory by computing localized single particle excitations and constructing approximate asymptotic states directly on the interacting vacuum.

\subsection{Local field space digitization}
\label{sec:II1}

Numerical simulation of scalar field theory requires truncating the continuum local bosonic degrees of freedom into a finite-dimensional representation. Adopting the Jordan–Lee–Preskill digitization framework~\cite{Jordan:2011ci,Jordan:2012xnu}, and building on previous study of scalar field digitization~\cite{NKdigi}, this work truncates the continuum local Hilbert space to a finite-dimensional subspace spanned by eigenvectors of the local field operator. 

When the local field is digitized, all quantities associated with the lattice scalar field are denoted by a tilde. For a digitization with $d$ field values, a local field cutoff $\phi_{\max}$ is introduced, and the field-basis vectors are uniformly sampled over the interval,
\begin{equation}
    \phi_{\beta}  =  -\phi_{\max} + \beta \delta_\phi , \quad \delta_\phi=
    \frac{2\phi_{\rm max}}{d-1} \ \ \   ,
\label{eq:phidigitzed}
\end{equation}
where $\beta = 0,1,\dots,d-1$. As such, the local field operator reads,
\begin{equation}
\scalemath{0.945}{\tilde{\phi}_j :=
\left( \bigotimes_{i<j}\mathbb{I} \right)
\otimes
\mathrm{diag}\left(
\phi_{0},
\dots,
\phi_{d-1}
\right)
\otimes
\left( \bigotimes_{i>j}\mathbb{I}\right) .}
\label{eq:phiop}
\end{equation}

In the finite-difference representation of local conjugate momentum operator in the local field basis, their derivative relations manifest as hopping structures between neighboring field-basis states, e.g., $-\delta_{\phi}^2 \Tilde{\pi}^2 := \text{circulant}\left[ -2, 1, 0, 0, \ldots, 0, 1\right]$. In this work, to ensure that the eigenvalues of the local conjugate momentum operator are distributed symmetrically about zero, and improve the finite-difference representation of derivatives, the exact local conjugate momentum operator with twisted boundary conditions in field space is adopted~\cite{NKdigi}. The corresponding eigenvalues read,
\begin{equation}
    k_{\beta}= -\frac{\pi}{\delta_{\phi}} + \left(\beta + \frac{1}{2}\right)\frac{2\pi}{d \delta_{\phi}} \ \ \ .
\label{eq:pidigitzed}
\end{equation} 
The eigenbases of the local field and conjugate momentum operators are related by quantum Fourier transform,
\begin{equation}
|k_{\beta}\rangle  =  \sum_{\beta'} \frac{e^{i  \beta' k_{\beta}\delta_{\phi}}}{\sqrt{d}} |\phi_{\beta'} \rangle \ \ \ ,
\label{eq:qufutrmap}
\end{equation}
which corresponds to the unitary transformation,
\begin{equation}
    U_{\mathrm{QFT}} = \sum_{\beta} |k_{\beta} \rangle \langle \phi_{\beta}| = \sum_{\beta' \beta} \frac{e^{i  \beta' k_{\beta}\delta_{\phi}}}{\sqrt{d}}
 |\phi_{\beta'} \rangle \langle \phi_{\beta}|  \ \ \ .
 \label{eq:qufutrU}
\end{equation}
As such, in the local field basis, the local conjugate momentum operator reads~\footnote{The matrix in the local conjugate field basis may be written in terms of the local field basis as, $\sum_{\alpha \beta} m_{\alpha \beta} |\Tilde{k}_{\alpha}\rangle \langle \Tilde{k}_{\beta}| = \sum_{\alpha \beta} m_{\alpha \beta}  \sum_{\alpha'} (U_{\mathrm{QFT}})_{\alpha' \alpha} |\Tilde{\phi}_{\alpha'} \rangle \sum_{\beta'} (U_{\mathrm{QFT}})_{\beta' \beta}^* \langle \Tilde{\phi}_{\beta'}|$, where the matrix element reads, $\sum_{\alpha \beta} (U_{\mathrm{QFT}})_{\alpha' \alpha} m_{\alpha \beta} (U_{\mathrm{QFT}})_{\beta' \beta}^*$.},
\begin{equation}
\begin{aligned}
\scalemath{0.945}{|k_{\beta}\rangle = \sum_{\beta'} (U_{\mathrm{QFT}})_{\beta' \beta} |\phi_{\beta'} \rangle, \quad
\Tilde{\pi}_j :=
 \left(\bigotimes_{i<j}\mathbb{I} \right)}\\ \scalemath{0.945}{\otimes
\left(U_{\mathrm{QFT}}
\,\mathrm{diag}\!\left(
k_{0},
\dots,
k_{d-1}
\right)
U_{\mathrm{QFT}}^{\dagger}\right) \otimes
 \left(\bigotimes_{i>j}\mathbb{I}\right) } \ \ \ .
\end{aligned}
\label{eq:piop}
\end{equation}
Updating recent implementations of the local conjugate momentum operator, e.g., Ref.~\cite{NKdigi}~\footnote{An additional transposition is applied to $U_{\mathrm{QFT}}$ in Ref.~\cite{NKdigi}. Although this leaves $\tilde{\pi}_j^2$ invariant, it alters $\tilde{\pi}_j$ such that the creation and annihilation operators are exchanged, thereby affecting the preparation of free wave packets with $\hat{a}_{\hat{k}}^{\dagger}$.}, transforming a quantum state from the position to the momentum representation requires an initial inverse quantum Fourier transform, as indicated from Eqs.~\eqref{eq:qufutrmap}--\eqref{eq:piop}. 

\section{Ground state and single particle states}
\label{sec:III}

For digitized lattice scalar field theory, ground states may be computed classically using the density matrix renormalization group (DMRG) with OBC~\cite{White:1992zz,White:1993zza,schollwock2011density}, or infinite DMRG (iDMRG) to approach the thermodynamic limit~\cite{mcculloch2008infinitesizedensitymatrix}. The resulting iDMRG ground state MPS can be used to  construct the quasiparticle ansatz, through which single particle momentum eigenstates are variationally optimized. Details of these calculations are provided in Appendices~\ref{app:mpsintro1} and~\ref{app:mpsintro2}. The MPS framework allows the convergence associated with Nyquist--Shannon sampling theorem to be examined on larger interacting lattices in Sec.~\ref{sec:III1}. The optimal digitization configurations are then used for the localized excitation and wave packet preparation in Sec.~\ref{sec:III2}.

\subsection{Identification of Nyquist-Shannon sampling theorem}
\label{sec:III1}

\begin{figure*}[t]
    \centering
    \includegraphics[width=.325\linewidth]{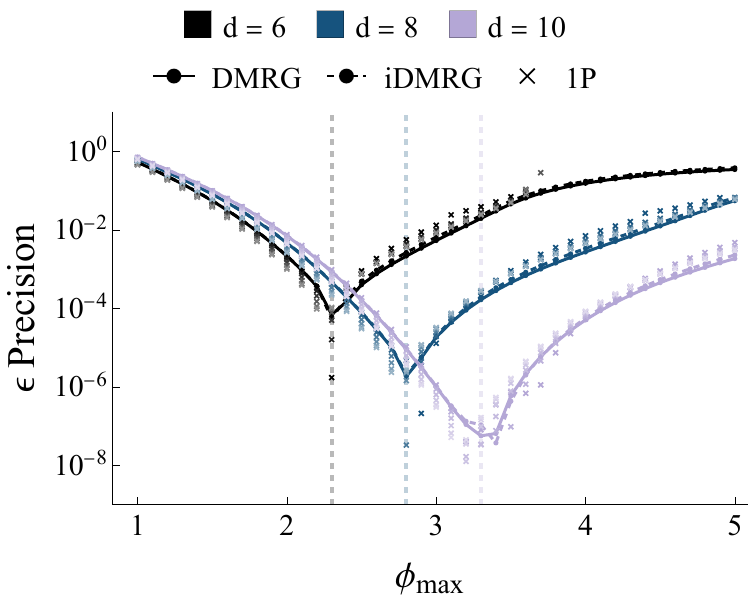}\hfill
    \includegraphics[width=.325\linewidth]{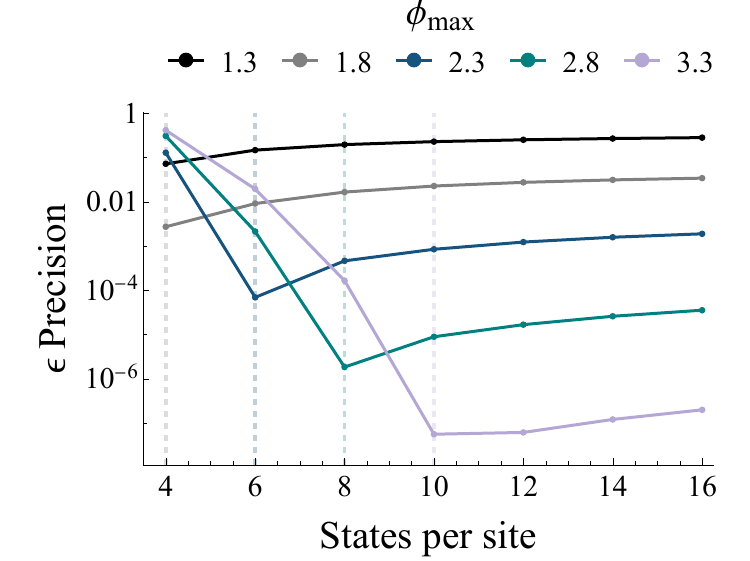}\hfill
    \includegraphics[width=.325\linewidth]{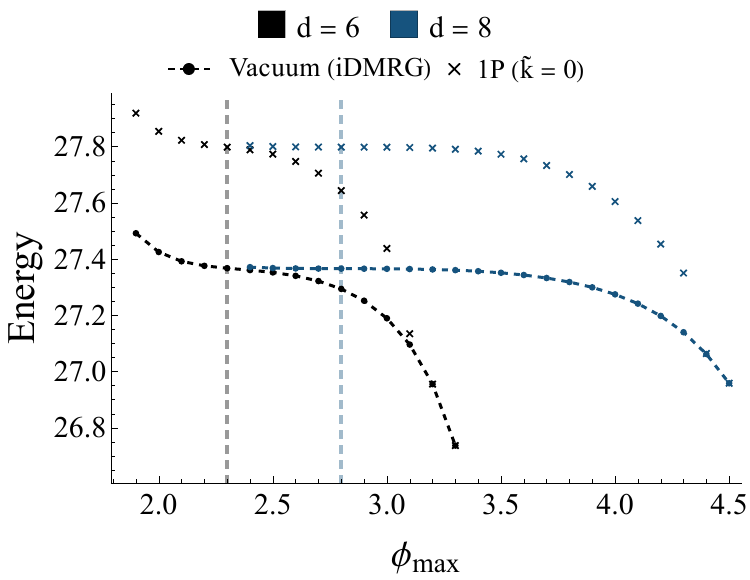}
    \caption{The precision of the energies $\epsilon$ of ground state and single particle momentum eigenstates for $\tilde{\lambda}_0=0.8$ and $\tilde{m}_0=0.1$. (Left) $\epsilon$ as a function of the local field cutoff $\phi_{\max}$ for $d=6,8,10$. DMRG and iDMRG calculations are shown with solid and dashed lines, while crosses denote single particle momentum eigenstate calculation at $\tilde{k}=0,0.6,1.2,1.8,2.4,3.0$, with lighter colors corresponding to larger momenta. The vertical dashed lines indicate saturation of the NS sampling bound for DMRG calculation at $\phi_{\max} = 2.3,2.8,3.3$ for $d=6,8,10$, with small deviations $\Delta\phi_{\max} \leq 0.1$ from the results of iDMRG and single particle momentum eigentates. (Middle) Ground state energy precision from DMRG as a function of $d$ for different $\phi_{\max}$. Vertical dashed lines in matching colors mark the corresponding saturation of NS sampling. (Right) iDMRG ground state and zero momentum single particle state energies for $d=6,8$ as a function of $\phi_{\max}$. Vertical dashed lines indicate the optimal sampling points, which remain within $\Delta\phi_{\max} \leq 0.1$ of the points where energy slopes are smallest in magnitude.}
    \label{fig:NSplot}
\end{figure*}

The Nyquist--Shannon (NS) sampling theorem guides the discretization of bosonic fields, enabling accurate calculations within a finite local Hilbert space~\cite{somma2015quantum,Macridin:2018oli,Macridin:2018gdw,NKdigi}. Studies of one and two site scalar field systems have shown that, with a suitable choice of $d$ and $\phi_{\max}$, digitization errors in low lying energy eigenstates can decrease exponentially with increasing local Hilbert space dimension $d$~\cite{NKdigi}. At a fixed $d$, the same sampling choice also gives the closest agreement with the local continuum energies. Assessing this convergence requires comparison with the local continuum energies, which are not known exactly for the interacting theory and must therefore be obtained from sufficiently converged numerical calculations as detailed in Appendix~\ref{app:mpsintro3}. This enables the subsequent analysis to be carried out for both the ground state and single particle momentum eigenstates on a periodic lattice with $\tilde{L}=41$ sites. 

Fig.~\ref{fig:NSplot} shows that the convergence observed in small systems~\cite{NKdigi} extends to larger lattices for both the ground state and single particle momentum eigenstates. These states attain their highest energy precision near a common sampling point, exhibiting exponential convergence with increasing $d$ when the field cutoff is adjusted accordingly. Remarkably, the optimal sampling points align with the locations where the energy slopes are smallest in magnitude. Extending the field range beyond these points at fixed $d$ can also lead to an artificial closing of the mass gap. For $\tilde{\lambda}_0=0.8$, $\tilde{m}_0=0.1$ and optimal $\phi_{\text{max}}$, the lattice physical mass is calculated to be $\tilde{m}_{\text{phy}}=0.43077$ for $d=6$, $\tilde{m}_{\text{phy}}=0.43238$ for $d=8$ and $\tilde{m}_{\text{phy}}=0.43223$ for $d=10$, indicating a large interaction induced correction to the bare mass. 

Estimates of leading finite volume effects indicate corrections to local expectation values on the order of $10^{-9}$, which is two orders of magnitude below the numerical precision of $d=10$. As also confirmed numerically, this suggests that the present digitization configurations agree with those on larger lattices and in the thermodynamic limit.

Lattice $\phi^4$ theories with different bare parameters but same lattice physical mass are observed to share similar optimal local field space cutoff. This observation is relevant to the application of adiabatic state preparation~\cite{Jordan:2011ci,Jordan:2012xnu}, where the interaction is gradually turned on. The physical mass can change substantially along this path, shifting the optimal local field cutoff and making the initial digitization inaccurate. Thus, adjusting the local field digitization during adiabatic state preparation or choosing an adiabatic path that maintains sampling accuracy as the interaction is turned on may be necessary.

\subsection{Localized excitation and scalable wave packet preparation}
\label{sec:III2}

\begin{figure*}[t]
    \centering
    \includegraphics[width=.325\linewidth]{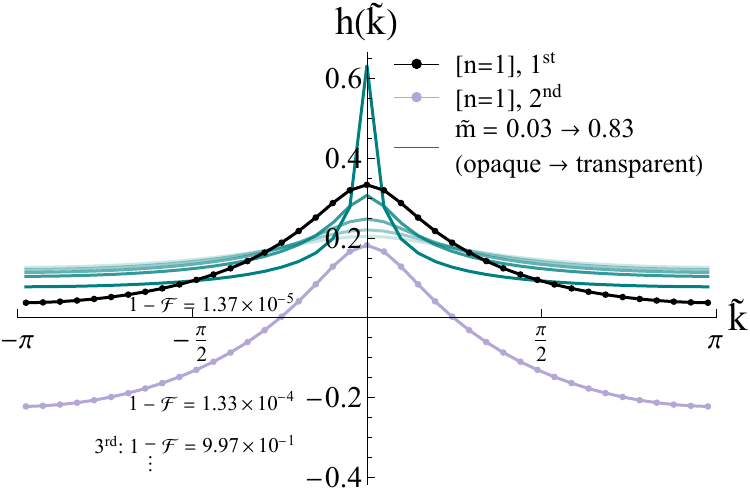}\hfill
    \includegraphics[width=.325\linewidth]{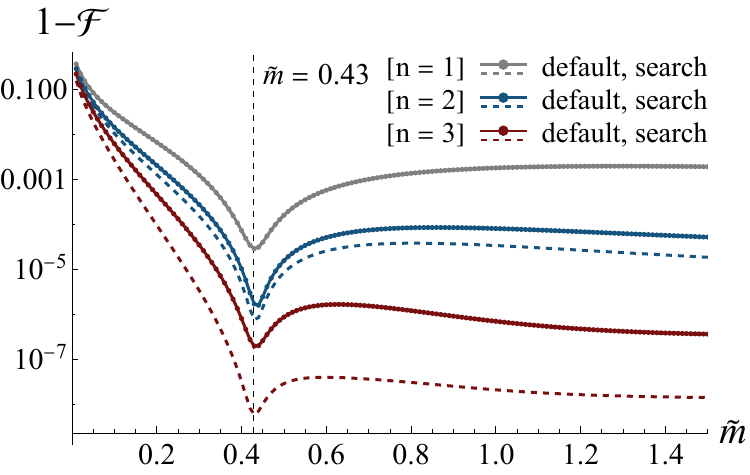}\hfill
    \includegraphics[width=.325\linewidth]{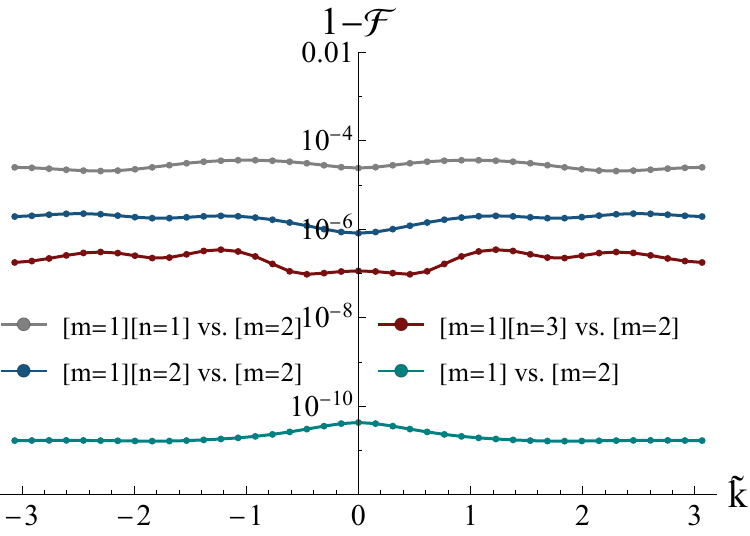}
    \caption{Localized excitation and scalable wave packet preparation for $\tilde{\lambda}_0=0.8$, $\tilde{m}_0=0.1$, $d=8$ and $\phi_{\text{max}}=2.9$. (Left) Leading and subleading eigenmodes of momentum space profile $h(\tilde{k})$, with their corresponding infidelities. Teal curves plot $h_{\tilde{m}}(\tilde{k})$ for $\tilde{m}=0.03$ to $0.83$ and $\delta \tilde{m} = 0.2$, with increasing transparency for larger masses. (Middle) Infidelity $1-\mathcal{F}$ as a function of $\tilde{m}$ for momentum space profile $h_{\tilde{m}_{\text{phy}}}(\tilde{k})$ and interpolating operators supported on $[n=1,2,3]$ sites. Solid and dashed curves denote default and search procedures for selecting the tolerance of pseudoinverse. The vertical dashed line indicates the mass parameter $\tilde{m}$ that minimizes the infidelity. The momentum space superposition profile for localized excitation preserves reflection symmetry about the origin. The $[n=2]$ case corresponds to an operator supported on the site at the origin and one adjacent site, while the $[n=3]$ case corresponds to an operator supported on three consecutive sites symmetrically arranged about the origin. (Right) Infidelity of single particle momentum eigenstates constructed using interpolating operators. Reference states are computed with two-site excitation tensors.}
    \label{fig:LEplot}
\end{figure*}

Single particle momentum eigenstates are computed with the quasiparticle ansatz on the same $\tilde{L}=41$ PBC lattice, using the optimal digitization configurations identified in Sec.~\ref{sec:III1}. Constructing localized excitations from superposing single particle momentum eigenstates first requires assessing their accuracy and fixing a consistent relative phase convention.

In the following, the single particle momentum eigenstates are denoted as $|\tilde{k}\rangle^{[m]}$, where $[m]$ is the number of consecutive lattice sites spanned by the excitation tensor. Increasing $[m]$ improves the description of single particle states, where the targeted eigenstate converges exponentially with a rate set by the spectral gaps below and above the targeted eigenvalue~\cite{PhysRevLett.111.080401}. Let $\mathcal{F}$ denotes the fidelity. The infidelity $1-\mathcal{F}$ between $[m=1]$ and $[m=2]$ in the right panel of Fig.~\ref{fig:LEplot} therefore estimates the error associated with the single site excitation, which is roughly three orders of magnitude below the numerical precision of the reported state preparation infidelity. Further details regarding the computation of arbitrary sized excitation tensor are provided in Appendix~\ref{app:mpsintro2}.

The relative phases between single particle momentum eigenstates are fixed using $\tilde{\phi}_0|\Omega\rangle$ as a common reference, where $\tilde{\phi}_0$ denotes the field operator at site $0$, and $|\Omega\rangle$ is the interacting vacuum. In the calculations below, site $0$ is chosen as the spatial origin and the reflection center of the lattice. Overall phase factor for each momentum eigenstate is calculated so that $\langle\tilde{k}|\tilde{\phi}_0|\Omega\rangle$ is real and positive. This choice is derived from a continuum Lorentz boost relation in Eq.~\eqref{eq:boost}, as discussed in Appendix~\ref{app:fieldintro1}.

Denote $\tilde{\Phi}_{j,[n]}^{[m]}$ the interpolating operator calculated from $| \tilde{k} \rangle^{[m]}$, which is centered at site $j$ and spans $n$ sites. The single site operator, $\tilde{\Phi}_{j,[1]}^{[m]}$, and the momentum space profile $h(\tilde{k})$ are jointly optimized,
\begin{equation}
\max_{\tilde{\Phi}_{j,[1]}^{[m]}, h(\tilde{k})}
\mathcal{F}
\left(
\tilde{\Phi}_{j,[1]}^{[m]} |\Omega\rangle ,
\frac{1}{\tilde{L}}
\sum_{\tilde{k}}
e^{-i\tilde{k}j}
h(\tilde{k})
| \tilde{k} \rangle^{[m]}
\right)  .
\label{eq:hphi}
\end{equation}
Without loss of generality, $j=0$ is chosen. As the $[n]$ increases, an increasing number of operator components that annihilate or nearly annihilate the vacuum render the optimization intractable. For $[n=1]$, however, the small optimization manifold avoids these flat directions in the optimization. As such, Eq.~\eqref{eq:hphi} is restricted to single site interpolating operators for numerical stability. This sets the spatial scale that defines the locality at $[n=1]$. The stability of the $[n=1]$ optimization across different numerical strategies is clearly illustrated in the middle panel of Fig.~\ref{fig:LEplot}.

While the choice $h(\tilde{k})=\frac{1}{\sqrt{2\tilde{E}_{\tilde{m}_{\text{phy}},\tilde{k}}}} \equiv h_{\tilde{m}_\text{phy}}(\tilde{k})$ is naturally motivated by the decay of correlations on the lattice and the correspondence with the free lattice scalar field, it is shown here that the same profile also emerges from the solution of Eq.~\eqref{eq:hphi}. As shown in Appendix~\ref{app:mpsintro4}, the optimization reduces to an eigenvalue problem whose eigenvalues give the preparation fidelities $\mathcal{F}$ and whose eigenvectors determine the momentum space profiles $h(\tilde{k})$. The momentum dependence of the excitation tensor enters through the matrix elements. The resulting eigenvalue spectrum is dominated by two leading eigenvalues, separated by several orders of magnitude from the remaining eigenvalues. When $\tilde{m}$ equals the lattice physical mass, $h_{\tilde{m}_{\text{phy}}}(\tilde{k})$ lies within this two dimensional subspace, with the squared norm of the component outside this subspace $7.1\times10^{-11}$. This profile is therefore used to construct the localized excitation. The left panel of Fig.~\ref{fig:LEplot} shows the $h(\tilde{k})$ eigenmodes along several choices of $h_{\tilde{m}}(\tilde{k})$. It is seen that a flat momentum space profile does not give the most accurate state that a single site operator can prepare. Note $h_{\tilde{m}_\text{phy}}(\tilde{k})$ recovers the relation between position and momentum space wavefunctions as in Eq.~\eqref{eq:freesingleparticlemain}. The localized excitation thus provides an analogue of a position eigenstate in interacting lattice field theory.

With $h_{\tilde{m}_{\text{phy}}}(\tilde{k})$, the interpolating operator can be evaluated for any $[n]$ following Appendix~\ref{app:mpsintro4}. Although the optimization procedure may encounter numerical instability and yield a suboptimal solution for $n \ge 2$, the resulting operator remains well defined. A parameter scan over $\tilde{m}$ is presented in the middle panel of Fig.~\ref{fig:LEplot}, which attains global minimum at $\tilde{m} = \tilde{m}_{\text{phy}}$ for all $[n]$, and independent of the numerical strategies. This suggests that the choice of $h_{\tilde{m}_{\text{phy}}}(\tilde{k})$ can extend beyond Eq.~\eqref{eq:hphi}, serving as a natural choice for localized excitation of interacting lattice scalar field theories. As the operator support increases from one to three sites, the infidelity decreases rapidly, motivating the conjecture that it decays exponentially with $[n]$.

Single particle momentum eigenstates may thus be constructed by,
\begin{equation}
    | \tilde{k} \rangle \sim | \tilde{k} \rangle_{[n]}^{[m]} =  \sqrt{2\tilde{E}_{\tilde{m}_{\text{phy}},\tilde{k}}} \sum_j e^{i\tilde{k}j} \tilde{\Phi}_{j,[n]}^{[m]}  | \Omega \rangle \ \ \ .
\label{eq:kint}
\end{equation}
The resulting localized excitation may be then superposed to construct wave packets with prescribed momentum space distributions. With,
\begin{equation}
g_j = \frac{1}{\tilde{L}} \sum_{\tilde{k}}  e^{i\tilde{k}j} f_{\tilde{k}}\sqrt{2\tilde{E}_{\tilde{m}_{\text{phy}},\tilde{k}}} \ \ \  ,
\label{eq:gfconnection}
\end{equation}
one obtains,
\begin{equation}
\scalemath{0.93}{\frac{1}{\tilde{L}} \sum_{\tilde{k}} f_{\tilde{k}}  |\tilde{k}\rangle \sim \frac{1}{\tilde{L}} \sum_{\tilde{k}} f_{\tilde{k}}  |\tilde{k}\rangle_{[m]}^{[n]} =  \sum_j  g_j \tilde{\Phi}_{j,[n]}^{[m]}  | \Omega \rangle}   \ \ \ ,
\label{eq:kwp}
\end{equation}
which enables the preparation of states with arbitrarily large spatial extent in the interacting theory. Eqs.~\eqref{eq:kint}--\eqref{eq:kwp} thus provide the interacting theory analog of Eqs.~\eqref{eq:freex}--\eqref{eq:fkgphi} for the free theory.

To estimate the wave packet infidelity that arises from finite spatial support of $[m]$ and $[n]$, one may first examine the infidelity of $|\tilde{k}\rangle_{[m]}^{[n]}$, as demonstrated by the explicit relations between $f_{\tilde{k}}$ and $g_j$ in Eqs.~\eqref{eq:kint}--\eqref{eq:kwp}. Leveraging the exponential convergence in $|\tilde{k}\rangle^{[m]}$~\cite{PhysRevLett.111.080401}, the infidelity of $\vert{}\tilde{k}\rangle_{[m]}^{[n]}$ is here benchmarked against the $[m=2]$ reference state. The right panel of Fig.~\ref{fig:LEplot} shows that the infidelity remains the same order of magnitude across $\tilde{k}$. Since preparation errors in different momentum sectors are orthogonal and the superposition is normalized, the wave packet has infidelity at a similar level. 

An approximate two particle asymptotic incoming state is then constructed by applying the same procedure a second time,
\begin{equation}   |\tilde{\psi}\rangle_{\text{in},[n]}^{[m]}
= \left( \sum_{j'}  g'_{j'} \tilde{\Phi}_{{j'},[n]}^{[m]} \right)  \left( \sum_j  g_j \tilde{\Phi}_{j,[n]}^{[m]} \right)  | \Omega \rangle ,
\label{eq:instate}
\end{equation}
where $g_j$ and $g'_{j'}$ are chosen such that the two wave packets are separated by a large distance. For example, the peaks of the two wave packets are separated by approximately 60 lattice sites (26 Compton wavelengths) for scattering simulations with $\tilde{\lambda}_0=0.8$ and $\tilde{m}_0=0.1$ in Sec.~\ref{sec:IV2}. The preparation errors of the wave packets are approximately independent, as follows from cluster decomposition.

With small $g_j$ truncated, the superposed interpolating operators are applied to the vacuum leveraging MPS compression~\cite{PhysRevLett.93.040502,PhysRevLett.93.207204,schollwock2011density}. The detailed procedures are given in Appendix~\ref{app:mpsintro5}. The implementation error is measured by the distance between the resulting MPS and the target state. Converting this distance into an infidelity shows that the MPS implementation error is roughly three orders of magnitude smaller than the finite support infidelity of $[n=3]$, and can therefore be neglected in the leading error estimate.

For a total infidelity $\epsilon$, general bounds on errors in expectation values and the Renyi-2 entropy scale as $\mathcal{O}(\sqrt{\epsilon})$. These bounds can be conservative, and errors in local observables such as the energy density may be substantially smaller in practice.

\section{Real-time simulation of scattering in \texorpdfstring{$\phi^4$}{phi4} theory}
\label{sec:IV}

The relative phase convention for single particle momentum eigenstates provides a common reference to scale $g_j$ or $f_{\tilde{k}}$ toward the continuum limit. This scaling is detailed in Appendix~\ref{app:scatteringcountlim}. With this prescription, Sec.~\ref{sec:IV1} examines the behavior of single particle Airy distortions. The conditions for suppressing these oscillations, as derived from their asymptotic form, hold increasingly well as the continuum limit is approached.

Projections of the late time state onto specified particle sectors~\cite{Milsted:2020jmf,Jha:2024jan} can help resolve individual scattering channels. When properly resolved and interpreted, additional structures in the excess energy density or cut entanglement entropy can also help identify inelastic scattering~\cite{Zemlevskiy:2026kpc}. As shown in this section, lattice and local field digitization artifacts can produce similar effects in both observables, complicating the identification of physical inelastic processes. Sec.~\ref{sec:IV2} examines how local field space digitization affects the excess energy density and cut Renyi-2 entropy. Establishing the presence of these artifacts, this work supports the ongoing development of detailed analyses for physical inelastic channels. 

\subsection{Airy distortion}
\label{sec:IV1}

Lattice induced single particle distortions can be analyzed through the time evolution of $g_{j}$. As shown in Fig.~\ref{fig:prepandscatterdiagram} and Appendix Fig.~\ref{fig:Adistortion}, the single particle dispersion relation in the $\phi^4$ theory is well approximated by Eq.~\eqref{eq:freedispersion} with the bare mass replaced by the lattice physical mass, consistent with earlier lattice Monte Carlo results~\cite{Rummukainen:1995vs,Guo:2018xbv}. In the $\tilde{L}\to\infty$ limit,
\begin{equation}
\scalemath{0.9}{g_j\left(t\right)  \simeq \int_{-\pi}^{\pi}\frac{d\tilde{k}}{2\pi} e^{i\left(\tilde{k}\frac{j}{t} -\tilde{E}_{\tilde{m}_{\text{phy}},\tilde{k}}  \right)t} e^{-\frac{(\tilde{k}-\tilde{\mu}_{\tilde{k}})^2}{2\tilde{\sigma}_{\tilde{k}}^2}} \sqrt{2\tilde{E}_{\tilde{m}_{\text{phy}},\tilde{k}}}} \, ,
\label{eq:positionprofiledynamics}
\end{equation}
where $\tilde{\mu}_{\tilde{k}}$ and $\tilde{\sigma}_{\tilde{k}}$ are the mean and standard deviation. The lattice and continuum group velocities read, 
\begin{equation}
\tilde{v}_{\tilde{m},\tilde{k}}= \frac{\sin\tilde{k}}{\sqrt{\tilde{m}^2 + 4 \sin^2 \frac{\tilde{k}}{2}}}, \quad \tilde{v}_{\tilde{m},\tilde{k}}^{\text{cont}} = \frac{\tilde{k}}{\sqrt{\tilde{m}^2 + \tilde{k}^2}}  \, .
\label{eq:derivativesofE1}
\end{equation}
Unlike the continuum dispersion, the lattice dispersion allows two distinct momenta to share the same group velocity, as shown in Appendix Fig.~\ref{fig:Adistortion}.

The propagation, broadening, and Airy distortion of the wave packet can be understood through the approximations of Eq.~\eqref{eq:positionprofiledynamics} developed in Appendix~\ref{app:latartifact}, where the conditions and validities are discussed in detail. At early times, or when $f_{\tilde{k}}$ is sufficiently narrow, the wave packet propagates and broadens while retaining an approximately Gaussian profile. When these conditions are not satisfied, contributions from stationary phase momenta, i.e., $\tilde{k}_{\pm}$ with $\tilde{v}_{\tilde{m}_{\text{phys}},\tilde{k}_{\pm}}=j/t$ become increasingly dominant~\cite{Copson_1965,NIST:DLMF}. In this regime, the wave packet profile may be described by an Airy function and its derivative~\cite{chester1957extension,bleistein1966uniform,NIST:DLMF}. 

The analytical expressions derived in Appendix~\ref{app:latartifact} give conditions for suppressing Airy distortions during wave packet propagation. Requiring higher order terms small in the expansion of phase is one way to preserve the Gaussian approximation. Balancing the contributions from the Airy function and its derivative is another way to cancel the oscillations. Intuitively, this cancellation requires the Gaussian momentum profile to favor one stationary point while carrying negligible weight on the other. The broadening of the wave packet must also remain small so that its spatial width does not significantly exceed the oscillation wavelength. Appendix Fig.~\ref{fig:Adistortion} compares evaluations of Eq.~\eqref{eq:positionprofiledynamics} using continuum and lattice dispersion relations, together with analytical approximations for the latter. For the wave packet dynamics considered in Sec.~\ref{sec:IV2}, it is shown that Airy distortions are significantly suppressed. As illustrated in the figure and predicted by the oscillation suppression conditions, Airy distortions weaken as the continuum limit is approached and disappear in the continuum, where $\tilde{v}_{\tilde{m},\tilde{k}}^{\text{cont}}$ is monotonic in $\tilde{k}$.

\subsection{Digitization induced inelastic processes}
\label{sec:IV2}

\begin{figure*}
 \begin{minipage}{1\textwidth}
 \includegraphics[width = \textwidth]{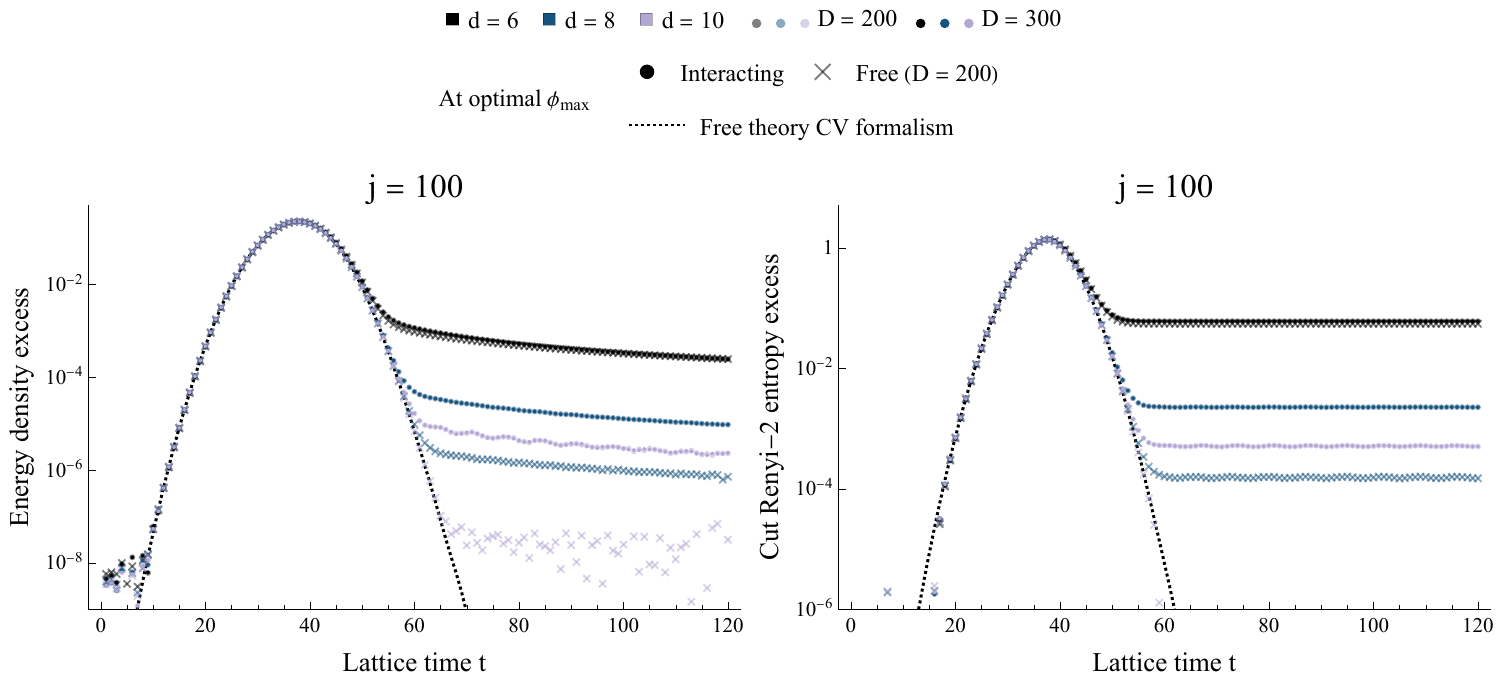}
  \end{minipage}\\
\caption{Free theory MPS simulation with $\tilde{m}_{0} = 0.43223$ and comparison with interacting theory $\tilde{m}_0 = 0.1$,  $\tilde{\lambda}_0 = 0.8$. Time evolution of (Left) energy density excess and (Right) cut Renyi-2 entropy excess at site $j=100$ are shown. Colors distinguish local Hilbert-space dimensions $d=6,8,10$. Dots show interacting theory MPS simulation with bond dimensions $D=200$ (lighter) and $D=300$ (darker), while crosses show free theory MPS simulation with $D=200$. The dotted curves show free theory calculations from the CV formalism. Same momentum space wave packet profiles are adopted for all scattering setups.}
\label{fig:j100scatterplot}
\end{figure*}

Following the procedure in Sec.~\ref{sec:III2}, $\Phi$ is computed on a lattice with $\tilde{L}=41$ PBC, and with $[n=3]$ using the default pseudoinverse. This operator is then used to prepare two well separated incoming wave packets on an $\tilde{L} = 200$ OBC chain, with central momenta $\pm\tilde{\mu}_{\tilde{k}}=\pm0.34\pi$ and momentum width $\tilde{\sigma}_{\tilde{k}}=0.06\pi$. The total energy at $\tilde{\mu}_{\tilde{k}}$ reads $E_{\text{total}}=2.21$, which is estimated from the free theory lattice dispersion relation using the physical mass. This is above the corresponding four-particle threshold $4\tilde{m}_{\text{phy}} = 1.73$. To study scattering in this regime, the simulations shown in Figs.~\ref{fig:j100scatterplot}, \ref{fig:freescatterplot}, and~\ref{fig:UVIRscatterplot} use the NS sampling parameters identified in Sec.~\ref{sec:III1}. Details of the subsequent MPS time evolution are provided in Appendix~\ref{app:mpsintro6}. Appendix~\ref{app:scatteringcountlim} discusses the observables, boundary effects, and the effects of decreasing the lattice spacing. It also includes figures illustrating these effects on the scattering observables at selected times.

Scattering channels may be examined by projecting the evolved state onto reference subspaces of well-separated quasiparticles~\cite{Milsted:2020jmf,Jha:2024jan,Belyansky:2023rgh}. In Fig.~\ref{fig:j100scatterplot}, after the dominant outgoing wave packets leave the collision center, the central excess energy remains nonzero and approaches a plateau over the simulated interval. This persistent remnant provides suggestive evidence for inelastic contributions, since the central excess energy significantly exceeds the contribution from the kinematically allowed asymptotic two particle outgoing states.

Coarse field sampling enhances the excess energy density between the outgoing wave packets. As shown in Fig.~\ref{fig:j100scatterplot} and Appendix Fig.~\ref{fig:freescatterplot}, similar inelastic signals appear in the digitized free theory, despite the absent of particle production in the underlying theory. As the field sampling is refined, the free theory MPS calculations approach the CV calculations, with diminishing inelasticity. This not only provides a cross check on the preparation and evolution methods present in this work, but supports the interpretation of these inelastic signals as digitization artifacts. The free theory CV formalism is elaborated in detail in Appendix~\ref{app:freefield}.

At $d=10$, the interacting theory shows excess of energy density and cut Renyi-2 entropy, significantly above that in the free theory, with latter hits the floor of numerical precision. These excesses are also shown in Appendix Figs.~\ref{fig:freescatterplot} and~\ref{fig:UVIRscatterplot}, which is conjectured to arise from four-particle final states with different momentum configurations. This gives rise to a complex spatial profile that are not yet fully understood. For example, the excess in Fig.~\ref{fig:prepandscatterdiagram} forms two outgoing peaks. These peaks are less developed in Appendix Fig.~\ref{fig:freescatterplot}, while additional shoulders appear near the wave packet tails in Appendix Fig.~\ref{fig:UVIRscatterplot}.

\section{Summary and Outlook}
\label{sec:IVI}
\vspace{0.2cm}

Simulating scatterings of fundamental particles on a quantum computer is expected to offer exponential advantages over classical methods, potentially accelerating the search for new physics. To accurately model full scattering processes and capture the emergence of nonlocal phenomena, it is essential to prepare extended physical sized wave packets. However, directly optimizing the entire wave packet can become increasingly intractable. This work is motivated by recently developed interacting wave packet construction~\cite{Davoudi:2024wyv,Davoudi:2025rdv}, where excitations are constructed directly with the Haag–Ruelle scattering framework~\cite{PhysRev.112.669,ruelle1962asymptotic,Duncan:2012aja,PRXQuantum.5.020311}. Rather than optimizing a prescribed operator ansatz whose support extends across the wave packet, this work directly identifies the local operator building blocks for wave packet preparation. This allows a quantum–classical implementation in which $\Phi$ is calculated classically to inform the design of an operator capable of creating a wave packet in the vacuum for a quantum simulation of size inaccessible to classical computers.

When utilizing the interpolating operator for a localized excitation to design a suitable circuit for wave packet creation in quantum simulations of quantum fields~\cite{Morgavi:2026elf}, this hybrid approach holds opportunity to systematically reduce quantum circuit depth. By omitting rotations with small angles, previous studies of state preparation and time evolution have reduced circuit depth while maintaining high fidelity~\cite{NKdesign,Davoudi:2024wyv,Davoudi:2025rdv,Ciavarella:2024fzw}. This motivates similar strategies for implementing nonunitary $\Phi$ and its weighted sums. The explicit form of $\Phi$ may enable systematic circuit transformations that reveal a hierarchy of suppressed contributions, analogous to the construction in Ref.~\cite{NKdesign}. Developing such transformations will likely require identifying and leveraging additional structures in $\Phi$. 

Approaches to characterizing scattering outcomes in quantum simulations include measuring return probabilities~\cite{Davoudi:2025rdv}, tracking outgoing wave packet profiles~\cite{Zemlevskiy:2026kpc}, and local detection through reduced density matrix overlaps~\cite{Morgavi:2026elf}.  Inspired from the present work and the classical calculations of Refs.~\cite{Milsted:2020jmf,Jha:2024jan,Belyansky:2023rgh}, the explicit form of $\Phi$ could also motivate a building block for detecting outgoing particles in quantum simulations. This may be achieved by optimizing a new operator, $\Psi$, with the desired annihilation properties on the vacuum and single particle states. The corresponding channel identification could then be expressed as expectation values of observables constructed from $\Psi$.

The localization considered here is based on optimizing how faithfully the state corresponding to a momentum profile can be prepared using a single site operator. Beyond its practical applications, this definition is also of theoretical interest, as the optimization leads to an analogue of a position eigenstate in the interacting theory. This raises the question of whether other localization criteria~\cite{PhysRevResearch.3.013078,Morgavi:2026elf} yield a similar analogue or whether this feature is specific to the definition adopted in this work. It would also be interesting to further explore the distinctive properties of the momentum space superposition profile that gives the highest fidelity in Eq.~\eqref{eq:hphi}.

The incoming wave packets considered here have a central total energy above four times the physical mass, making the $2\to4$ channel kinematically accessible. The residual floor in the excess energy density, which persists after the dominant outgoing packets have separated, is conjectured to reflect slowly moving particles produced in higher multiplicity channels. The structure of this remnant warrants further investigation. Perturbative calculations based on Feynman diagrams~\cite{Feynman:1949zx,Peskin:1995ev,weinberg1995quantum} may offer initial insight or guide a detection procedure that isolates inelastic channels. Meanwhile, Fig.~\ref{fig:j100scatterplot} and Appendix Fig.~\ref{fig:UVIRscatterplot} indicate that extending the current simulations to $d=12$ will likely be necessary to reliably identify physical inelastic channels in $\phi^4$ theory.

The small energy slopes between configurations near the optimal sampling points suggest a practical strategy for calculating the recommended local field digitization, although the reliability of this criterion requires further analytical justification. If validated, minimizing $|\partial E/\partial\phi_{\max}|$ could guide the search of the NS sampling configuration in the interacting theory, complementary to the present techniques of Appendix~\ref{app:mpsintro3}. 

The observed digitization artifacts also offer an opportunity to study how local field resolution affects quantum correlations in scattering. It is shown that increasing $d$ reduces the excess entanglement entropy. Complementing the study of vacuum entanglement and its sensitivity to the spatial latticization~\cite{NKentsphere}, this work reveal a dynamical connection between local field resolution and long distance quantum correlations.

The structure of MPS can guide quantum circuit design~\cite{Schoen:2005nnw,Smith:2022nbd,Smith:2024nbh}. Recently, efficient and systematic calculations have been reported for arbitrary sized excitation tensors~\cite{Osborne:2024oqg}. Relating this computational cost to state preparation circuits, e.g., with variational energy minimization~\cite{farrell2026digital}, may help estimate how the quantum resources required for scattering simulations scale toward the continuum limit.

\begin{acknowledgments}
The authors thank Roland C. Farrell, Nikita A. Zemlevskiy for valuable and insightful discussions. The authors also thank Tom Milledge, Edward McLaurin III, and the Duke Computing Cluster (DCC) for computational resources and technical support. The MPS calculations were performed on CPU nodes of the DCC over a two month production period using independently developed code written in Julia~\cite{doi:10.1137/141000671}. This work was supported in part by the NSF STAQ Program PHY-2325080 and National Science Foundation CAREER award number PHY-2541830.
\end{acknowledgments}

\bibliography{biblio}

\onecolumngrid
\newpage
\appendix

\section{Lattice scalar field and the continuum limit}
\label{app:fieldintro}

This appendix introduces the lattice formulation of scalar field theory, focusing on the continuum limit and the calibration of single particle momentum eigenstates. To make the relation between the lattice and continuum theories explicit, the lattice spacing is first retained in the Hamiltonian before the theory is rewritten in lattice unit for numerical computation. This explicitly helps clarify the the connection between lowest lying momentum eigenstates on the lattice, as further illustrated in Appendix~\ref{app:fieldintro1}, as well as different scattering schemes considered in this work.

The discretized Hamiltonian for 1+1 dimensional real scalar field is~\cite{Montvay:1994cy,Gattringer:2010zz,Jordan:2011ci,Jordan:2012xnu,NKdigi},
\begin{equation}
H = a \sum_{j}^{} \left[ \frac{1}{2}\pi_j^2 + \frac{1}{2} m_{0}^2 \phi_j^2 + \frac{1}{2}\left(\frac{\phi_{j+1}-\phi_j}{a}\right)^2 + \frac{\lambda_0}{4!}\phi_j^4 \right] \ \ \ ,
\label{eq:Hamiltonianwitha}
\end{equation}
with j integer and $a$ the lattice spacing. The discretized field operators are defined as $\phi_j := \phi(ja)$, $\pi_j := \pi(ja)$, and the canonical commutation relations (CCR) reads $[\phi_i, \pi_j] = \frac{i}{a} \delta_{ij}$, $[\phi_i, \phi_j] = [\pi_i, \pi_j] = 0$. With PBC, the length of the lattice with $\hat{L}$ sites is given by $L = \hat{L} a$, and the allowed momenta reads, $k = \frac{2\pi n}{\hat{L}a}$, where integer $n \in (-\frac{\hat{L}}{2},\frac{\hat{L}}{2}]$. The momentum spacing reads, $\delta k=\frac{2\pi}{\hat{L}a}$, and maximum momentum $k_{\mathrm{\text{max}}}=\frac{\pi}{a}$. 

In the free theory, the dispersion relation is given by,
\begin{equation}
    E_{m_0,k} \equiv E_k = \sqrt{m_0^2 + \frac{4}{a^2}\sin^2{\left(\frac{ak}{2}\right)}} \ \ \ ,
\label{eq:Ekwitha}
\end{equation}
where the continuum dispersion relation is recovered in the limit $a\to0$. The spectrum of free theory Hamiltonian can be solved exactly upon identifying the creation and annihilation operators $a_k^{\dagger}$ and $a_k$,
\begin{equation}
    a_k^{\dagger} = a \sum_j e^{ikaj} \left(\sqrt{\frac{E_k}{2}} \phi_j - i \sqrt{\frac{1}{2E_k}} \pi_j\right) \ \ \ ,
\label{eq:akwitha}
\end{equation}
where $[a_p, a_k^{\dagger}] = L \delta_{p,k}$, $[a_p, a_k] = 0$. The free theory Hamiltonian thus reads,
\begin{equation}
    H_{0} = \sum_k E_k \left( \frac{1}{L} a^{\dagger}_k a_k + \frac{1}{2}  \right) \ \ \ ,
\label{eq:H0withaa}
\end{equation}
and the field and conjugate field operators can thus be expressed as,
\begin{align}
\phi_j &= \frac{1}{L} \sum_k \sqrt{\frac{1}{2 E_k}} \left( a_k e^{i k a j} + a_k^\dagger e^{-i k a j} \right) \ \ \ , 
\label{eq:phiwitha} \\
\pi_j &= \frac{1}{L} \sum_k (-i) \sqrt{\frac{E_k}{2}} \left( a_k e^{i k a j} - a_k^\dagger e^{-i k a j} \right)  \ \ \ .
\label{eq:piwitha}
\end{align}
As $a \to 0$, $a \sum_j \to \int dx$, and as $L \to \infty$, $[a_p, a_k^{\dagger}]  = L \delta_{p,k} = \left(2 \pi \right) \frac{\delta_{p,k}}{\delta k} \to \left(2 \pi \right)\delta\left(p-k\right)$ and $\frac{1}{L} \sum_k = \sum_k \frac{\delta k}{2 \pi} \to \int\frac{dk}{2\pi}$. With $|0\rangle$ denoting the free vacuum, $a_{k}|0\rangle = 0$, the single particle momentum eigenstates in the free theory are,
\begin{equation}
| k \rangle_0 = a_k^\dagger | 0 \rangle, \quad _0 \langle p |k \rangle_0 = L \delta_{p,k} \ \ \ .
\label{eq:freesingleparticle}
\end{equation}
The single particle wave packet is given by,
\begin{equation}
|\psi\rangle_{1P}
= \frac{1}{L} \sum_k f_{k} |k\rangle, \quad \frac{1}{L} \sum_k |f_{k}|^2 = 1, \quad _{1P} \langle \psi | \psi\rangle_{1P}  = 1   \ \ \ ,    
\label{eq:wpnormalization}
\end{equation}
such that $f_{k}$ becomes a normalized distribution in the continuum limit (c.f. Ref.~\cite{Peskin:1995ev}). 

By introducing the dimensionless variables $\hat{\phi}:=\phi$, $\hat{\pi}:=a\pi$, $\hat{H}:=aH$, $\hat{m}_0:=a m_0$, and $\hat{\lambda}_0:=a^{2}\lambda_0$, Eq.~\eqref{eq:Hamiltonianwitha}--Eq.~\eqref{eq:wpnormalization} may be rewritten in unit of the lattice spacing, yielding Eq.~\eqref{eq:Hamiltonianmain}--Eq.~\eqref{eq:wpnormalizationmain} in the main text. With PBC, the allowed lattice momenta reads $\hat{k} := a k = \frac{2\pi n}{\hat{L}}$. Note that the corresponding dimensionless ladder operator satisfies $\hat{a}_{\hat{k}}^{\dagger} = \frac{1}{\sqrt{a}} a_{k}^{\dagger}$, such that $| \hat{k} \rangle_0 = a^{-\frac{1}{2}}| k \rangle_0$. Toward the continuum limit, the dimensionless lattice quantities become increasingly small as the ultraviolet cutoff is pushed to increasingly high energy scales and each lattice site becomes responsible for resolving physics at increasingly shorter distances.

\subsection{Calibration of single particle momentum eigenstates}
\label{app:fieldintro1}

Unlike the free theory where analytic solution exists, to construct localized excitations from superpositions of interacting single particle momentum eigenstates, the relative phases of the individual momentum eigenstates must first be fixed. This section introduces the additional conditions required to ensure phase consistency when superposing eigenstates with different momenta.

In the continuum, single particle momentum eigenstates with different momenta are related by Lorentz boosts, while the vacuum remains invariant~\cite{weinberg1995quantum,Peskin:1995ev},
\begin{equation}
U(\Lambda_k)\sqrt{2E_{k=0}}|k=0\rangle =  \sqrt{2E_{k}}|k\rangle, 
\quad 
U(\Lambda_k)|\Omega\rangle = |\Omega\rangle, \quad a \to 0, \quad L \to \infty \ \ \ .
\label{eq:boost}
\end{equation}
Here, $E_k$ denotes the interacting single particle energy, which in the continuum takes the free theory form with the bare mass replaced by the physical mass. This can be equivalently characterized by the matrix element of local field operators between vacuum and single particle states,
\begin{equation}
\sqrt{2E_k}\langle k|\phi_0|\Omega\rangle =
\sqrt{2E_{k=0}}\langle k=0|
U^\dagger(\Lambda_k)
\phi_0
U(\Lambda_k)
|\Omega\rangle =
\sqrt{2E_{k=0}}\langle k=0|\phi_0|\Omega\rangle, \quad a \to 0 , \quad L \to \infty  \ \ \ ,
\label{eq:phaseconventioncont}
\end{equation}
where in the continuum limit, $\phi_0$ is Lorentz invariant. Note the matrix element considered here is non-vanishing~\cite{Duncan:2012aja,PRXQuantum.5.020311}. By fixing the phases of the vacuum state and the single particle momentum eigenstates at $k=0$, the matrix element $\langle k=0|\phi_0|\Omega\rangle$ is well defined and can be chosen to be real and positive. All other single particle momentum eigenstates can then be calibrated through Eq.~\eqref{eq:phaseconventioncont}.

Therefore, to ensure a smooth continuum limit, the phase ambiguity of the lattice single particle momentum eigenstates is fixed by requiring,
\begin{equation}
    \langle \hat{k} | \hat{\phi}_0 | \Omega \rangle \in \mathcal{R}_{>0},
    \quad \forall \hat{k} \ \ \ .
\label{eq:phaseconvention}  
\end{equation}
Writing in the physical unit, $\langle \hat{k} | \hat{\phi}_0 | \Omega \rangle \sim a^{-\frac{1}{2}} \langle k | \phi_0 | \Omega \rangle $~\footnote{This can be seen from dimensional analysis, and may be derived in the free theory where $_0 \langle \hat{k}|\hat{\phi}_0|0\rangle = \sqrt{\frac{1}{2\hat{E}_{\hat{k}}}}$.}. The reflection operator $\mathbb{R}$ maps a momentum eigenstate to one with opposite momentum up to a phase. The phase convention in Eq.~\eqref{eq:phaseconvention} fixes this phase so that $\mathbb{R}|\hat{k}\rangle=|-\hat{k}\rangle$. This is aligned with the adiabatic evolution~\cite{Jordan:2011ci,Jordan:2012xnu}, by noting that Eq.~\eqref{eq:phaseconvention} is satisfied by the analytically known single particle momentum eigenstates in the free theory, and the scalar field Hamiltonian preserves reflection symmetry. On the other hand, from the definition of translation operator, the complex conjugation of the single particle momentum eigenstates also reverses their direction up to a phase. The phase convention in Eq.~\eqref{eq:phaseconvention} fixes this phase so that $|\hat{k}\rangle=|-\hat{k}\rangle^{*}$. In small systems, the adiabatic evolution from analytically known free theory is numerically seen to preserves this relation in the adiabatic limit.

It is worth emphasizing that the calibration basis state, $\hat{\phi}_0 | \Omega \rangle$, adopted in Eq.~\eqref{eq:phaseconvention} is physically motivated. In contrast, the approach outlined in Refs.~\cite{PhysRevResearch.3.013078,Morgavi:2026elf} resolves the phase ambiguity automatically in the optimization, or by choosing an arbitrary calibration basis state that has nonzero overlap with all $| \hat{k} \rangle$ and calibrate the phase such that the overlapping amplitudes are real~\cite{Belyansky:2023rgh,farrell2026digital}. In this prescription, however, single particle momentum eigenstates will acquire relative phases that are calibration basis state dependent, which can become inconsistent with Eq.~\eqref{eq:boost} in the continuum. As an example, consider first resolving the phase ambiguity using Eq.~\eqref{eq:phaseconvention}, which leads to $|\hat{k}\rangle = |-\hat{k}\rangle^{*}$. One may resolve the phase ambiguity again using, e.g., a computational basis state with complex overlapping amplitude, where in order for the selected component to be real, a nonvanishing relative phase between $|\hat{k}\rangle$ and $|-\hat{k}\rangle$ generally needs to be enforced.

Finally, note that in order for this phase calibration, $\phi$ must be located at position zero. As such, when calculating single particle momentum eigenstates in the following sections, this work chooses the system that respects reflection symmetry, ensuring that $\phi_0$ is well defined.

\section{MPS formalism for scattering in lattice scalar field}
\label{app:mpsintro}

This appendix elaborates on the MPS framework used to target ground states, single particle momentum eigenstates, localized single particle excitations, incoming wave packets at large separation, and time evolution. With this framework, real-time scattering in interacting scalar field theory is studied, as also discussed in Sec.~\ref{sec:IV}. The methods introduced in this appendix are cross validated with the phase space formulation of the free lattice scalar field theory, as introduced in Appendix~\ref{app:freefield}. The calculations are carried out for lattice scalar field theories with digitized local field degrees of freedom, as introduced in Sec.~\ref{sec:II1}.

\subsection{Determining the ground state}
\label{app:mpsintro1}

This section overviews the techniques used in this work for the variational ground state optimization with local field space digitization.

The first step in applying DMRG to scalar field theory is to construct the matrix product operator (MPO) representation of the lattice scalar field Hamiltonian, which is a series of operator valued matrix, $\tilde{W}_{bb'} = \sum_{\tilde{\phi} \tilde{\phi}'} W_{bb'}^{\tilde{\phi} \tilde{\phi}'} |\tilde{\phi}\rangle \langle\tilde{\phi}'|$. With local field space digitization and open boundary conditions, Eq.~\eqref{eq:Hamiltonianmain} for a system of $\tilde{L}$ lattice sites may be written in terms of the MPO as,
\begin{equation}
\tilde{W}^{[1]}=
\begin{pmatrix}
\tilde{H}^{[1]} & -\tilde{\phi}_1 & I
\end{pmatrix}, \quad \tilde{W}^{[j]}=
\begin{pmatrix}
I & 0 & 0 \\
\tilde{\phi}_j & 0 & 0 \\
\tilde{H}^{[j]} & -\tilde{\phi}_j & I
\end{pmatrix},  \quad \tilde{W}^{[{\tilde{L}}]}=
\begin{pmatrix}
I \\
\tilde{\phi}_{\tilde{L}} \\
\tilde{H}^{[{\tilde{L}}]} 
\end{pmatrix}, \quad j = 2,\dots,\tilde{L}-1 \ \ \ ,
\label{eq:MPO}
\end{equation}
where,
\begin{equation}
    \tilde{H}^{[j]} = (1+\frac{\tilde{m}_0^2}{2})\tilde{\phi}^2_j + \frac{1}{2} \tilde{\pi}^2_j + \frac{\tilde{\lambda}_0}{4!}\tilde{\phi}^4_j , \quad j = 1,\dots,\tilde{L} \ \ \ .
\end{equation}
It is easy to verify that,
\begin{equation}
    \tilde{H}_{OBC} = \sum_{j=1}^{\tilde{L}-1} \left[\tilde{H}^{[j]} - \tilde{\phi}_j \tilde{\phi}_{j+1} \right] + \tilde{H}^{[\tilde{L}]} = \prod_{j=1}^{\tilde{L}} \tilde{W}^{[j]} \ \ \ ,
\label{eq:obcH}
\end{equation}
where the contractions between the MPOs are performed on the vertual index, $b$. As such, the MPO bond dimension of the lattice scalar field Hamiltonian is $D_{\tilde{W}}=3$.

With Eq.~\eqref{eq:MPO}, the DMRG procedure can be performed by iteratively~\cite{White:1992zz,White:1993zza,schollwock2011density}. In this work, starting from the two site Hamiltonian and the corresponding ground state, the MPS chain is grown iteratively by inserting a two site MPO into the energy expectation value and optimizing the corresponding two site MPS tensor such that the ground state energy is minimized. A single sweep is used to refine the calculation, and the results agree with those obtained using two sweeps at the $\sim 10^{-14}$ level.

Instead of performing finite system sweeps, iDMRG keeps growing the chain until the state reaches a thermodynamic fixed point, where reduced density matrix of the half chain converge between successive iterations, and a unit cell tensor can be identified~\cite{mcculloch2008infinitesizedensitymatrix,schollwock2011density}. In this work, the MPS chain is grown through two site insertions, with a single site insertion performed only in the final step. Convergence of the half-chain and central reduced density matrices is observed at the $\sim 10^{-10}$ and $\sim 10^{-14}$ levels, respectively.

A different gauge convention is imposed compared with Ref.~\cite{mcculloch2008infinitesizedensitymatrix} in order to allow exchange symmetry of the virtual indices, $\mathcal{A}_{bb'}^{\tilde{\phi}} = \mathcal{A}_{b'b}^{\tilde{\phi}}$. As a result, the left fixed-point tensor, $\ell_{ab}$, obtained from the dominant left eigenvector of the transfer operator $E \equiv \left(\sum_{\tilde{\phi}}\mathcal{A}^{\tilde{\phi}}\mathcal{A}^{*\tilde{\phi}}\right)_{(ab),(a'b')}$, is identical to the right fixed-point tensor obtained from the corresponding dominant right eigenvector. The ground state in the thermodynamic limit may be represented with PBC. In light of Appendix.~\ref{app:fieldintro1}, the sites are labeled from $-\tilde{\ell}$ to $\tilde{\ell}$ counterclockwise, with $\tilde{\ell} \equiv \frac{\tilde{L}-1}{2}$ and odd $\tilde{L}$,
\begin{equation}
|\Omega_{\text{PBC}}\rangle = \sum_{\tilde{\phi}_{-\tilde{\ell}}, \dots, \tilde{\phi}_{\tilde{\ell}}} \Tr \left[ \mathcal{A}^{\tilde{\phi}_{-\tilde{\ell}}} \cdots \mathcal{A}^{\tilde{\phi}_{\tilde{\ell}}} \right] | \tilde{\phi}_{-\tilde{\ell}}, \dots, \tilde{\phi}_{\tilde{\ell}}\rangle := \sum_{\{\tilde{\phi}\}} \Tr \left[ \mathcal{A}^{\tilde{\phi}_{-\tilde{\ell}}} \cdots \mathcal{A}^{\tilde{\phi}_{\tilde{\ell}}} \right] |\{\tilde{\phi}\}\rangle \ \ \ ,
\label{eq:pbcgs}
\end{equation}
where $\tilde{\phi}_j$ denotes the local field basis at j-th site. The contractions between the virtual indices, $b$, are omitted in Eq.~\eqref{eq:pbcgs}. 

For the results presented in Figs.~\ref{fig:NSplot} and~\ref{fig:NSsearch}, a singular value cutoff of $10^{-6}$ is used in both DMRG and iDMRG. At optimal $\phi_{\text{max}}$ for $d=6,8,10$, this yields a maximum bond dimension $D=21$ in DMRG and a bond dimension $D=21$ for the iDMRG unit cell. As detailed in Appendix~\ref{app:mpsintro2}, the single site excitation tensor in the quasiparticle ansatz is optimized using this iDMRG ground state MPS and thus shares its bond dimension.

Convergence with respect to this cutoff is assessed by comparing with calculations using the looser cutoff, $10^{-5}$. With the same local field space digitization, both DMRG and iDMRG yield $D=13,13,14$, respectively. The corresponding energy precisions in DMRG, iDMRG, and single particle calculations are $\lesssim 10^{-9}$. Comparison with Fig.~\ref{fig:NSplot} indicates that the $10^{-6}$ cutoff provides sufficient accuracy for the calculations presented here.


\subsection{Determining single particle momentum eigenstates}
\label{app:mpsintro2}

This section overviews the techniques used in this work for the variational optimization of single particle momentum eigenstates. 

Constructing states from superpositions of momentum eigenstates play a central role in the establishment of this work. Unlike previous works, where the single particle ansatz is formulated directly in the thermodynamic limit within the uMPS framework~\cite{Vanderstraeten:2019voi}, this work keeps both the number of lattice sites finite and the PBC explicit, as shown in Eq.~\eqref{eq:pbcgs}~\footnote{Although finite, the system size is chosen sufficiently large such that the error in the fixed-point relation is $\sim 10^{-10}$, justifying the subsequent use of the left and right fixed-point tensors for the computation.}. In contrast to the thermodynamic limit, where momentum eigenstates form a continuous spectrum, the present formulation yields a discrete set of momentum eigenstates.

The quasiparticle ansatz~\cite{Haegeman:2011lcd,Pirvu:2011ijm,PhysRevLett.111.080401,Vanderstraeten:2019voi} in PBC is given by,
\begin{equation}
|\tilde{k}\rangle^{[m]} = \sum_{j} e^{i\tilde{k}j} \sum_{\{\tilde{\phi}\}} \Tr \left[ \mathcal{A}^{\tilde{\phi}_{-\tilde{\ell}}} \cdots \mathcal{A}^{\tilde{\phi}_{j-1}} \mathcal{B}_{\tilde{k}}^{\tilde{\phi}_j,[m]} \mathcal{A}^{\tilde{\phi}_{j+m}} \cdots \mathcal{A}^{\tilde{\phi}_{\tilde{\ell}}} \right] | \{\tilde{\phi}\}\rangle \ \ \ ,
\label{eq:pbck}
\end{equation}
where $\mathcal{B}_{\tilde{k}}^{\tilde{\phi}_j,[m]}$ denotes the variational tensor with support starting at site $j$ and extending over $m$ sites. This corresponds to a localized excitation embedded in Eq.~\eqref{eq:pbcgs}. A coherent superposition over spatial translations, $\sum_{j} e^{i\tilde{k}j}$, gives rise to a quasiparticle ansatz with definite momentum $\tilde{k}$, which also induces the long-range correlations across the lattice. 

The calculation of the energy expectation value for single particle momentum eigenstates can be simplified by noting that the PBC Hamiltonian may be written in terms of the translation operator as,
\begin{equation}
    \tilde{H}_{\text{PBC}} = \sum_{j} T_{j} \, \tilde{h}_{\text{PBC}} \, T_{j}^\dagger, \quad \tilde{h}_{\text{PBC}} = \frac{\tilde{m}_0^2}{4}\tilde{\phi}^2_{\tilde{\ell}} +\frac{\tilde{m}_0^2}{4}\tilde{\phi}^2_{-\tilde{\ell}} + \frac{1}{4} \tilde{\pi}^2_{\tilde{\ell}} + \frac{1}{4} \tilde{\pi}^2_{-\tilde{\ell}} + \frac{1}{2}\frac{\tilde{\lambda}_0}{4!}\tilde{\phi}^4_{\tilde{\ell}} + \frac{1}{2}\frac{\tilde{\lambda}_0}{4!}\tilde{\phi}^4_{-\tilde{\ell}} + \frac{1}{2} \left( \tilde{\phi}_{\tilde{\ell}} -\tilde{\phi}_{-\tilde{\ell}} \right)^2 \ \ \ ,
\label{eq:pbcH}
\end{equation}
since single particle momentum eigenstates are eigenstates of the translation operator,
\begin{equation}
^{[m]} \langle \tilde{k} | \tilde{H}_{\text{PBC}} |\tilde{k}\rangle^{[m]} = ^{[m]} \langle \tilde{k} | \left( \sum_{j} T_{j} \, \tilde{h}_{\text{PBC}} \, T_{j}^\dagger \right) |\tilde{k}\rangle^{[m]} = \tilde{L} \, \, ^{[m]} \langle \tilde{k} | \tilde{h}_{\text{PBC}} |\tilde{k}\rangle^{[m]} \ \ \ .
\label{eq:Eexpsimple}
\end{equation}
By observing that $\tilde{L}$ is a constant factor, the optimization problem becomes,
\begin{equation}
    \min_{|\tilde{k}\rangle^{[m]}} \frac{^{[m]} \langle \tilde{k} | \tilde{h}_{\text{PBC}} |\tilde{k}\rangle^{[m]}}{^{[m]} \langle \tilde{k}  |\tilde{k}\rangle^{[m]}} = \min_{\mathcal{B}_{(bb'\tilde{\phi}_1 \cdots \tilde{\phi}_m)}} \frac{\left( \mathcal{B}_{(bb'\tilde{\phi}_1 \cdots \tilde{\phi}_m)} \right)^{\dagger} H_{eff}  \left( \mathcal{B}_{(bb'\tilde{\phi}_1 \cdots \tilde{\phi}_m)} \right) }{\left( \mathcal{B}_{(bb'\tilde{\phi}_1 \cdots \tilde{\phi}_m)} \right)^{\dagger} N_{eff}  \left( \mathcal{B}_{(bb'\tilde{\phi}_1 \cdots \tilde{\phi}_m)} \right) }  \ \ \ .
\label{eq:quasiopt}
\end{equation}
With the convention established under Eq.~\eqref{eq:pbcgs}, $H_{eff}$ may be represented with MPS diagram,
\begin{equation}
H_{eff} :=  \sum_{j_1,j_2} e^{i\tilde{k}(j_1 - j_2)}
\begin{tikzpicture}[scale=0.95,baseline=(current bounding box.center),trim left=-0.25cm,trim right=7.69cm]

\foreach \x in {0}{
    \draw[fill=white] (\x,1) rectangle +(0.45,0.45);
}

\foreach \x in {5,7}{
    \draw[fill=white] (\x,1) rectangle +(0.45,0.45);
}

\node at (0.22,1.22) {$\mathcal{A}$};
\node at (1.22,1.22) {$j_1$};
\node at (3.22,1.22) {$\cdots$};
\node at (3.22,0) {$\cdots$};
\node at (5.22,1.22) {$\mathcal{A}$};
\node at (6.22,1.22) {$\cdots$};
\node at (6.22,0) {$\cdots$};
\node at (7.22,1.22) {$\mathcal{A}$};

\draw (0.45,1.22)--(0.75,1.22);
\draw (4.75,1.22)--(5,1.22);
\draw (5.45,1.22)--(5.7,1.22);
\draw (6.75,1.22)--(7,1.22);

\draw[out=135,in=45] (0,1.22) to (0,1.22) to (7.46,1.22) to (7.46,1.22);

\foreach \x in {4,5,7}{
    \draw[fill=white] (\x,-1.45) rectangle +(0.45,0.45);
}

\node at (0.22,-1.22) {$j_2$};
\node at (3.22,-1.22) {$\cdots$};
\node at (4.22,-1.22) {$\mathcal{A}$};
\node at (5.22,-1.22) {$\mathcal{A}$};
\node at (6.22,-1.22) {$\cdots$};
\node at (7.22,-1.22) {$\mathcal{A}$};

\draw (3.75,-1.22)--(4,-1.22);
\draw (4.45,-1.22)--(5,-1.22);
\draw (5.45,-1.22)--(5.7,-1.22);
\draw (6.75,-1.22)--(7,-1.22);

\draw[out=-135,in=-45] (-0.1,-1.22) to (-0.1,-1.22) to (7.45,-1.22) to (7.45,-1.22);

\foreach \x in {5.22,7.22}{
    \draw (\x,1)--(\x,-1);
}

\foreach \x in {2.22}{
    \draw (\x,0.6)--(\x,-0.7);
}

\foreach \x in {4.22}{
    \draw (\x,0.6)--(\x,-1);
}

\draw[fill=white] (-0.05,-0.22) rectangle +(1.55,0.45);

\node at (0.72,0) {$\tilde{h}_{\text{PBC}}$};

\draw (0.22,1)--(0.22,0.23);
\draw (1.22,0.625)--(1.22,0.23);
\draw (0.22,-0.23)--(0.22,-0.625);
\draw (1.22,-0.23)--(1.22,-0.625);

\end{tikzpicture} ,
\label{eq:Heff}
\end{equation}
with first $\cdots$ denoting a span of $m$ lattice sites. In Eq.~\eqref{eq:Heff}, the ket may be placed on top and the bra on the bottom. The resulting indices from the top and bottom layers are then flattened to form the effective matrix. Note $N_{eff}$ can be obtained analogously by replacing $\tilde{h}_{\text{PBC}}$ in Eq.~\eqref{eq:Heff} with the identity operator. Once $\mathcal{B}$ is determined, the state may be normalized as $ ^{[m]} \langle \tilde{p} |\tilde{k} \rangle^{[m]} = \tilde{L} \delta_{\tilde{p},\tilde{k}}$.

The quasiparticle ansatz contains gauge redundancies for $\mathcal{B}$~\cite{PhysRevLett.111.080401,Vanderstraeten:2019voi}. For $\mathcal{B}_{\tilde{k}}^{\tilde{\phi}_j,[m]}$ that spans $m$ sites, the resulting quasiparticle MPS remains invariant under the transformation,
\begin{equation}
\begin{tikzpicture}[scale=0.95,baseline=-1ex,trim left=1cm,trim right=1cm]
\draw[fill=white] (-1,-0.22) rectangle +(3,0.45);
\node at (0.5,0) {$\mathcal{B}$};
\draw (-1.25,0)--(-1,0);
\draw (2,0)--(2.25,0);
\draw (0.5,-0.23)--(0.5,-0.625);
\draw (1.5,-0.23)--(1.5,-0.625);
\draw (-0.5,-0.23)--(-0.5,-0.625);
\node at (0.53,-0.875) {$\cdots$};
\end{tikzpicture} \quad \quad \quad \rightarrow \begin{tikzpicture}[scale=0.95,baseline=-0.95ex,trim left=-1.25cm,trim right=1.69cm]
\draw[fill=white] (-1,-0.22) rectangle +(3,0.45);
\node at (0.5,0) {$\mathcal{B}$};
\draw (-1.25,0)--(-1,0);
\draw (2,0)--(2.25,0);
\draw (0.5,-0.23)--(0.5,-0.625);
\draw (1.5,-0.23)--(1.5,-0.625);
\draw (-0.5,-0.23)--(-0.5,-0.625);
\node at (0.53,-0.875) {$\cdots$};
\end{tikzpicture}  \quad \, \, +
\begin{tikzpicture}[scale=0.95,baseline=-0.9ex,trim left=-1.3cm,trim right=1.69cm]
\draw[fill=white] (-1,-0.22) rectangle +(2,0.45);
\node at (0,0) {$\mathcal{Y}$};
\draw (-1.25,0)--(-1,0);
\draw (1,0)--(1.25,0);
\draw (0.5,-0.23)--(0.5,-0.625);
\draw (-0.5,-0.23)--(-0.5,-0.625);
\foreach \x in {0}{
    \draw[fill=white] (1.25,-0.21) rectangle +(0.45,0.45);
}
\node at (1.5,0) {$\mathcal{A}$};
\draw (1.5,-0.23)--(1.5,-0.625);
\draw (1.7,0)--(1.95,0);
\node at (0.53,-0.875) {$\cdots$};
\end{tikzpicture} \, \, \, \, - e^{i\tilde{k}} \begin{tikzpicture}[scale=0.95,baseline=-0.85ex,trim left=-1.1cm,trim right=1.69cm]
\draw[fill=white] (0,-0.22) rectangle +(2,0.45);
\node at (1,0) {$\mathcal{Y}$};
\draw (-1,0)--(-0.75,0);
\draw (-0.3,0)--(0,0);
\draw (0.5,-0.23)--(0.5,-0.625);
\draw (-0.5,-0.23)--(-0.5,-0.625);
\foreach \x in {0}{
    \draw[fill=white] (-0.75,-0.21) rectangle +(0.45,0.45);
}
\node at (-0.5,0) {$\mathcal{A}$};
\draw (1.5,-0.23)--(1.5,-0.625);
\draw (2,0)--(2.25,0);
\node at (1.53,-0.875) {$\cdots$};
\end{tikzpicture}  \quad \ \ \ ,
\label{eq:Bgauge}
\end{equation}
where $\mathcal{Y}$ is a arbitrary $(D \times \overbrace{d \times \cdots \times d}^{m-1} \times D)$ tensor, whereas $\mathcal{B}$ is a $(D \times  \overbrace{d \times \cdots \times d}^{m} \times D)$ tensor. It can be checked that, when the $\mathcal{B}$ tensor is chosen to be equal to the last two terms of Eq.~\eqref{eq:Bgauge}, the quasiparticle ansatz reduces to a zero state, thereby rendering the optimization problem in Eq.~\eqref{eq:quasiopt} ill-defined. To exclude such null vectors from the variational manifold, gauge-fixing condition needs to be imposed such that,
\begin{equation}
\begin{tikzpicture}[scale=0.95,baseline=-7.5ex,trim left=-2cm,trim right=0cm]
\draw[fill=white] (-1,-0.22) rectangle +(3,0.45);
\node at (0.5,0) {$\mathcal{B}$};
\draw (2,0)--(2.25,0);
\draw (0.5,-0.23)--(0.5,-0.625);
\draw (1.5,-0.23)--(1.5,-0.625);
\draw (-0.5,-0.23)--(-0.5,-2);
\node at (0.53,-0.875) {$\cdots$};

\foreach \x in {0}{
    \draw[fill=white] (-0.75,-2.21) rectangle +(0.45,0.45);
}
\node at (-0.5,-2) {$\mathcal{A}$};
\draw (-0.05,-2)--(-0.3,-2);
\draw (-1,0) .. controls (-2,-0.5) and (-2,-1.9) .. (-0.75,-2);
\foreach \x in {0}{
    \draw[fill=white] (-1.9,-1.2) rectangle +(0.45,0.45);
}
\node at (-1.7,-1) {$\ell$};
\end{tikzpicture}  \qquad \qquad \qquad = 0 \ \ \ ,
\label{eq:gaugefix}
\end{equation}
where $\ell$ denotes the left fixed-point tensor, identical to the right fixed-point tensor from the construction of $\mathcal{A}$. As such, Eq.~\eqref{eq:gaugefix} imposes $d^{m-1}D^2$ constraint equations. Identify the $D\times dD$ dimensional matrix, $M_{b',(a\tilde{\phi})} := \sum_{b}\ell^{\frac{1}{2}}_{ab} \mathcal{A}_{bb'}^{\tilde{\phi}}$, and the $D(d-1)\times dD$ dimensional null space, $\mathcal{V}_{ab''}^{\tilde{\phi}} := \left( \nullspace\left(M\right) \right)_{b'',(a\tilde{\phi})}$, $\mathcal{B}$ can therefore be written as,
\begin{equation}
\begin{tikzpicture}[scale=0.95,baseline=-1ex,trim left=-2cm,trim right=1cm]
\draw[fill=white] (-1,-0.22) rectangle +(3,0.45);
\node at (0.5,0) {$\mathcal{B}$};
\draw (-1.25,0)--(-1,0);
\draw (2,0)--(2.25,0);
\node at (0.53,-0.875) {$\cdots$};
\draw (0.5,-0.23)--(0.5,-0.625);
\draw (1.5,-0.23)--(1.5,-0.625);
\draw (-0.5,-0.23)--(-0.5,-0.625);
\end{tikzpicture} \quad \quad \quad =  \begin{tikzpicture}[scale=0.95,baseline=-0.92ex,trim left=-1.7cm,trim right=1.69cm]
\draw[fill=white] (0,-0.22) rectangle +(2,0.45);
\node at (1,0) {$\mathcal{X}$};
\draw (-1,0)--(-0.75,0);
\draw (-0.3,0)--(0,0);
\draw (0.5,-0.23)--(0.5,-0.625);
\draw (-0.5,-0.23)--(-0.5,-0.625);
\foreach \x in {0}{
    \draw[fill=white] (-0.75,-0.21) rectangle +(0.45,0.45);
}
\node at (-0.5,0) {$\mathcal{V}$};
\draw (1.5,-0.23)--(1.5,-0.625);
\draw (2,0)--(2.25,0);

\foreach \x in {0}{
    \draw[fill=white] (-1.45,-0.21) rectangle +(0.45,0.45);
}
\node at (-1.22,0) {$\scalemath{0.6}{\ell^{-\frac{1}{2}}}$};
\draw (-1.45,0)--(-1.7,0);

\draw (2.5,0)--(2.95,0);
\foreach \x in {0}{
    \draw[fill=white] (2.25,-0.21) rectangle +(0.45,0.45);
}
\node at (2.47,0) {$\scalemath{0.6}{\ell^{-\frac{1}{2}}}$};
\node at (1.53,-0.875) {$\cdots$};
\end{tikzpicture}  \quad \quad \quad \ \ \ ,
\end{equation}
such that Eq.~\eqref{eq:gaugefix} is automatically satisfied. This removes the gauge related null modes and improves the conditions of the generalized eigenvalue problem. The resulting variational tensor, $\mathcal{X}$, has dimension, $(D(d-1) \times \overbrace{d \times \cdots \times d}^{m-1} \times D)$.

\subsection{Identification of the Nyquist-Shannon sampling theorem}
\label{app:mpsintro3}

As discussed in Sec.~\ref{sec:III1}, there exists an optimal choice of $\phi_{\max}$, denoted as $\phi_{\max,\mathrm{opt}} (d)$, such that the energy is best approximated, and approaches the local continuum limit exponentially with respect to $d$~\cite{NKdigi}. For a given $d_0$, the calculation of $\phi_{\max,\mathrm{opt}} (d_0)$ in the free lattice scalar field theory is simple, where the exact energy can be calculated analytically, and $\phi_{\max,\mathrm{opt}} (d_0)$ can be determined by scanning over $\phi_{\max}$. In the interacting theory, however, the exact energy cannot be calculated analytically. This section presents the algorithm for determining $\phi_{\max,\mathrm{opt}} (d_0)$ in an interacting lattice scalar field theory, based on the properties of the NS sampling theorem for scalar field identified in Ref.~\cite{NKdigi}.

\begin{figure*}[t]
    \centering
    \includegraphics[width=.325\linewidth]{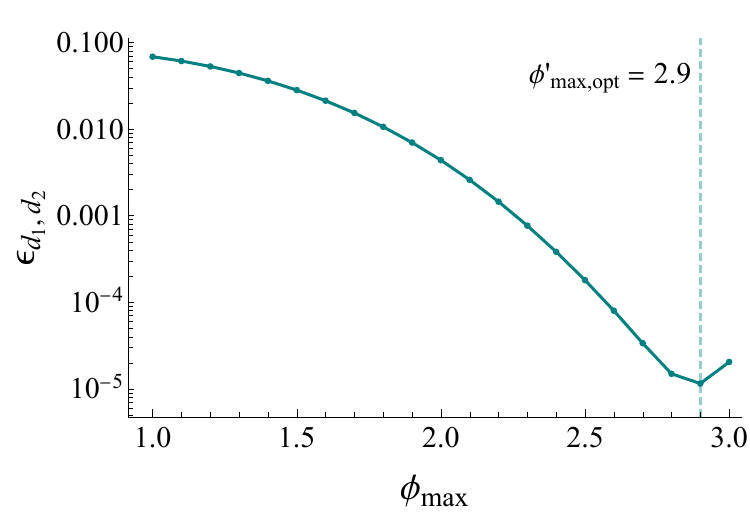}\hfill
    \includegraphics[width=.325\linewidth]{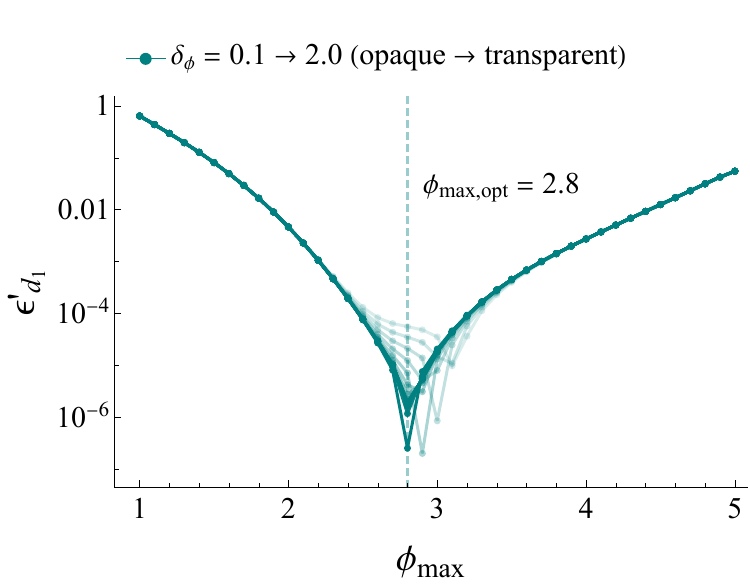}\hfill
    \includegraphics[width=.325\linewidth]{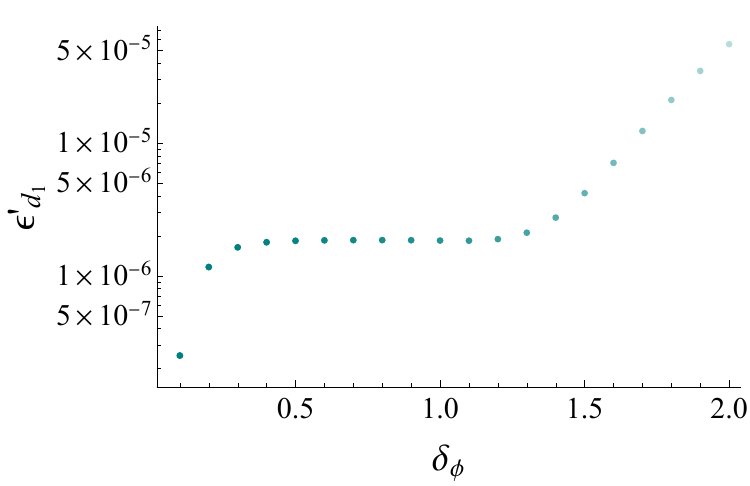}
    \caption{Algorithm for determining $\phi_{\max,\mathrm{opt}} (d_{1}=8)$ for DMRG in the interacting theory using the relative precision $\epsilon_{d_1,d_2}(\phi_{\max})$ defined in Eq.~\eqref{eq:relprecision} (with $d_1=8$ and reference $d_2=12$). (Left) Relative precision $\epsilon_{d_1,d_2}$ as a function of $\phi_{\max}$. The minimum indicates the initial estimation $\phi'_{\max,\mathrm{opt}}=2.9$. (Middle) $d_1$ precision as a function of $\phi_{\text{max}}$. Here, $d_2$ is evaluated at $\phi_{\max} = \phi'_{\max,\mathrm{opt}}(d_1)+\delta_{\phi}$, with $\delta_\phi$ range from $0.1$ to $2.0$ (transitioning from opaque to transparent curves). (Right) Precision as a function of $\delta_\phi$ at $\phi_{\text{max}} = 2.8$.}
    \label{fig:NSsearch}
\end{figure*}

As shown in the left panels of Figs.~\ref{fig:NSplot} and~\ref{fig:NSsearch}, by scanning over $\phi_{\max}$, the energy values for both digitization levels initially converge toward the exact energy. However, once $\phi_{\max} > \phi_{\max,\mathrm{opt}} (d_1)$, further increasing $\phi_{\max}$ causes the energy for $d_1$ to deviate from the exact energy, while the energy for $d_2$ continues to converge towards the exact energy. This behavior allows $\phi_{\max,\mathrm{opt}} (d_1)$ to be estimated. Define the precisions,
\begin{equation}
\epsilon_{d1,d2} (\phi_{\max}) = \frac{\lvert E_{d_1}(\phi_{\max}) - E_{d_2}(\phi_{\max}) \rvert}{E_{d_2}(\phi_{\max})}, \quad \epsilon_{d_0} (\phi_{\max}) = \frac{\lvert E_{d_0}(\phi_{\max}) - E_{\mathrm{exact}} \rvert}{E_{\mathrm{exact}}}\ \ \ ,
\label{eq:relprecision}
\end{equation}
where the first quantity is the relative precision between $d_1$ and $d_2$, while the second quantity is defined relative to the exact energy. The value of $\phi_{\max}$ where $\epsilon_{d1,d2}(\phi_{\max})$ attains its minimum, denoted as $\phi'_{\max,\mathrm{opt}} (d_1)$, can serve as an initial estimation of $\phi_{\max,\mathrm{opt}} (d_1)$.

The above result provides only an estimation because in general $\epsilon_{d_1} (\phi_{\max}) \neq \epsilon_{d_2} (\phi_{\max})$, and the onset of the deviation may not be identified immediately once $\phi_{\max}$ exceeds $\phi_{\max,\mathrm{opt}} (d_1)$. To refine the initial estimation, the energies for $d_2$ are evaluated at several values of $\delta_{\phi}$ such that $\phi_{\max} = \phi'_{\max,\mathrm{opt}}(d_1)+\delta_{\phi}$ in order to approximate the exact energy. For each approximation of the exact energy, $\epsilon'_{d_1}(\phi_{\max})$ maybe computed, as shown in the middle panel of Fig.~\ref{fig:NSsearch}. The final $\phi_{\text{max,opt}}$ is determined through a convergence analysis of all $\epsilon'_{d_1}(\phi_{\max})$, where the flat window represents the region where the precision is robust against the change of $\delta_{\phi}$, indicating a reliable convergence plateau for determining the optimal digitization scheme. This is shown in the right panel of Fig.~\ref{fig:NSsearch}.

To ensure that the result is representative of the thermodynamic limit, the above procedures can be repeated for a larger system size to verify the convergence of $\phi_{\max,\mathrm{opt}}(d_1)$.

\subsection{Joint optimization of the wave packet and the interpolating operator}
\label{app:mpsintro4}

Building on the calibration procedure for single particle momentum eigenstates detailed in Appendix~\ref{app:fieldintro1}, this section jointly optimizes both the optimal interpolating operator and the superposition profile.

With the total Hilbert space $\mathcal{H} = \mathcal{H}_C \otimes \mathcal{H}_D$, it is convenient to define the following auxiliary functions,
\begin{equation}
\scalemath{0.985}{\mathcal{I}_{[n]}(\mathcal{B}_{\tilde{k}}^{\tilde{\phi}_j,[m]})^{(\tilde{\phi}_{-c}, \dots, \tilde{\phi}_{c}),(\tilde{\phi}'_{-c}, \dots, \tilde{\phi}'_{c})} \equiv  \sum_{\{\tilde{\phi}\}}  \Tr \left[{\mathcal{A}^{\tilde{\phi}'_{-\tilde{\ell}}}}^{*} \cdots {\mathcal{A}^{\tilde{\phi}'_{j-1}}}^{*} {\mathcal{B}_{\tilde{k}}^{\tilde{\phi}'_j,[m]}}^{*} {\mathcal{A}^{\tilde{\phi}'_{j+m}}}^{*} \cdots {\mathcal{A}^{\tilde{\phi}'_{\tilde{\ell}}}}^{*} \right]  \Tr_{D} \left[ |\Omega_{\text{PBC}}\rangle \langle \{\tilde{\phi}\} |\right]}   ,
\label{eq:Ifunction}
\end{equation}
and the vacuum reduced density matrix,
\begin{equation}
    \scalemath{0.985}{\mathcal{I}_{[n]}^{(\tilde{\phi}_{-c}, \dots, \tilde{\phi}_{c}),(\tilde{\phi}'_{-c}, \dots, \tilde{\phi}'_{c})} = \Tr_{D} \left[ |\Omega_{\text{PBC}}\rangle \langle \Omega_{\text{PBC}}|\right]} \ \ \ ,
\end{equation}
which is Hermitian and positive semidefinite. Here, subsystem $D$ is traced and $C$ corresponds to the central $n$ sites from $-c$ to $c$. With $\mathcal{A}$ and $\mathcal{B}$ constructed in Appendix~\ref{app:mpsintro1}--Appendix~\ref{app:mpsintro3}, $\mathcal{I}$ can be evaluated numerically, and serves as the primary quantity computed in this section.

The phase fixing procedure, i.e., Eq.~\eqref{eq:phaseconvention}, for single particle momentum eigenstates can thus be implemented through,
\begin{equation}
    {}^{[m]}\langle \tilde{k} | \tilde{\phi}_0 | \Omega \rangle = \sum_{j=-\ell}^{\ell} e^{-i\tilde{k}j} \Tr \left[ \tilde{\phi}_0 \mathcal{I}_{[1]}(\mathcal{B}_{\tilde{k}}^{\tilde{\phi}_j,[m]}) \right], \quad |\tilde{k}\rangle^{[m]} \to \frac{{}^{[m]}\langle \tilde{k} | \tilde{\phi}_0 | \Omega \rangle}{\abs{{}^{[m]}\langle \tilde{k} | \tilde{\phi}_0 | \Omega \rangle}} |\tilde{k}\rangle^{[m]} \ \ \ .
\label{eq:phaseconventionpractical}
\end{equation}
In practice, this complex phase can be absorbed into the $\mathcal{B}$ tensor. In the following, the notations $|\tilde{k}\rangle^{[m]}$ and $\mathcal{B}_{\tilde{k}}^{\tilde{\phi}_j,[m]}$ will refer to the calibrated single particle momentum eigenstates obtained through Eq.~\eqref{eq:phaseconventionpractical}.

The optimization problem originally involves both the interpolating operator and the momentum-space superposition profile, with the goal that the superposition of single particle momentum eigenstates is maximally approximated by a localized operator acting on the vacuum. Here, the coupled optimization will be decoupled into sequential optimization steps.

Without loss of generality, $j=0$ may be assumed. Given the profile $h(\tilde{k})$, the interpolating operator then maximizes the fidelity,
\begin{equation}
\max_{\tilde{\Phi}_{0,[n]}^{[m]}} \frac{
    \abs{ \mathrm{Tr}\!\left[\tilde{\Phi}_{0,[n]}^{[m]} \mathcal{J}_{[n]}^{[m]} \right]}^2
}{
    \mathrm{Tr}\!\left[
        \mathcal{I}_{[n]} \tilde{\Phi}_{0,[n]}^{[m]\dagger} \tilde{\Phi}_{0,[n]}^{[m]}
    \right]
}, \quad \mathcal{J}_{[n]}^{[m]} = \frac{1}{\tilde{L}} \sum_{\tilde{k}} h^{*}(\tilde{k}) \sum_{j} e^{-i\tilde{k}j} \mathcal{I}_{[n]}(\mathcal{B}_{\tilde{k}}^{\tilde{\phi}_j,[m]})  \ \ \ .
\label{eq:optimizationappc4}
\end{equation}
The subscripts and superscripts will be omitted in the following unless required for clarity. Eq.~\eqref{eq:optimizationappc4} can be simplified to a generalized eigenvalue problem,
\begin{equation}
   \max_{\tilde{\Phi}} \frac{
    \abs{ \mathrm{Tr}\!\left[\tilde{\Phi} \mathcal{J} \right]}^2
}{
    \mathrm{Tr}\!\left[
        \mathcal{I} \tilde{\Phi}^{\dagger} \tilde{\Phi}
    \right] 
} = \max_{\vct{\tilde{\Phi}}} \frac{
   \vct{\tilde{\Phi}}^{\dagger} \vct{\mathcal{J}^{\dagger}} \vct{\mathcal{J}^{\dagger}}^{\dagger}  \vct{\tilde{\Phi}}
}{
   \vct{\tilde{\Phi}}^{\dagger} \left( \mathbb{I} \otimes \mathcal{I}^{T} \right)  \vct{\tilde{\Phi}} 
} \ \ \ ,
\label{eq:optsimp}
\end{equation}
where $\vct{\cdot}$ denotes vectorization. Since $\vct{\mathcal{J}^{\dagger}} \vct{\mathcal{J}^{\dagger}}^{\dagger}$ is rank-one, the optimal $\vct{\tilde{\Phi}}$ can be found analytically by multiplying the pseudoinverse of $\left( \mathbb{I} \otimes \mathcal{I}^{T} \right)$ to $\vct{\mathcal{J}^{\dagger}}$, i.e., $\left( \mathbb{I} \otimes \mathcal{I}^{T} \right)^{-1}\vct{\mathcal{J}^{\dagger}} = \vct{\mathcal{J}^{\dagger} \mathcal{I}^{-1}}$. Since $\mathcal{I}$ is real and positive semidefinite (PSD), working in its eigenbasis further simplifies the expressions for the operator that yields the normalized state, $\tilde{\Phi}_{\mathrm{opt}}|\Omega_{\mathrm{PBC}}\rangle$, and the corresponding fidelity,
\begin{equation}
\tilde{\Phi}_{\text{opt}} = \frac{\mathcal{J}^{\dagger} \mathcal{I}^{-1}}{\sqrt{\Tr\left[\mathcal{J} \mathcal{J}^{\dagger} \mathcal{I}^{-1}\right]}}, \quad \max_{\tilde{\Phi}} \frac{
    \abs{ \mathrm{Tr}\!\left[\tilde{\Phi} \mathcal{J} \right]}^2
}{
    \mathrm{Tr}\!\left[
        \mathcal{I} \tilde{\Phi}^{\dagger} \tilde{\Phi}
    \right] 
} =  \Tr\left[\mathcal{J} \mathcal{J}^{\dagger} \mathcal{I}^{-1}\right] \ \ \ .
\label{eq:optsolu1}
\end{equation}
With Eq.~\eqref{eq:optsolu1}, the remaining optimization over the momentum-space superposition profile reads,
\begin{equation}
\scalemath{0.9}{\mathcal{K}_{[n],\tilde{k},\tilde{k}'}^{[m]} \equiv \Tr\left[\left(\sum_{j} e^{-i\tilde{k}j} \mathcal{I}_{[n]}(\mathcal{B}_{\tilde{k}}^{\tilde{\phi}_j,[m]})\right) \left(\sum_{j'} e^{i\tilde{k}'j'} \mathcal{I}_{[n]}^{\dagger}(\mathcal{B}_{\tilde{k}'}^{\tilde{\phi}_{j'},[m]})\right) \mathcal{I}^{-1}\right], \quad \frac{1}{\tilde{L}}\max_{h(\tilde{k})} \frac{\sum_{\tilde{k},\tilde{k}'} h^{*}(\tilde{k}) \mathcal{K}_{[n],\tilde{k},\tilde{k}'}^{[m]}  h(\tilde{k}')}{\sum_{\tilde{k}} h^{*}(\tilde{k}) h(\tilde{k}) } \ \ \ ,}
\label{eq:optsolu2}
\end{equation}
which is an eigenvalue problem for $\mathcal{K}$, where the eigenvalues yield the fidelities, and the eigenvectors determine the profile functions, $h(\tilde{k})$. In addition to the explicit Fourier phases, the entries of $\mathcal{K}_{[n]}^{[m]}$ depend on momentum through the excitation tensors. To see the positive semidefinite property of $\mathcal{K}_{[n]}^{[m]}$, note that it can be reformulated as a matrix of vector inner products of $\vct{\left(\sum_{j'} e^{i\tilde{k}'j'} \mathcal{I}_{[n]}^{\dagger}(\mathcal{B}_{\tilde{k}'}^{\tilde{\phi}_{j'},[m]})\right) \mathcal{I}^{-\frac{1}{2}}}$. Furthermore, due to the phase calibration in Sec.~\ref{app:fieldintro1}, $\mathcal{K}_{[n]}^{[m]}$ is real.

For $n\geq 2$, the small eigenvalues of $\mathcal{I}$ can make the numerical evaluation of $\Tr[\mathcal{J}\mathcal{J}^{\dagger}\mathcal{I}^{-1}]$ unreliable. The fidelity is therefore evaluated using the full ratio in Eq.~\eqref{eq:optsolu1}. These small eigenvalues correspond to operator components that annihilate or nearly annihilate the vacuum, and thus also responsible for the small eigenvalues of $J$. Compensating the amplification introduced by $\mathcal{I}^{-1}$, the suppression of $\mathcal{J}^{\dagger}$ along the null directions of $\mathcal{I}$ leaves the combination $\mathcal{J}^{\dagger}\mathcal{I}^{-1}$ numerically stable.

\subsection{Constructing the asymptotic incoming state}
\label{app:mpsintro5}

With the construction of $g_j$, $\tilde{\Phi}_{j,[n]}^{[m]}$, and $| \Omega \rangle$ established in Sec.~\ref{sec:III1}, Appendix~\ref{app:mpsintro4} and Appendix~\ref{app:mpsintro1}, this section presents the detailed implementation of Eq.~\eqref{eq:instate} within the MPS framework.

To construct a single localized excitation, $\tilde{\Phi}_{j,[n]}^{[m]}| \Omega \rangle$, the interpolating operator is converted into an MPO with support across $n$ sites via successive singular value decompositions (SVDs), and contracted with the vacuum MPS over sites $\overbrace{j_{\min}, \dots, j_{\max}}^{n}$. Truncating the growth of bound dimension after applying the MPO is crucial to preserve computational tractability. The conventional compression scheme consists of a sweep across the entire chain (SVD compression)~\cite{PhysRevLett.93.040502,schollwock2011density}. Here, since the spatial support of the MPO is restricted to $n$ sites, SVD compression is preformed locally. In practice, MPS is first gauge transformed into left-canonical form (denoted by $L$) from site $1$ to $j_{\max}-1$ and into right-canonical form (denoted by $R$) from site $\tilde{L}$ to $j_{\max}+1$. Local SVD compression is then performed sequentially from site $j_{\max}$ to site $j_{\min}+1$,
\begin{equation}
\begin{aligned}
&\xrightarrow{\text{gauge transformation}}\sum_{\{\tilde{\phi}\}} L^{\tilde{\phi}_1} \cdots L^{\tilde{\phi}_{j_{\min}}} \cdots L^{\tilde{\phi}_{j_{\max}-1}} M^{\tilde{\phi}_{j_{\max}}} R^{\tilde{\phi}_{j_{\max}+1}} \cdots R^{\tilde{\phi}_{\tilde{L}}} | \{\tilde{\phi}\} \rangle  \\ &\xrightarrow{\text{compression}} \sum_{\{\tilde{\phi}\}} L^{\tilde{\phi}_1} \cdots 
L^{\tilde{\phi}_{j_{\min}}} \cdots L^{\tilde{\phi}_{j_{\max}-1}} \Bar{U}\Bar{\Lambda}\Bar{R}^{\tilde{\phi}_{j_{\max}}} R^{\tilde{\phi}_{j_{\max}+1}} \cdots R^{\tilde{\phi}_{\tilde{L}}} | \{\tilde{\phi}\} \rangle \\ &\xrightarrow{\text{sweep}} \sum_{\{\tilde{\phi}\}} L^{\tilde{\phi}_1} \cdots \left(L^{\tilde{\phi}_{j_{\min}}} \, \Bar{U}'\Bar{\Lambda'}\right)   \Bar{R}^{\tilde{\phi}_{j_{\min}+1}} \cdots \Bar{R}^{\tilde{\phi}_{j_{\max}}}  R^{\tilde{\phi}_{j_{\max}+1}} \cdots R^{\tilde{\phi}_{\tilde{L}}} | \{\tilde{\phi}\} \rangle \ \ \   .
\end{aligned}
\label{eq:svdcompress1}
\end{equation}
Here, SVD of $M^{\tilde{\phi}_{j_{\max}}}$ yields $U\Lambda R^{\tilde{\phi}_{j_{\max}}}$. Truncation then produces $\Bar{U}\Bar{\Lambda}\Bar{R}^{\tilde{\phi}_{j_{\max}}}$ by discarding the singular values in $\Lambda$ that are smaller than $\epsilon_0$. Subsequently, $L^{\tilde{\phi}_{j_{\max}-1}}$ is contracted with $\Bar{U}\Bar{\Lambda}$ that updates the local tensor at site $j_{\max}-1$, setting up for next round of compression. In a mixed-canonical form such as Eq.~\eqref{eq:svdcompress1}, SVD truncation minimizes the Hilbert space norm of the difference between the original and truncated states.

The superposition of localized excitation is realized through MPS addition~\cite{schollwock2011density}. From Eq.~\eqref{eq:gfconnection}, the Gaussian profile in momentum space yields exponentially decaying $g_j$ in position space. Terms with $|g_j| < \epsilon_0$ are thus discarded, and the resulting state is normalized to unity for later convenience. As a result of the local SVD compression in Eq.~\eqref{eq:svdcompress1}, the localized excitation share the same environmental tensors, allowing their MPS addition to be performed locally,
\begin{equation}
|\psi \rangle + |\psi \rangle' = \sum_{\{\tilde{\phi}\}} L^{\tilde{\phi}_1} \cdots L^{\tilde{\phi}_{i_{1}-1}} 
\begin{pmatrix} M_i^{\tilde{\phi}_{i_{1}}} & M^{'\tilde{\phi}_{i_{1}}} \end{pmatrix}
\begin{pmatrix} M^{\tilde{\phi}_{i_{1}+1}} & 0 \\ 0 & M^{'\tilde{\phi}_{i_{1}+1}} \end{pmatrix} \cdots 
\begin{pmatrix} M^{\tilde{\phi}_{i_{2}}} \\ M^{'\tilde{\phi}_{i_{2}}} \end{pmatrix} 
R^{\tilde{\phi}_{i_{2}+1}} \cdots R^{\tilde{\phi}_{\tilde{L}}} | \{\tilde{\phi}\} \rangle  \ \ \  ,
\label{eq:localaddition}
\end{equation}
where MPS tensors differ only within a consecutive region $[i_1, i_2]$. As such, the expansion of bond dimension is restricted exclusively to the differing sites. SVD truncation as described by Eq.~\eqref{eq:svdcompress1} is then performed between $i_1$ to $i_2$, perserving the environmental tensors and setting up the MPS for the next round of local addition.

The resulting MPS exhibits a bond dimension profile that is localized around the wave packet center, with spatially decaying tails that mirror the corresponding cut entropy distribution. This bond dimension profile captures the high bipartite entanglement towards the wave packet center, where the singular value spectrum decays more slowly. The accuracy of this procedure is evaluated through the distance between the resulting MPS and the target state, $|\tilde{\psi}\rangle_{\text{in},[n]}^{'[m]}$, with $\epsilon_{d} = \left\| \, |\tilde{\psi}\rangle_{\text{in},[n]}^{'[m]} - |\tilde{\psi}\rangle_{\text{in},[n]}^{[m]}\, \right\|_2^2$.

To further improve the MPS representation of the incoming state, the MPS tensors are optimized to minimize $\epsilon_{d}$. As detailed in Ref.~\cite{PhysRevLett.93.207204,schollwock2011density}, the optimization problem can be approximated by a sequence of linear local optimization (iterative compression). In this work, iterative compression is performed by optimizing one site tensor at a time to minimized $\epsilon_{d}$. The optimization is restricted the support of truncated $g_j$. The variational sweep is conducted once from left to right and then from right to left across this localized spatial region. 

In this work, coefficients with $|g_j|<10^{-6}$ are first discarded. The subsequent MPS compression yields $\epsilon_d\sim10^{-10}$.

\subsection{Time evolution with MPS}
\label{app:mpsintro6}

This section summarizes the methods used for the real-time simulations presented in Sec.~\ref{sec:IV2}. In particular, the errors arise from MPS truncations are discussed. 

The time evolution with MPS is performed using time evolving block decimation (TEBD) algorithm~\cite{PhysRevLett.93.040502} with OBC, which applies two-site gates using the Suzuki--Trotter formula~\cite{Trotter:1959ytf,Suzuki:1976be} and truncates the updated MPS after applying each gate. The accuracy is controlled jointly by the time step and the truncation threshold. In this work, a second-order Suzuki--Trotter formula with $\delta t=0.02$ is used. The singular value cutoff is initially set to $10^{-9}$, with maximum bound dimension $D=200$ or $300$.

TEBD truncation discards a component orthogonal to the retained state after applying each gate, where the truncation error may be characterized by the discarded weight at each step. The accumulated loss of squared norm, $1-\langle \tilde{\psi}(t)|\tilde{\psi}(t)\rangle$, provides a practical estimate of the truncation error, as also used in Ref.~\cite{Zemlevskiy:2026kpc} for scattering simulations. For a single truncation, this loss is equivalent to the infidelity relative to the untruncated state. For successive truncations, it provides an accurate estimate of the infidelity when truncation errors are random and uncorrelated~\cite{Zhou:2020kkm}. 

For $\tilde{m}_0=0.1$ and $\tilde{\lambda}_0=0.8$, Table~\ref{tab:TEBDerrors} summarizes this norm loss for the scattering simulations at the optimal NS sampling configurations. The small accumulated norm loss may reflect limited entanglement growth in the scattering of $\phi^4$ theory. Increasing the bond dimension from $D=200$ to $300$ reduces this loss modestly over the simulated interval.

\begin{table*}[t]
    \centering
    \caption{Accumulated loss of squared norm, $1-\langle \tilde{\psi}(t)|\tilde{\psi}(t)\rangle$, during TEBD evolution for $\tilde{m}_0=0.1$ and $\tilde{\lambda}_0=0.8$.}
    \label{tab:TEBDerrors}
    \small
    \setlength{\tabcolsep}{4pt}
    \begin{tabular}{cc*{6}{c}}
        \hline
        $d$ & $D$
        & $t=20$ & $t=40$ & $t=60$
        & $t=80$ & $t=100$ & $t=120$ \\
        \hline
        6  & 200 & $3.2356 \times 10^{-8}$ & $7.20246\times10^{-8}$ & $1.33495\times10^{-7}$ & $2.05433\times10^{-7}$ & $2.78583\times10^{-7}$ & $3.47113\times10^{-7}$ \\
        6  & 300 & $2.29465\times10^{-8}$ & $5.47666\times10^{-8}$ & $9.54205\times10^{-8}$ & $1.39968\times10^{-7}$ & $1.81215\times10^{-7}$ & $2.16551\times10^{-7}$ \\
        8  & 200 & $1.04018\times10^{-8}$ & $2.75621\times10^{-8}$ & $5.91976\times10^{-8}$ & $9.66672\times10^{-8}$ & $1.35812\times10^{-7}$ & $1.72122\times10^{-7}$ \\
        8  & 300 & $6.67082\times10^{-9}$ & $1.82112\times10^{-8}$ & $3.73076\times10^{-8}$ & $5.88667\times10^{-8}$ & $7.93389\times10^{-8}$ & $9.79271\times10^{-8}$ \\
        10 & 200 & $6.05901\times10^{-8}$ & $1.51159\times10^{-7}$ & $2.63691\times10^{-7}$ & $3.59751\times10^{-7}$ & $4.35789\times10^{-7}$ & $4.99217\times10^{-7}$ \\
        10 & 300 & $3.6164\times10^{-8}$ & $1.05108\times10^{-7}$ & $1.91889\times10^{-7}$ & $2.70104\times10^{-7}$ & $3.30762\times10^{-7}$ & $3.78325\times10^{-7}$ \\
        \hline
    \end{tabular}
\end{table*}

\section{Simulations of scattering: one step toward the thermodynamic and continuum limits}
\label{app:scatteringcountlim}

\begin{figure*}
 \begin{minipage}{1\textwidth}
 \includegraphics[width = \textwidth]{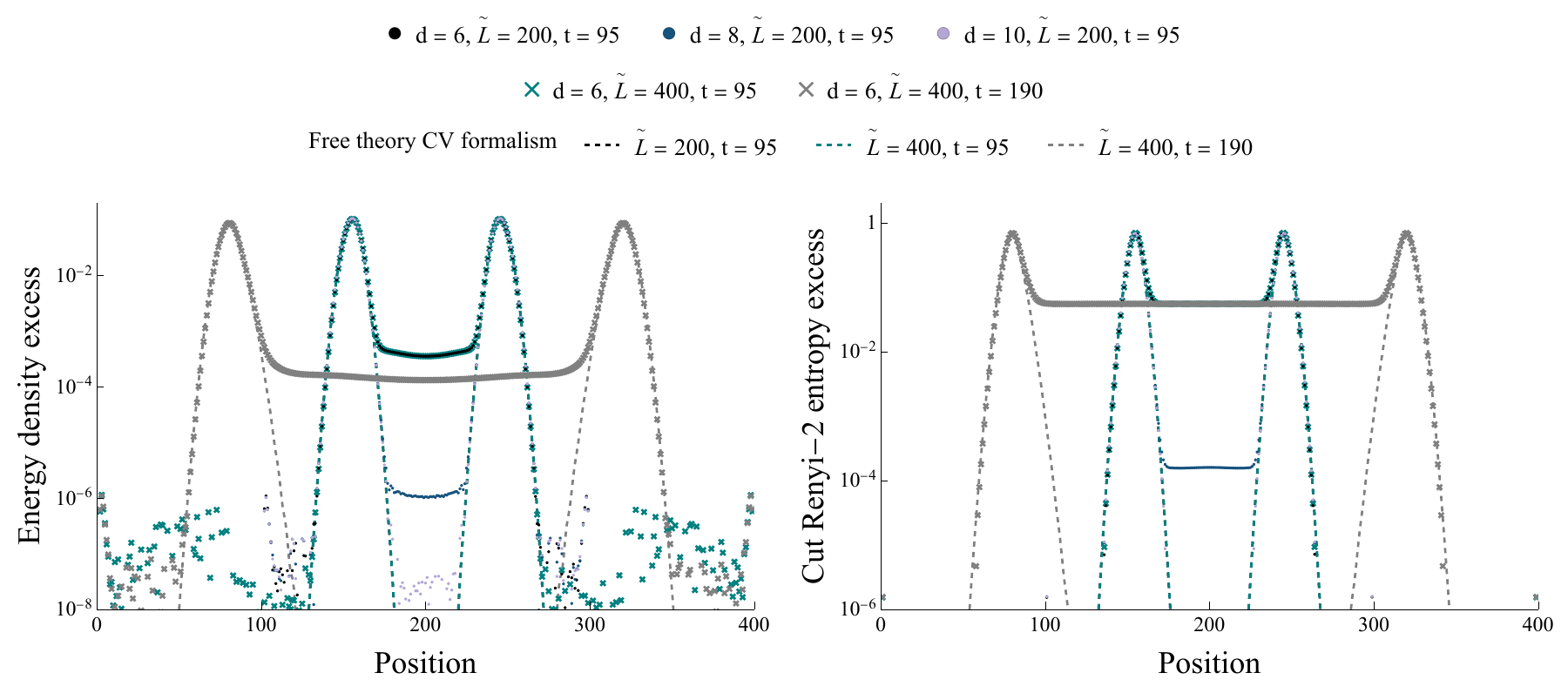}
  \end{minipage}\\
\caption{Free theory MPS simulation with $\tilde{m}_{\text{phy}} = 0.43223$ at optimal digitization configurations. $\Phi$ is calculated from a PBC lattice with $\tilde{L}=41$. (Left) Energy density excess and (Right) cut Renyi-2 entropy excess at given time slices. Scattering simulations with different lattice sizes, $\tilde{L}$, are compared with the same momentum space wave packet profile. The initial wave packet positions are adjusted to align their separation and collision center. Dots and crosses show simulations with $\tilde{L}=200$ and $\tilde{L}=400$. The dotted curves show free theory predictions from the CV formalism, using the same lattice physical mass and momentum space wave packet profiles with their corresponding MPS simulations.}
\label{fig:freescatterplot}
\end{figure*}

As discussed in the main text, fixing the relative phases of the single particle momentum eigenstates allows $g_j$ and $f_{\tilde{k}}$ to be scaled consistently toward the continuum limit. This section describes the procedure for increasing the volume and reducing the lattice spacing in the scattering simulations and examines the effects of taking one step toward the thermodynamic and continuum limits.

Throughout this work, simulations are performed at fixed physical volume, chosen sufficiently large that finite-volume effects are negligible (see Fig.~\ref{fig:freescatterplot}). Computations are performed directly in the dimensionless lattice formulation. The dynamics are characterized by the energy density $\tilde{h}_{OBC,j}$,
\begin{equation}
    \tilde{H}_{OBC} = \sum_{j}  \tilde{h}_{OBC,j} , \quad \tilde{h}_{OBC,j} = \frac{1}{2}\tilde{\pi}_j^2 + \frac{1}{2} \tilde{m}_{0}^2 \tilde{\phi}_j^2 + \frac{\tilde{\lambda}_0}{4!} \tilde{\phi}_j^4 + \frac{1}{4}\left(\tilde{\phi}_{j+1}-\tilde{\phi}_j\right)^2 + \frac{1}{4}\left(\tilde{\phi}_{j}-\tilde{\phi}_{j-1}\right)^2 \ \ \ ,
\label{eq:EdensityOBC1}
\end{equation}
where $j = 2,\dots,\tilde{L}-1$. The boundary term reads,
\begin{equation}
 \tilde{h}_{OBC,1} = \frac{1}{2}\tilde{\pi}_1^2 + \frac{1}{2} \tilde{m}_{0}^2 \tilde{\phi}_1^2 + \frac{\tilde{\lambda}_0}{4!} \tilde{\phi}_1^4 + \frac{1}{4}\left(\tilde{\phi}_{2}-\tilde{\phi}_{1}\right)^2 + \frac{1}{2} \tilde{\phi}_{1}^2 \ \ \ ,
\label{eq:EdensityOBC2}
\end{equation}
and similar for $\tilde{h}_{OBC,\tilde{L}}$. The entanglement is characterized by the Renyi-2 entropy across the cut between sites $j$ and $j+1$,
\begin{equation}
    \hat{S}_2(j,t) = -\log \operatorname{Tr}\!\left[\rho_{[1,j]}(t)^2\right],
\end{equation}
where $\rho_{[1,j]}(t)$ is the reduced density matrix of sites $1,\dots,j$. 

The wave packets are constructed as Gaussian superpositions of single particle momentum eigenstates. Let $\mu$, $\sigma$, and $\Delta$ denote the mean, standard deviation, and translation. The superposition may be expressed as,
\begin{equation}
\hat{\Delta} \equiv \frac{\Delta}{a}, \quad \hat{\mu}_{\hat{k}} \equiv a \mu_{k}, \quad \hat{\sigma}_{\hat{k}} \equiv a \sigma_{k}, \quad |\hat{\psi}\rangle_{1P}
= 
\frac{1}{\hat{L}}\sum_{\hat{k}} \mathcal{N} e^{-i\hat{k} \hat{\Delta}} \, e^{-\frac{(\hat{k}-\hat{\mu}_{\hat{k}})^2}{2\hat{\sigma}_{\hat{k}}^2}} |\hat{k}\rangle \ \ \ ,
\label{eq:momentumprofilen}
\end{equation}
where $\mathcal{N}$ normalizes the momentum space profile following Eq.~\eqref{eq:wpnormalizationmain}.

To take one step towards the thermodynamic limit, the physical volume is scaled while the lattice spacing is kept fixed. This corresponds to $\hat{L}\rightarrow\gamma\hat{L}$ and $a\rightarrow a$, leaving the bare mass and the coupling unchanged. The momenta that the lattice supports scale as, $\delta k\rightarrow \frac{\delta k}{\gamma}$ while $k_{\mathrm{\text{max}}}\rightarrow k_{\mathrm{\text{max}}}$. The lattice mean, standard deviation and translation are unchanged. In this limit, the position space profile of the interpolating operator yields Eq.~\eqref{eq:positionprofiledynamics}. Fig.~\ref{fig:freescatterplot} shows the simulation results as the lattice volume is increased, where observables between the $\tilde{L}=200$ and $400$ lattices agree, and the larger lattice effectively adds more sites on each side. This indicates that boundary effects are negligible for the observables and evolution times considered in this work.

\begin{figure*}
 \begin{minipage}{1\textwidth}
 \includegraphics[width = \textwidth]{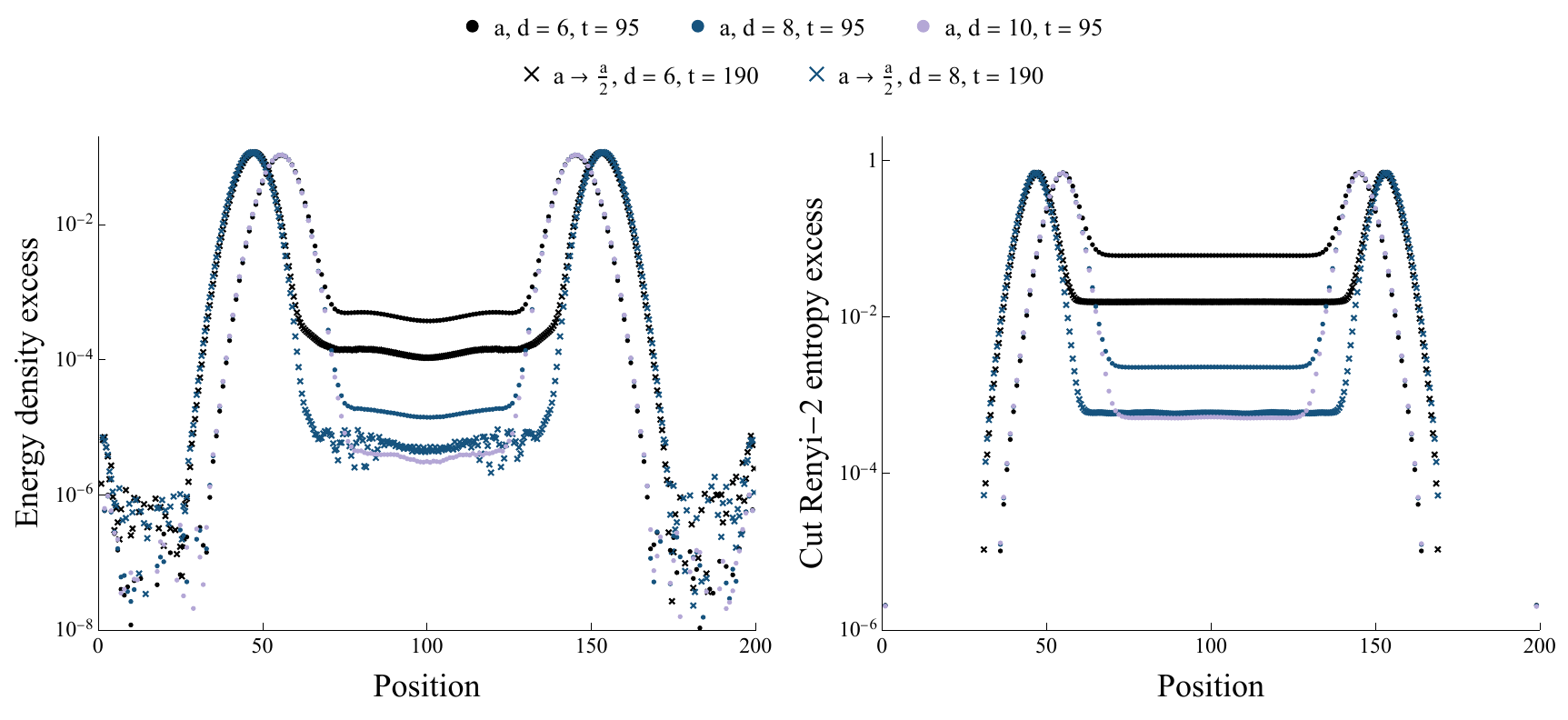}
  \end{minipage}\\
\caption{Interacting theory MPS simulation with $\tilde{m}_0 = 0.1$, $\tilde{\lambda}_0 = 0.8$ and $\tilde{m}_0 = 0.05$, $\tilde{\lambda}_0 = 0.2$ at optimal digitization configurations. $\Phi$ is calculated from a PBC lattice with $\tilde{L}=41$ and $\tilde{L}=81$. (Left) Energy density excess and (Right) cut Renyi-2 entropy excess at given time slices. Scattering simulations with different lattice spacing are compared with the same wave packet profile in physical momentum space. The initial wave packet positions are adjusted to align their physical position. Dots and crosses show simulations with $a$ and $a\to \frac{a}{2}$. The wave packet profiles show disagreement at the same physical time due to the course initial lattice discretization.}
\label{fig:UVIRscatterplot}
\end{figure*}

To take the continuum limit, the lattice spacing is reduced by a factor of $\gamma'$ while the physical volume is kept fixed. This corresponds to $\hat{L}\rightarrow\gamma'\hat{L}$ and $a\rightarrow \frac{a}{\gamma'}$, such that $\hat{m}_0 \to \frac{\hat{m}_0}{\gamma'}$ and $\hat{\lambda}_0 \to \frac{\hat{\lambda}_0}{\gamma'^{2}}$. The momenta that the lattice supports scale as $\delta k\rightarrow \delta k$ while $k_{\mathrm{\text{max}}}\rightarrow \gamma' k_{\mathrm{\text{max}}}$. For a Gaussian superposition in momentum space, $\hat{\mu}_{\hat{k}}\rightarrow \frac{\hat{\mu}_{\hat{k}}}{\gamma'}$, $\hat{\sigma}_{\hat{k}}\rightarrow \frac{\hat{\sigma}_{\hat{k}}}{\gamma'}$ and $\hat{\Delta}\rightarrow \gamma' \hat{\Delta}$. The scaling of the Hamiltonian implies that, at fixed physical time, the lattice time increases as $1/a$ toward the continuum limit. Vacuum subtraction is performed to extract finite physical observables for energy density, as well as Renyi-2 entropy~\cite{PhysRevLett.121.170602}. The local energy densities in lattice and physical units are related by
$\hat{h}_{\mathrm{OBC},j}=a^2 h_{\mathrm{OBC},j}$ while the Renyi-2 entropy is dimensionless. This allows a simple scaling $\hat{h}_{\mathrm{OBC},j}\rightarrow \gamma'^{2}\hat{h}_{\mathrm{OBC},j}$ and $\hat{S}_2 \to \hat{S}_2$. Fig.~\ref{fig:UVIRscatterplot} shows one step towards the continuum limit, $a\to a/2$, for the scattering setup considered in this work. The interacting theory results at $a$ are shown at the same lattice time as those in Fig.~\ref{fig:freescatterplot}. It is seen that nonzero energy density and Renyi-2 entropy excess are presented at $d=10$ in the interacting theory. 

\section{Airy distortion in single particle wave packet dynamics}
\label{app:latartifact}

To accurately characterize particle production in wave packet collisions, it is important to understand the particle-like distortions that arise during the scattering processes. Such distortions have been recently observed under the context of two-particle scattering and are described as a ``checkerboard pattern''~\cite{farrell2026digital}. To clarify the origin of these distortions and derive conditions under which they are suppressed, this section considers single particle wave packet dynamics. Conditions on the evolution time and momentum space distribution profile are then derived to suppress these distortions. This analysis requires higher order derivatives of the lattice dispersion relation,
\begin{equation}
\tilde{s}_{\tilde{m},\tilde{k}} \equiv \frac{\partial^2\tilde{E}_{\tilde{m},\tilde{k}}}{\partial\tilde{k}^2} = \frac{\cos\tilde{k}}{\tilde{E}_{\tilde{m},\tilde{k}}} - \frac{\tilde{v}_{\tilde{m},\tilde{k}}^2}{\tilde{E}_{\tilde{m},\tilde{k}}}, \quad
\tilde{u}_{\tilde{m},\tilde{k}} \equiv \frac{\partial^3\tilde{E}_{\tilde{m},\tilde{k}}}{\partial\tilde{k}^3} = -\tilde{v}_{\tilde{m},\tilde{k}}\left(1+\frac{3\tilde{s}_{\tilde{m},\tilde{k}}}{\tilde{E}_{\tilde{m},\tilde{k}}}\right) \ \ \ ,
\label{eq:derivativesofE}
\end{equation}
where $\tilde{m}_{\mathrm{phys}}$ is denoted by $\tilde{m}$ throughout this appendix.

For short evolution times, it suffices to examine the Gaussian integral,
\begin{equation}
\mathcal{I} \xrightarrow[
    \frac{t}{6} |\tilde{u}_{\tilde{m},\tilde{\mu}_{\tilde{k}}}| \tilde{\sigma}_{\tilde{k}}^3 \ll 1
]{
    \frac{|\sin\tilde{\mu}_{\tilde{k}}|\,\tilde{\sigma}_{\tilde{k}}} {2\tilde{E}_{\tilde{m},\tilde{\mu}_{\tilde{k}}}^{2}} \ll 1
}e^{i\left(\tilde{\mu}_{\tilde{k}}j-\tilde{E}_{\tilde{m},\tilde{\mu}_{\tilde{k}}}t\right)}\sqrt{2\tilde{E}_{\tilde{m},\tilde{\mu}_{\tilde{k}}}} \int_{-\infty}^{\infty} \frac{d\tilde{k}}{2\pi}e^{i\left( \left(j-t\tilde{v}_{\tilde{m},\tilde{\mu}_{\tilde{k}}}\right)(\tilde{k}-\tilde{\mu}_{\tilde{k}}) - \frac{t}{2}\tilde{s}_{\tilde{m},\tilde{\mu}_{\tilde{k}}}(\tilde{k}-\tilde{\mu}_{\tilde{k}})^2 \right)} e^{-\frac{(\tilde{k}-\tilde{\mu}_{\tilde{k}})^2}{2\tilde{\sigma}_{\tilde{k}}^2}} \ \ \  ,
\label{eq:positionprofiledynamicssimp1}
\end{equation}
where $\frac{|\sin\tilde{\mu}_{\tilde{k}}|\,\tilde{\sigma}_{\tilde{k}}} {2\tilde{E}_{\tilde{m},\tilde{\mu}_{\tilde{k}}}^{2}} \ll 1$ allows $\sqrt{2\tilde{E}_{\tilde{m},\tilde{k}}}$ to be approximated by its value at $\tilde{\mu}_{\tilde{k}}$. The localized Gaussian momentum distribution allows the integration range to be extended to $(-\infty,\infty)$. With $\tilde{u}_{\tilde{m},\tilde{k}}\neq 0$ for $\tilde{k}\in \left(0,\pi\right)$, the condition $\frac{t}{6} |\tilde{u}_{\tilde{m},\tilde{\mu}_{\tilde{k}}}| \tilde{\sigma}_{\tilde{k}}^3 \ll 1$ ensures that the second order expansion of the phase about $\tilde{\mu}_{\tilde{k}}$ is sufficiently accurate. Approximation of the short time dynamics thus reads,
\begin{equation}
\mathcal{I}
\simeq
\frac{
\tilde{\sigma}_{\tilde{k}}\sqrt{2\tilde{E}_{\tilde{m},\tilde{\mu}_{\tilde{k}}}}
}{\sqrt{2\pi}
\sqrt{1+it\tilde{\sigma}_{\tilde{k}}^2\tilde{s}_{\tilde{m},\tilde{\mu}_{\tilde{k}}}}
}
\exp\left[
i\left(\tilde{\mu}_{\tilde{k}}j-\tilde{E}_{\tilde{m},\tilde{\mu}_{\tilde{k}}}t\right)
-
\frac{
\tilde{\sigma}_{\tilde{k}}^2
\left(j-t\tilde{v}_{\tilde{m},\tilde{\mu}_{\tilde{k}}}\right)^2
}{
2\left(1+it\tilde{\sigma}_{\tilde{k}}^2\tilde{s}_{\tilde{m},\tilde{\mu}_{\tilde{k}}}\right)
}
\right]
\label{eq:shorttimeG}
\ \ \ ,
\end{equation}
where the position space Gaussian envelope propagates at the group velocity and broadens at a rate determined by the curvature of the dispersion relation,
\begin{equation}
\mu_j(t)
=
t\tilde{v}_{\tilde{m},\tilde{\mu}_{\tilde{k}}},
\quad
\sigma_j(t)
=
\frac{
\sqrt{1+t^2\tilde{\sigma}_{\tilde{k}}^4
\tilde{s}_{\tilde{m},\tilde{\mu}_{\tilde{k}}}^2}
}{
\tilde{\sigma}_{\tilde{k}}
}
\ \ \ .
\label{eq:shorttimebroaden}
\end{equation}
It is seen that the envelope develops no oscillatory structure within this approximation. To capture oscillations of the wave packet creation profile, it is required to retain as least the cubic term in the phase expansion.

At long evolution times, rapid phase oscillations in Eq.~\eqref{eq:positionprofiledynamics} cause contributions from most of the integration domain to cancel, leaving the dominant contributions near stationary points, where the phase varies slowly~\cite{Copson_1965,NIST:DLMF}. Contributions of the stationary phase can interfere to produce oscillatory Airy distortions. The stationary phase condition for the lattice is,
\begin{equation}
\left.
\frac{\partial}{\partial\tilde{k}}
\left(\tilde{k}j-\tilde{E}_{\tilde{m},\tilde{k}}t\right)
\right|_{\tilde{k}=\tilde{k}_{\pm}}
=0
\quad\Longrightarrow\quad\tilde{v}_{\tilde{m},\tilde{k}_{\pm}} = \frac{j}{t}, \quad -\tilde{v}_{\tilde{m},\tilde{k}_c} \leq \tilde{v}_{\tilde{m},\tilde{k}_{\pm}} \leq \tilde{v}_{\tilde{m},\tilde{k}_c}, \quad \tilde{v}_{\tilde{m},\tilde{k}_c}
=\frac{\sqrt{\tilde{m}^2+4}-\tilde{m}}{2} \ \ \ ,
\label{eq:latticesaddle}
\end{equation}
where $\tilde{v}_{\tilde{m},\tilde{k}_c}$ is the maximum group velocity supported by the lattice. The corresponding momenta are,
\begin{equation}
\tilde{k}_{\pm} =  \left(2\theta\left(\frac{j}{t}\right)-1\right)
\arccos\left(
\frac{j^2}{t^2}
\mp \sqrt{\frac{j^4}{t^4}-(\tilde{m}^2+2)\frac{j^2}{t^2}+1}
\right) \ \ \ ,
\label{eq:restationarypoints}
\end{equation}
where $\theta\left(\cdot\right)$ is the Heaviside step function, and $|\tilde{k}_{+}| \geq |\tilde{k}_{-}|$. The two solutions coalesce when $|j/t|=\tilde{v}_{\tilde{m},\tilde{k}_c}$. It is the nonmonotonic momentum dependence of the lattice group velocity that gives rise to these two stationary points. When the two stationary points are well separated, each contribution can be approximated using a quadratic phase expansion over an interval $t^{-1/2} \ll \delta \ll t^{-1/3}$~\cite{Copson_1965}. While the stationary points approach each other such that $|\tilde{k}_{+}-\tilde{k}_{-}| \lesssim t^{-1/2}$, the contributions of stationary points can no longer be treated separately. To account for their coalescence, a uniform approximation involving the Airy function and its derivative is used, which remains valid for both well separated and coalescing stationary points~\cite{chester1957extension,bleistein1966uniform,NIST:DLMF}. The construction begins by introducing a new coordinate, $\tilde{p}$, that brings the phase into cubic form,
\begin{equation}
\tilde{k} \frac{j}{t} -\tilde{E}_{\tilde{m},\tilde{k}} = \frac{1}{3} \tilde{p}^3 - \tilde{p} \xi^2 + \eta , \quad \frac{\partial \tilde{k}}{\partial \tilde{p}} =\frac{\tilde{p}^{2}-\xi^2}{\frac{j}{t}-\tilde{v}_{\tilde{m},\tilde{k}}}, \quad \left(\frac{\partial \tilde{k}}{\partial \tilde{p}}\right)^2 =\frac{2\tilde{p}}{-\tilde{s}_{\tilde{m},\tilde{k}}} - \frac{\frac{j}{t}-\tilde{v}_{\tilde{m},\tilde{k}}}{-\tilde{s}_{\tilde{m},\tilde{k}}} \frac{\partial^2 \tilde{k}}{\partial \tilde{p}^2} \ \ \  ,
\label{eq:positionprofiledynamicscubic}
\end{equation}
where $\xi$ and $\eta$ are functions of $\tilde{m}$ and $j/t$. Here, the saddle points are matched to ensure that the coordinate transformation is uniformly regular at the saddle points~\cite{chester1957extension}. For analytical integrand, it is $\partial\tilde{k}/\partial\tilde{p} \neq 0$ that ensures monotonic relation between $\tilde{p}$ and $\tilde{k}$, which allows a well defined change of integration variable. The coefficients are obtained through matching vanishing first derivatives at stationary points, $\tilde{k}\left(\tilde{p} = \xi\right) = \tilde{k}_{+}$ and $\tilde{k}\left(\tilde{p} = -\xi\right) = \tilde{k}_{-}$, yielding,
\begin{equation}
\xi^3 = \frac{3\left(\tilde{k}_{-}\frac{j}{t} -\tilde{E}_{\tilde{m},\tilde{k}_{-}}  -\tilde{k}_{+}\frac{j}{t} +\tilde{E}_{\tilde{m},\tilde{k}_{+}} \right)}{4}, \quad \eta = \frac{\tilde{k}_{-}\frac{j}{t} -\tilde{E}_{\tilde{m},\tilde{k}_{-}}  + \tilde{k}_{+}\frac{j}{t} -\tilde{E}_{\tilde{m},\tilde{k}_{+}} }{2} \ \ \ .
\end{equation}
With $\tilde{v}^{'}_{\tilde{m},\tilde{k}_c} \equiv (\sqrt{\tilde{m}^2+4}+\tilde{m})/2$, the behavior separates into the following regimes:
\begin{enumerate}
\item For $|j/t| \leq \tilde{v}_{\tilde{m},\tilde{k}_c}$, $\tilde{k}_{\pm}$ are given by Eq.~\eqref{eq:restationarypoints} with $\xi, \eta\in\mathbb{R}$, and since $|\tilde{k}_{+}| \geq |\tilde{k}_{-}|$, $\xi \geq 0$.
\item For $\tilde{v}_{\tilde{m},\tilde{k}_c}<|j/t|<\tilde{v}^{'}_{\tilde{m},\tilde{k}_c}$, Eq.~\eqref{eq:restationarypoints} is analytically continued~\footnote{Beyond $\tilde{v}_{\tilde{m},\tilde{k}_c}$, no real stationary points exist. The matching of stationary points persist through analytic continuation, as they govern the leading exponential decay according to the method of steepest descent. See also Sec. 36.12 in Ref.~\cite{NIST:DLMF}. Here, the relations in Eq.~\eqref{eq:derivativesofE} and~\eqref{eq:positionprofiledynamicscubic} continue to hold for complex arguments.} with stationary points satisfy $\tilde{k}_{+}=\tilde{k}_{-}^{*}$, hence $i\xi^3, \eta\in\mathbb{R}$. For $\tilde{k} \in \mathbb{R}$ as in Eq.~\eqref{eq:positionprofiledynamics}, the condition $i\xi\in\mathbb{R}$ is imposed, which allows $\tilde{p} \in \mathbb{R}$ with $\partial\tilde{k}/\partial\tilde{p} \neq 0$, hence Eq.~\eqref{eq:positionprofiledynamics} to be written as an integral along real $\tilde{p}$ with a cubic phase.

\item For $|j/t| \geq \tilde{v}^{'}_{\tilde{m},\tilde{k}_c}$, define $\tilde{k}_{\pm} \equiv  \pm i \left(2\theta\left(\frac{j}{t}\right)-1\right) \kappa_{+}$ with $\tilde{\kappa}_{\pm} \equiv \mathrm{arcosh}\left(
\frac{j^2}{t^2}
\mp \sqrt{\frac{j^4}{t^4}-(\tilde{m}^2+2)\frac{j^2}{t^2}+1}
\right)$~\footnote{As $|j/t|\to\tilde{v}^{'}_{\tilde{m},\tilde{k}_c}$, the argument of $\arccos$ in the analytically continued Eq.~\eqref{eq:restationarypoints} approaches its branch cut on the real axis, with the resulting $\tilde{k}_{\pm}$ approach the branch cut of $\tilde{E}_{\tilde{m},\tilde{k}}$ on the imaginary axis. When $|j/t|>\tilde{v}^{'}_{\tilde{m},\tilde{k}_c}$, the pair $\tilde{k}_{\pm}$ splits into four saddle points, $\pm i\tilde{\kappa}_{\pm}$, each understood as a one-sided limit approaching the branch cut (the same one-sided limits apply to $P$ and $Q$ constructed below). In this regime, the cubic phase transformation in Eq.~\eqref{eq:positionprofiledynamicscubic} cannot match all four saddle points and is therefore constructed by matching the conjugate pair $\tilde{k}_{\pm}$. Taking the limit toward the branch cut yields,
\begin{equation}
\xi^3 = \frac{3i}{2} \left(-\left(2\theta\left(\frac{j}{t}\right)-1\right)\tilde{\kappa}_{+}\frac{j}{t}+\sqrt{-\tilde{m}^2+4\sinh^2{\frac{\tilde{\kappa}_{+}}{2}}}\right) , \quad \eta = 0, \quad |j/t| > \tilde{v}^{'}_{\tilde{m},\tilde{k}_c} \ \ \ ,
\label{eq:otherregimexieta}
\end{equation}
which not only ensures a continuous transition between the regimes but captures the dominant decaying contribution when the four saddles are well separated. The absolute error when $i\tilde{\kappa}_{+}$ and $i\tilde{\kappa}_{-}$ are not well separated is bounded above by $\mathcal{O}(t^{-1})$, subleading relative to the $\mathcal{O}(t^{-1/2})$ contribution from isolated real stationary points.}. This choice yields a pure imaginary conjugate pair, allowing the same construction in 2 to extend to this regime.
\end{enumerate}
The integration thus becomes,
\begin{equation}
g_j\left(t\right)  \simeq \int_{\tilde{p}\left(-\pi\right)}^{\tilde{p}\left(\pi\right)}\frac{d\tilde{p}}{2\pi} e^{i\left(\frac{1}{3} \tilde{p}^3 - \tilde{p} \xi^2 + \eta \right) t}e^{-\frac{(\tilde{k}\left(\tilde{p}\right)-\tilde{\mu}_{\tilde{k}})^2}{2\tilde{\sigma}_{\tilde{k}}^2}} \sqrt{2\tilde{E}_{\tilde{m},\tilde{k}\left(\tilde{p}\right)}} \frac{\partial \tilde{k}\left(\tilde{p}\right)}{\partial \tilde{p}} \ \ \ ,
\label{eq:positionprofiledynamicscubicint}
\end{equation}
where the derivative can be evaluated with Eq.~\eqref{eq:positionprofiledynamicscubic}. For $j/t \geq 0$, $\frac{dk}{dp}>0$, thus $\tilde{p}\left(\pi\right) > \tilde{p}\left(-\pi\right)$. Similarly, for $j/t < 0$, $\frac{dk}{dp}<0$, thus $\tilde{p}\left(\pi\right) < \tilde{p}\left(-\pi\right)$. The derivative at the stationary points can be evaluated via $\left(\partial\tilde{k}/\partial\tilde{p}\right)^2$ in Eq.~\eqref{eq:positionprofiledynamicscubic}, or similarly $\left(\partial\tilde{k}/\partial\tilde{p}\right)^3$ at coalescence, with the latter typically avoided for lattice evaluations. Define the auxiliary functions,
\begin{equation}
G\left(\tilde{p}\right) \equiv e^{-\frac{(\tilde{k}\left(\tilde{p}\right)-\tilde{\mu}_{\tilde{k}})^2}{2\tilde{\sigma}_{\tilde{k}}^2}} \sqrt{2\tilde{E}_{\tilde{m},\tilde{k}\left(\tilde{p}\right)}} \sqrt{\frac{2\tilde{p}}{-\tilde{s}_{\tilde{m},\tilde{k}}\left(\tilde{p}\right)}}, \quad P \equiv\frac{G\left(\xi\right) + G\left(-\xi\right)}{2}, \quad Q \equiv\frac{G\left(\xi\right) - G\left(-\xi\right)}{2\xi} \ \ \ ,
\label{eq:auxiliaryG}
\end{equation}
where $P, Q \in \mathbb{R} \quad \forall |j/t|$. For $v\geq0$, $G\left(\tilde{p}\right)$ equals to $P + Q \tilde{p}$ at the stationary phase points. For $v<0$, $G(\tilde{p})$ accounts for the reversed integration limits $\tilde{p}(\pi)<\tilde{p}(-\pi)$, which, together with the assumption of a localized Gaussian momentum distribution, allows the integration range to be extended to $(-\infty,\infty)$ for all $v$. An uniform approximation of the long time dynamics therefore reads,
\begin{equation}
\mathcal{I} \xrightarrow{t \gg 1} e^{i\eta t}  \int_{-\infty}^{\infty}\frac{d\tilde{p}}{2\pi} e^{i\left(\frac{1}{3} \tilde{p}^3 - \tilde{p} \xi^2 \right)t} \left(P + Q \tilde{p}\right) = e^{i\eta t}\left(
\frac{P}{t^{1/3}}\operatorname{Ai}(-t^{2/3}\xi^2)
-i\frac{Q}{t^{2/3}}\operatorname{Ai}'(-t^{2/3}\xi^2)
\right) \ \ \ ,
\label{eq:AiAiprime}
\end{equation}
where $\operatorname{Ai}(\cdot)$ is the Airy function of the first kind and $\operatorname{Ai}'(\cdot)$ denotes its derivative with respect to its argument. Here, $P$, $Q$, $\xi$, and $\eta$ are functions of $\tilde{m}$ and $j/t$. The dependence on $j/t$ sets a spatial scale that grows linearly with time, while the prefactors $t^{-1/3}$ and $t^{-2/3}$ suppress the amplitudes of the two Airy contributions. Meanwhile, the factor $t^{2/3}$ in the Airy argument expands the range sampled within the dynamically visible region, revealing more of its oscillatory structure. Together, these explicit time dependences modify the linear spatial scaling to describe the evolving amplitude and shape of the propagating wave packet. Note that the argument of the Airy function and its derivative is invariant under $v \to -v$, whereas $P$ and $Q$ depend on the sign of $v$, which determines propagation direction of the wave packet. Relative to this direction, the squared amplitude, 
\begin{equation}
|\mathcal{I}|^2 \simeq \frac{P^2}{t^{2/3}} \operatorname{Ai}^2(-t^{2/3}\xi^2) + \frac{Q^2}{t^{4/3}} \operatorname{Ai}^{'2}(-t^{2/3}\xi^2) \ \ \ ,
\end{equation}
transitions from oscillatory behavior to exponential decay. This transition reflects the change from real to complex saddle points as $|j/t|$ crosses $\tilde{v}_{\tilde{m},\tilde{k}_c}$, which can also be seen from the asymptotic forms of the Airy function and its derivative (see Sec.~9.7 of Ref.~\cite{NIST:DLMF}),
\begin{equation}
\begin{aligned}
\operatorname{Ai}(x)
&\sim \frac{e^{-\frac{2}{3}x^{3/2}}}{2\sqrt{\pi}x^{1/4}},
&
\operatorname{Ai}'(x)
&\sim -\frac{x^{1/4}e^{-\frac{2}{3}x^{3/2}}}{2\sqrt{\pi}},
\\
\operatorname{Ai}(-x)
&\sim \frac{\cos\left(\frac{2}{3}x^{3/2}-\frac{\pi}{4}\right)}
{\sqrt{\pi}\,x^{1/4}},
&
\operatorname{Ai}'(-x)
&\sim \frac{x^{1/4}\sin\left(\frac{2}{3}x^{3/2}-\frac{\pi}{4}\right)}{\sqrt{\pi}},
\end{aligned}
\qquad x\gg 1 \ \ \ .
\end{equation}
These asymptotic forms exhibit a common exponential decay factor when $-t^{2/3}\xi^2 \gg 1$, i.e., corresponding to the leading side $|j/t| > \tilde{v}_{\tilde{m},\tilde{k}_c}$. In contrast, for $t^{2/3}\xi^2 \gg 1$, the asymptotic forms are oscillatory, corresponding to the trailing side $|j/t| < \tilde{v}_{\tilde{m},\tilde{k}_c}$. The squared profile in this regime reads,
\begin{equation}
\begin{aligned}
|\mathcal{I}|^2
\xrightarrow{t^{2/3}\xi^2\gg1}& \frac{1}{\pi t\xi}
\left(
P^2\cos^2\left(\frac{2}{3}t\xi^3-\frac{\pi}{4}\right)
+\xi^2Q^2\sin^2\left(\frac{2}{3}t\xi^3-\frac{\pi}{4}\right)
\right)
\\
=& \frac{1}{2\pi t\xi}
\left(
P^2+\xi^2Q^2
+\left(P^2-\xi^2Q^2\right)
\sin\left(\frac{4}{3}t\xi^3\right)
\right)
\\
=& \frac{1}{4\pi t\xi}
\left(
G\left(\xi\right)^2+G\left(-\xi\right)^2+2 G\left(\xi\right)G\left(-\xi\right)
\sin\left(\frac{4}{3}t\xi^3\right)
\right) \ \ \ ,
\end{aligned}
\end{equation}
where the simplification in the third step follows from Eq.~\eqref{eq:auxiliaryG}, and $G\left(\xi\right), G\left(-\xi\right) > 0$. To suppress oscillations in the wave packet creation profile, one saddle contribution must dominate over the other, i.e., $G(\xi)\gg G(-\xi)$ or $G(-\xi)\gg G(\xi)$. This may be achieved by placing the center of Gaussian distribution far from the coalescence point, $\tilde{\sigma}_{\tilde{k}} \lesssim |\tilde{\mu}_{\tilde{k}}-\sign\left(\tilde{\mu}_{\tilde{k}}\right)\tilde{k}_c|$, so that the contribution from the other stationary point is exponentially suppressed. Moreover, the dependence of the envelope on $j/t$ implies a spatial width proportional to $t$ at late times, consistent with the $t \gg 1$ scaling of Eq.~\eqref{eq:shorttimebroaden}. In contrast, the local oscillation wavelength remains independent of time at fixed $j/t$, $\lambda_j=2\pi/|\partial_j(\frac{4}{3}t\xi^3)|=2\pi/|4\xi^2\,d\xi/d(j/t)|=2\pi/|\tilde{k}_+-\tilde{k}_-|$. The expanding wave packet therefore accommodates an increasing number of oscillations, allowing even weak oscillations to become more apparent. To limit this effect, requiring the width in Eq.~\eqref{eq:shorttimebroaden} to remain comparable to its initial value yields $t\lesssim 1/(\tilde{\sigma}_{\tilde{k}}^2|\tilde{s}_{\tilde{m},\tilde{\mu}_{\tilde{k}}}|)$.

\begin{figure*}
 \begin{minipage}{1\textwidth}
 \includegraphics[width = \textwidth]{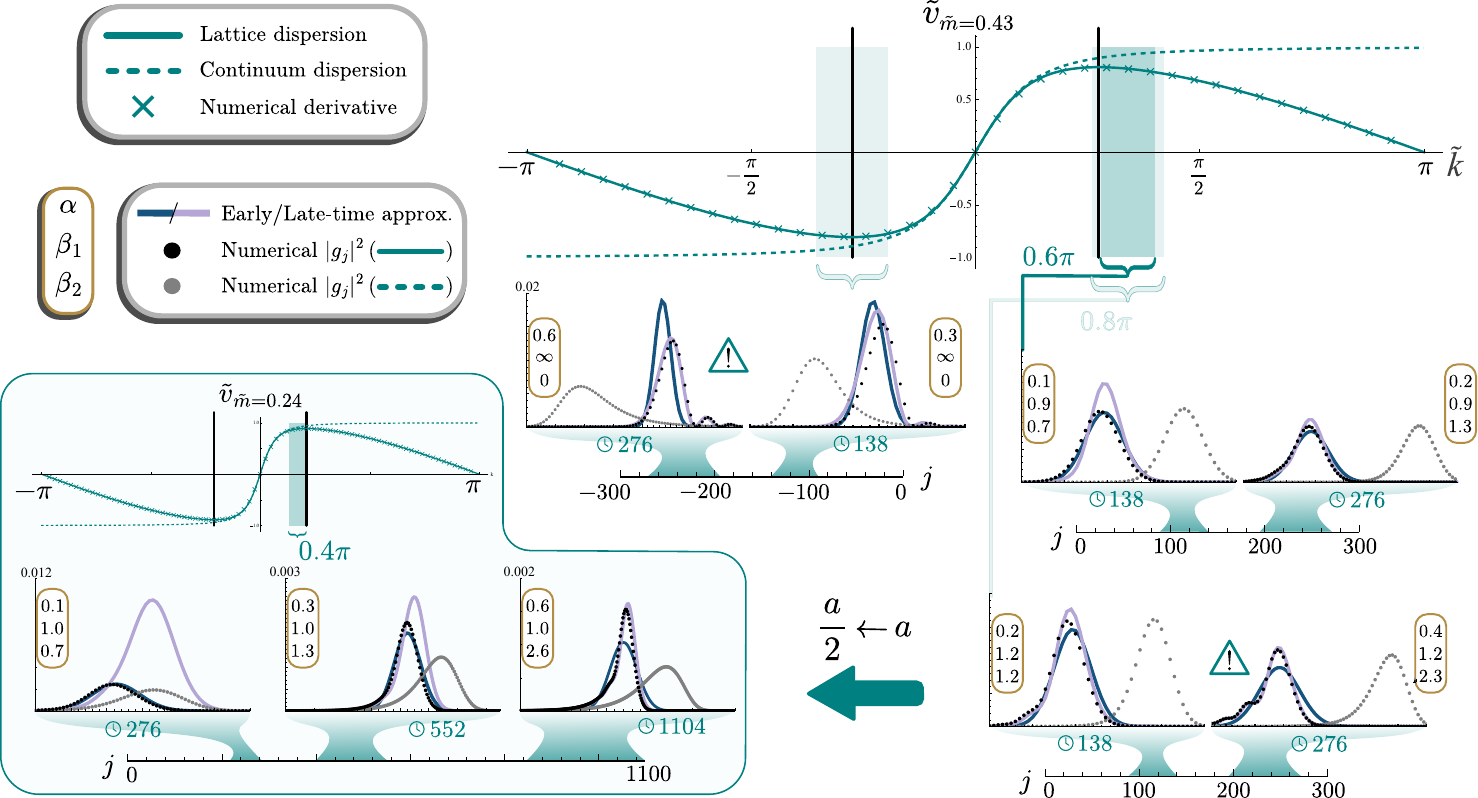}
  \end{minipage}\\
  \caption{Wave packet propagation and its distortion for different momentum widths $\tilde{\sigma}_{\tilde{k}}$, central momenta $\tilde{\mu}_{\tilde{k}}$, and evolution times $t$. The two group velocity plots compare the derivatives of the lattice (solid teal) and continuum (dotted teal) dispersion relations, evaluated using Eq.~\eqref{eq:derivativesofE} with $m$ replaced by the physical masses $\tilde m=0.43223$ and $0.23816$. Numerical derivatives of the single particle energies (teal crosses) agree well with the predictions obtained from the lattice dispersion relation using the physical mass. The initial wave packet parameters are indicated in the two group velocity plots where dotted vertical lines indicates $\tilde{\mu}_{\tilde{k}}$ and shaded bands spanning $\tilde{\mu}_{\tilde{k}}\pm\tilde{\sigma}_{\tilde{k}}$. Solid vertical lines indicate the extrema of the lattice group velocity at $\pm\tilde k_c$. The wave packets are characterized by the squared magnitude of the position space creation profile, $|g_j|^2$. Numerical results for the lattice (black dots) and continuum (gray dots) dispersions are compared with the short-time Gaussian (blue) and long-time Airy (purple) approximations, Eqs.~\eqref{eq:shorttimeG} and~\eqref{eq:AiAiprime}. Each panel lists the evolution time and the parameters $\alpha$, $\beta_1$, and $\beta_2$ characterizing the suppression of Airy distortion Eq.~\eqref{eq:negligibleA}.}
\label{fig:Adistortion}
\end{figure*}

To characterize the regime of negligible Airy distortion, define
\begin{equation}
\alpha = \frac{t}{6}
|\tilde{u}_{\tilde{m},\tilde{\mu}_{\tilde{k}}}|
\tilde{\sigma}_{\tilde{k}}^3,
\quad
\beta_1 =
\frac{\tilde{\sigma}_{\tilde{k}}}{|\tilde{\mu}_{\tilde{k}}-\sign(\tilde{\mu}_{\tilde{k}})\tilde{k}_c|},
\quad
\beta_2 = t \tilde{\sigma}_{\tilde{k}}^2
|\tilde{s}_{\tilde{m},\tilde{\mu}_{\tilde{k}}}| \ \ \ .
\label{eq:negligibleA}
\end{equation}
The conditions $\alpha\ll 1$ and $\beta_{1,2}\lesssim 1$ are motivated from Eqs.~\eqref{eq:shorttimeG} and~\eqref{eq:AiAiprime}. As shown in Fig.~\ref{fig:Adistortion}, for $\tilde{\sigma}_{\tilde{k}}=0.06\pi$ and $\tilde{\mu}_{\tilde{k}}=0.34\pi$, the wave packet exhibits negligible distortion up to $t=138$, consistent with the short-time approximation. This parameter regime is therefore adopted in the scattering simulations presented in the main text. At later times, the numerical profile begins to deviate from the short-time approximation, entering a crossover regime toward the long-time approximation. Further increasing the momentum width to $0.08\pi$ increases $\alpha$, $\beta_1$, and $\beta_2$, making Airy distortion more apparent at late times. The oscillations become even stronger when the packet is centered at $\tilde{\mu}_{\tilde{k}}=\pm\tilde{k}_c$ with the same momentum width, where the two stationary points contribute with comparable amplitudes. This lattice artifact can be suppressed by approaching the continuum limit. In this process, the mass in lattice units decreases to zero, causing $\tilde{k}_c$ to approach zero, as shown in Eq.~\eqref{eq:latticesaddle} and Fig.~\ref{fig:Adistortion}. However, the momentum of the wave packet approach zero more rapidly, resulting $\pm \tilde{k}_c$ farther into the tail of the localized momentum distribution. This behavior is illustrated in the bottom row of Fig.~\ref{fig:Adistortion}, where one step towards the continuum limit is taken. At matched physical times, Airy distortion is suppressed. The delayed onset of oscillations suggests that approaching the continuum limit allows the wave packet to retain its shape for longer time. This can also be seen by replacing the lattice dispersion relation with continuum dispersion relation, where the phase admits only one real stationary point and thus lacks this oscillatory interference in the continuum.

\section{Counter-propagating wave packets in lattice free scalar field}
\label{app:freefield}

To enable clear study of lattice and digitization artifacts on wave packet dynamics and entanglement with relevance extending to the simulation of scattering processes within interacting fields, the following develops a phase-space formalism for pure single- and distinct-momentum two-particle states in free lattice scalar fields. Formulated directly for locally continuous fields, this formalism bypasses qudit field digitization and is applicable to any bilinear Hamiltonian. As for Gaussian states~\cite{BraunsteinGaussiareview,WeedbrookGaussiareview,serafini2017quantum}, the computational cost of this formalism is found to scale polynomially with the number of field modes, in contrast to the exponential scaling of generic quantum many-body simulations. 

\subsection{Phase space formalism for pure single- and two-particle states}
\label{app:freefield1}

This section develops a phase-space formalism for simulating the dynamics of pure single- and two-particle states of distinct momenta in a lattice free scalar field theory with PBC.

Following Eq.~\eqref{eq:Hamiltonianmain}, the Hamiltonian for a single-dimensional free scalar field on a lattice with unit spacing can be expressed as,
\begin{equation}
    \hat{H}_{\text{PBC}} = \frac{1}{2} \sum_{j} \hat{\pi}_{j}^{2} + \frac{1}{2} \sum_{i,j} \hat{\phi}_{i} ( \hat{m}_0^{2} \mathbb{I} -\nabla^2)_{ij} \hat{\phi}_{j} \ \ \ ,
    \label{eq:discreteHfree}
\end{equation}
where $\mathbb{I}$ denotes the identity matrix and $\nabla^2$ denotes the discrete Laplacian matrix. With PBC, a leading-order finite-difference approximation yields circulant matrix $\nabla^2 = \text{circulant}\left[ -2, 1, 0, 0, \ldots, 0, 1\right]$. Equation~\eqref{eq:discreteHfree} may be expressed in quadratic form,
\begin{equation}
   \hat{H}_{\text{PBC}} = \frac{1}{2} \begin{pmatrix}
      \hat{\boldsymbol{\phi}}^T & \hat{\boldsymbol{\pi}}^T
   \end{pmatrix} \begin{pmatrix}
      \hat{m}_0^{2} \mathbb{I} -\nabla^2 & 0 \\ 0 & \mathbb{I}
   \end{pmatrix} \begin{pmatrix}
      \hat{\boldsymbol{\phi}}  \\ \hat{\boldsymbol{\pi}} 
   \end{pmatrix} = \frac{1}{2} \hat{\boldsymbol{r}}^T\mathcal{H}_{\text{PBC}} \hat{\boldsymbol{r}} \ \ \ ,
    \label{eq:discreteHfree1}
\end{equation}
where $\mathcal{H}_{\text{PBC}}$ is the Hamiltonian matrix and the vector of phase-space operators numbered from left-to-right in one spatial dimension reads $\hat{\boldsymbol{r}} = \left( \hat{\phi}_{-\hat{\ell}} \dots \hat{\phi}_{\hat{\ell}},\hat{\pi}_{-\hat{\ell}} \dots \hat{\pi}_{\hat{\ell}} \right)^{T} \equiv \begin{pmatrix}
      \hat{\boldsymbol{\phi}} & \hat{\boldsymbol{\pi}}
\end{pmatrix}$. The CCR for the ordered operators, $\hat{\boldsymbol{r}}$, can be described using the symplectic matrix~\cite{BraunsteinGaussiareview,WeedbrookGaussiareview,serafini2017quantum},
\begin{equation}
    \Omega_{2\hat{L}} = \begin{pmatrix} 
        0 & \mathbb{I}_{\hat{L}\times\hat{L}} \\ 
        -\mathbb{I}_{\hat{L}\times\hat{L}} & 0 
    \end{pmatrix} , \quad [\hat{\boldsymbol{r}},\hat{\boldsymbol{r}}^{T}]=i\Omega_{2\hat{L}} \ \ \ ,
\label{eq:ccr}
\end{equation}
which remains invariant under any symplectic transformation $S$ such that $S \Omega S^{T} = \Omega$. Being unitary in the Hilbert space, symplectic transformations in phase space are generated by the time evolution operator of a bilinear Hamiltonian.

Any positive definite (PD) matrix can be transformed into its Williamson normal form~\cite{williamsonnormalform} by a global symplectic transformation. Applying this transformation symplectically diagonalizes the Hamiltonian of Eq.~\eqref{eq:discreteHfree1} into a basis of decoupled momentum oscillators~\footnote{Note that, in contrast to Eq.~\eqref{eq:freesingleparticlenormmain} and~\eqref{eq:wpnormalizationmain}, single particle momentum eigenstates in the phase space formalism are normalized to unity, $\langle 0| \tilde{a}_{\hat{k}} \tilde{a}^{\dagger}_{\hat{k}} |0\rangle = 1$. The rescaled creation operator $\tilde{a}^{\dagger}_{\hat{k}} \equiv \hat{a}^{\dagger}_{\hat{k}}/\sqrt{L}$ is therefore used throughout this appendix, with $\sum_{\hat{k}} |\tilde{f}(\hat{k})|^2 = 1$ for the momentum space distribution of the wave packet.},
\begin{equation}
\hat{H}_{\text{PBC}} = \sum_{\hat{k}} \hat{E}_{\hat{k}} \left( \frac{1}{2}\hat{\pi}^2_{\hat{k}} + \frac{1}{2}\hat{\phi}^2_{\hat{k}} \right), \quad \hat{\pi}_{\hat{k}} = -i \frac{1}{\sqrt{2}} \left(\tilde{a}_{\hat{k}} - \tilde{a}^{\dagger}_{\hat{k}} \right), \quad \hat{\phi}_{\hat{k}} = \frac{1}{\sqrt{2}} \left(\tilde{a}_{\hat{k}} + \tilde{a}^{\dagger}_{\hat{k}} \right) \ \ \ .
\end{equation}
While numerical techniques have been developed to calculate valid normal mode symplectic transforms~\cite{serafini2017quantum,NKvolumemeasure,gao2024partialtransposeguided,gao2024detectingspacelikevacuumentanglement}, an approach with greater physical significance is here constructed via understanding of the Fourier transform and momentum mode creation operators $\tilde{a}^\dagger_k$. In particular, $\hat{\phi}_k$ and $\hat{\pi}_k$ have expansions in terms of phase space operators for position oscillators as 
\begin{subequations}
\begin{align}
&\hat{\pi}_{\hat{k}} = \frac{1}{\sqrt{\hat{L}}} \sum_j \left( -\sqrt{\hat{E}_{\hat{k}} }\sin \left(\hat{k}j\right) \hat{\phi}_j + \frac{1}{\sqrt{\hat{E}_{\hat{k}}}}\cos\left( \hat{k}j \right) \hat{\pi}_j \right)  \ \ \ ,  \\
&\hat{\phi}_{\hat{k}} = \frac{1}{\sqrt{\hat{L}}} \sum_j \left( \sqrt{\hat{E}_{\hat{k}} } \cos \left(\hat{k}j\right) \hat{\phi}_j + \frac{1}{\sqrt{\hat{E}_{\hat{k}}}} \sin\left( \hat{k}j \right) \hat{\pi}_j \right)  \ \ \ ,
\end{align}
\label{eq:symplecticrelation}%
\end{subequations}
specifying a symplectic transformation $S_{FT}$ to well-defined momentum modes
\begin{equation}
S_{FT} = \frac{1}{\sqrt{\hat{L}}} 
\begin{pmatrix} C^{(11)} & D^{(12)} \\ D^{(21)} & C^{(22)} \end{pmatrix} \ \ \ ,
\label{eq:symplecticSmat}
\end{equation}
where each block is an $\hat{L} \times \hat{L}$ matrix,
\begin{equation}
    C^{(11)}_{\hat{k} j} = \sqrt{\hat{E}_{\hat{k}} } \cos(\hat{k} j), \quad C^{(22)}_{\hat{k} j} = \frac{1}{\sqrt{\hat{E}_{\hat{k}} }}\cos(\hat{k} j), \quad D^{(12)}_{\hat{k} j} = \frac{1}{\sqrt{\hat{E}_{\hat{k}}}} \sin(\hat{k} j), \quad D^{(21)}_{\hat{k} j} = -\sqrt{\hat{E}_{\hat{k}}} \sin(\hat{k} j) \; ,
\end{equation}
indexed by $\hat{k}$ and $j$ with ranges defined in Sec.~\ref{sec:II}. Constructing this transformation analytically resolves the degeneracies in the symplectic spectrum associated with opposite momentum directions, ensuring that each normal mode corresponds to a physical momentum sector. The Hamiltonian matrix thus may be faithfully transformed $S_{\text{FT}}^{-T} \mathcal{H}_{\text{PBC}} S_{\text{FT}}^{-1} = \text{diag}\left(\hat{\mathbf{E}}, \hat{\mathbf{E}}\right)$ to the basis of phase space operators for the independent momentum modes, $\hat{\mathbf{r}}_{\text{FT}} = S_{\text{FT}}\hat{\mathbf{r}}$. It is from the perspective of this basis that single particle wave packets will be reliably constructed.

The characteristic function provides a direct mapping from the density matrix to a complex-valued function~\cite{serafini2017quantum}. For the $j$-th mode, the characteristic function is defined over the two-dimensional phase space,
\begin{equation}
    \chi_{j}\left(\xi_j\right) = \Tr \left[ \hat{D}_j(\xi_j) \rho_j \right], \quad \hat{D}_j(\xi_j)  = e^{i\xi_j^T \hat{\mathbf{r}}_j}, \quad \xi_j \equiv \Omega_{2}\begin{pmatrix}
        \langle \hat{\phi}_j\rangle & \langle \hat{\pi}_j\rangle
    \end{pmatrix}^T, \quad \hat{\boldsymbol{r}}_j = \begin{pmatrix}
        \hat{\phi}_{j} ,\hat{\pi}_{j}  
    \end{pmatrix}^T \ \ \ ,
\end{equation}
where $\hat{D}_j\left( \xi_j\right) = \text{exp}\left( \alpha_j \tilde{a}_j^\dagger - \alpha_j^* \tilde{a}_j\right)$ is the displacement operator at mode $j$ such that $\xi$ relates to the usual parametrization of displacement as $\alpha_j \equiv \langle \tilde{a}_j \rangle \equiv \frac{\langle \hat{\phi}_j\rangle +i\langle \hat{\pi}_j\rangle}{\sqrt{2}}$ for expectation values in the displaced vacuum. The displacement operator for a multimode system, $\hat{D}\left(\xi\right) = \bigotimes_j \hat{D}_j\left(\xi_j\right) = e^{i \xi^T \hat{\mathbf{r}}}$, is defined in terms of $\hat{\boldsymbol{r}}$ and
\begin{equation}
\xi = \Omega_{2\hat{L}} \left( \langle \hat{\phi}_{-\hat{\ell}} \rangle \dots \langle \hat{\phi}_{\hat{\ell}} \rangle,\langle\hat{\pi}_{-\hat{\ell}} \rangle \dots \langle \hat{\pi}_{\hat{\ell}} \rangle \right)^{T} = \left(\langle\hat{\pi}_{-\hat{\ell}} \rangle \dots \langle \hat{\pi}_{\hat{\ell}} \rangle, -\langle \hat{\phi}_{-\hat{\ell}} \rangle \dots -\langle \hat{\phi}_{\hat{\ell}} \rangle \right)^{T} \ \ \ ,
\label{eq:xi}
\end{equation}
which then defines the multimode characteristic function via the trace of its product with the density matrix, $\chi\left( \xi \right) = \text{Tr}\left[ \hat{D}\left(\xi\right) \rho\right]$. 

The characteristic function for a single particle wavepacket (1P) with (normalized) momentum space superposition profile $\tilde{f}_{1}$ can be written as a weighted sum of displacement operator matrix elements,
\begin{equation}
\chi_{\text{1P}} \left( \xi_{\text{FT}} \right) = \sum_{\hat{k}, \hat{k'}} \tilde{f}_{1\hat{k}} \tilde{f}^*_{1\hat{k}'} \langle \hat{k}'| \hat{D}\left(\xi_{\text{FT}}\right) | \hat{k}\rangle \ \ \ ,
\label{eq:chi1PxiFT}
\end{equation}
where $\xi_{\text{FT}} = S_{\text{FT}}^{-T} \xi$ characterizes displacement in the phase space of momentum modes.
As multimode displacement is a tensor product operation, further evaluation may be performed via the number basis matrix elements for displacement of a single mode~\cite{Cahill:1969it}, 
\begin{equation}
     _j \langle m | \hat{D}_j(\xi_j) | n \rangle_j = \sqrt{\frac{n!}{m!}} \alpha_j^{m-n} e^{-\frac{1}{4} \xi_j^T \xi_j} L_n^{m-n} \left(\frac{1}{2} \xi_j^T \xi_j\right) \ \ \ ,
\label{eq:Dmatelement}
\end{equation}
where $L_n^{m-n}$ are the associated Laguerre polynomials~\cite{sonine1880recherches}. For this single particle sector in particular, the following set of matrix elements, 
\begin{equation}
\begin{aligned}
{}_j\langle 0 | \hat{D}_j(\xi_j) | 0 \rangle_j
&= e^{-\frac{1}{4}\xi_j^T\xi_j},
&{}_j\langle 1 | \hat{D}_j(\xi_j) | 0 \rangle_j
&= \alpha_j e^{-\frac{1}{4}\xi_j^T\xi_j},
\\
{}_j\langle 0 | \hat{D}_j(\xi_j) | 1 \rangle_j
&= -\alpha_j^* e^{-\frac{1}{4}\xi_j^T\xi_j},
&{}_j\langle 1 | \hat{D}_j(\xi_j) | 1 \rangle_j
&= e^{-\frac{1}{4}\xi_j^T\xi_j}
   \left(1-\frac{1}{2}\xi_j^T\xi_j\right) \ \ \ ,
\end{aligned}
\label{eq:Dmatelement1P}
\end{equation}
allows evaluation of Eq.~\eqref{eq:chi1PxiFT} in the form of,
\begin{equation}
\begin{aligned}
    \chi_{1P}\left(\xi_{FT}\right) &= e^{-\frac{1}{4} \xi_{FT}^T \xi_{FT}} \left(\sum_{\hat{k}} |\tilde{f}_{1\hat{k}}|^2 \left(1-\frac{1}{2} \xi_{FT,\hat{k}}^T \xi_{FT,\hat{k}} \right)+ \sum_{\hat{k}}\sum_{\hat{k}' \neq \hat{k} }  \tilde{f}_{1\hat{k}} \tilde{f}^*_{1\hat{k}'} \left(- \alpha_{FT,\hat{k}}^* \alpha_{FT,\hat{k}'}   \right) \right) \\ 
    &= e^{-\frac{1}{4} \xi_{FT}^T \xi_{FT}} \left(1-\sum_{\hat{k}} \tilde{f}_{1\hat{k}} \tilde{f}^*_{1\hat{k}} \left(\frac{1}{2} \xi_{FT,\hat{k}}^T \xi_{FT,\hat{k}} \right) \right. \\
    &\quad \left. - \sum_{\hat{k}}\sum_{\hat{k}' > \hat{k} }  \left(\Re \left[ \tilde{f}_{1\hat{k}} \tilde{f}^*_{1\hat{k}'} \right]  \left( \langle \hat{\phi}_{\hat{k}}\rangle\langle \hat{\phi}_{\hat{k}'}\rangle + \langle \hat{\pi}_{\hat{k}}\rangle\langle \hat{\pi}_{\hat{k}'}\rangle \right)  -  \Im \left[ \tilde{f}_{1\hat{k}} \tilde{f}^*_{1\hat{k}'} \right]  \left( \langle \hat{\phi}_{\hat{k}}\rangle\langle \hat{\pi}_{\hat{k}'}\rangle -\langle \hat{\pi}_{\hat{k}}\rangle\langle \hat{\phi}_{\hat{k}'}\rangle \right)\right) \right)  \\ &\equiv e^{-\frac{1}{4} \xi_{FT}^T \sigma_{FT,G} \xi_{FT}} \left(1-\frac{1}{4} \xi_{FT}^T \sigma_{FT,1} \xi_{FT} \right)  \ \ \ ,
\end{aligned}
\end{equation}
where the last line has defined, 
\begin{equation}
    \sigma_{FT,G} = \mathbb{I}, \quad \sigma_{FT,1} = 2\begin{pmatrix} 
        \Re \left[\tilde{f}_1 \tilde{f}_1^{\dagger}\right] & \Im \left[\tilde{f}_1 \tilde{f}_1^{\dagger}\right]^T \\ 
        \Im \left[\tilde{f}_1 \tilde{f}_1^{\dagger}\right] & \Re \left[\tilde{f}_1 \tilde{f}_1^{\dagger}\right] 
    \end{pmatrix} \ \ \ ,
\label{eq:momentumspacemats}
\end{equation}
by identification within the structure of Eq.~\eqref{eq:xi} for phase space operators of momentum modes. Because the real(imaginary) part of $\tilde{f}_1 \tilde{f}_1^\dagger$ is symmetric(antisymmetric), $\sigma_{\text{FT},1}$ is symmetric. While this single particle state is not Gaussian, its characteristic function is fully captured by a pair of vacuum and single particle real symmetric matrices.

The characteristic function for a pure two-particle state can be constructed analogously, provided that the two momentum space superposition profiles, $f_1$ and $f_2$, have complementary support (2Pc), $|\psi_{\text{2Pc}}\rangle = \sum_{\hat{k} \hat{p}} \tilde{f}_{1\hat{k}} \tilde{f}_{2 \hat{p}} \tilde{a}^\dagger_{\hat{k}} \tilde{a}^\dagger_{\hat{p}} |0\rangle$ with $\tilde{f}_{1\hat{k}} \tilde{f}_{2\hat{k}} = 0 \quad \forall \hat{k}$. This is a natural approximation for scattering states consisting of two counter-propagating wave packets whose momentum space distributions are localized around opposite momenta, such that the weight of each packet in the other packet's momentum sector is exponentially suppressed. Given this separation, the 2Pc characteristic function contains an additional summation over the second particle's excitation location in the tensor product momentum modes with a matrix element in each term of the form $\langle \hat{k}' \hat{p}' | \hat{D}\left(\xi_{\text{FT}} \right)|\hat{k} \hat{p}\rangle$, where $\hat{k}^{(\prime)} \neq \hat{p}^{(\prime)}$ allows continued evaluation solely with the matrix elements of Eq.~\eqref{eq:Dmatelement1P}. Direct evaluation thus yields a 2Pc characteristic function
\begin{equation}
    \chi_{2Pc}\left(\xi_{FT}\right) =e^{-\frac{1}{4} \xi_{FT}^T \sigma_{FT,G} \xi_{FT}} \left(1-\frac{1}{4} \xi_{FT}^T \sigma_{FT,1} \xi_{FT} \right) \left(1-\frac{1}{4} \xi_{FT}^T \sigma_{FT,2} \xi_{FT} \right) \ \ \ ,
\end{equation}
where $\sigma_{FT,2}$ is defined analogously as $\sigma_{FT,2} = 2\begin{pmatrix} 
        \Re \left[\tilde{f}_2 \tilde{f}_2^{\dagger}\right] & \Im \left[\tilde{f}_2 \tilde{f}_2^{\dagger}\right]^T \\ 
        \Im \left[\tilde{f}_2 \tilde{f}_2^{\dagger}\right] & \Re \left[\tilde{f}_2 \tilde{f}_2^{\dagger}\right] 
    \end{pmatrix}$.

In the basis of spatial lattice phase space operators, the characteristic function for 2Pc states thus reads,
\begin{equation}
    \chi_{2Pc}\left(\xi\right) = e^{-\frac{1}{4} \xi^T \sigma_{G} \xi} \left(1-\frac{1}{4} \xi^T \sigma_{1} \xi \right) \left(1-\frac{1}{4} \xi^T \sigma_{2} \xi \right), \,\,\, \sigma_{G} = S_{FT}^{-1} S_{FT}^{-T}, \,\,\, \sigma_{i} = S_{FT}^{-1} \sigma_{FT,i} S_{FT}^{-T}, \,\,\, i\in\{1,2\} .
\label{eq:1Pcha}
\end{equation}
By their construction, characteristic functions contain all moments of phase space operators embedded in analysis of their derivatives. For example, the expansion $\chi\left(\xi\right) = \langle e^{i\xi^T \hat{\boldsymbol{r}}}\rangle = 1 + i\sum_j \xi_j \langle\hat{r}_j \rangle - \frac{1}{2} \sum_{jj'} \xi_j \xi_{j'} \langle \hat{r}_j \hat{r}_{j'} \rangle+\mathcal{O}(\xi^3)$ followed by differentiation, 
\begin{equation}
\scalemath{0.92}{\frac{\partial  \chi(\xi)}{\partial  \xi_k}  = i \langle \hat{r}_k \rangle - \frac{1}{2} \sum_{jj'} \left(\delta_{kj}\xi_{j'}+\xi_j\delta_{kj'}\right)\langle \hat{r}_j \hat{r}_{j'} \rangle + \mathcal{O}(\xi^2), \quad \frac{\partial^2 \chi(\xi)}{\partial  \xi_{k} \partial  \xi_{k'}} =   - \frac{1}{2} \sum_{jj'} \left(\delta_{k'j}\delta_{kj'}+\delta_{kj}\delta_{k'j'} \right)\langle \hat{r}_j \hat{r}_{j'} \rangle + \mathcal{O}(\xi),}
\end{equation}
obtains the first and second moments by evaluation at the origin.  For the 2Pc states of Eq.~\eqref{eq:1Pcha}, this yields spatial lattice correlators of,
\begin{equation}
\scalemath{0.945}{\langle\hat{r}_k \rangle_{2Pc} =  -i\left.\frac{\partial  \chi_{2Pc}(\xi)}{\partial  \xi_k} \right|_{\xi = 0} = 0, \quad \left(\sigma_{2Pc}\right)_{kk'}= \langle\hat{r}_k \hat{r}_{k'} +\hat{r}_{k'} \hat{r}_{k} \rangle_{2Pc} = -2\left. \frac{\partial^2 \chi_{2Pc}(\xi)}{\partial  \xi_{k} \partial  \xi_{k'}} \right|_{\xi = 0} = \left(\sigma_{G}+\sigma_{1}+\sigma_{2}\right)_{kk'} ,}
\label{eq:cmfor2Pc}
\end{equation}
where $\sigma_G$ and $\sigma_{2Pc}$ are the covariance matrices (CM) for the vacuum state and 2Pc state, respectively. In this way, $\sigma_{1,2}$ and $\sigma_1 + \sigma_2$ capture the vacuum subtracted second moments of the 1P and 2Pc states, respectively.

In addition to providing a compact phase-space representation of pure single- and complementary two-particle states, the characteristic function also offers an efficient description of their real-time evolution. Denote unitary time evolution generated by $\hat{H}_{\text{PBC}}$ as $\hat{U}_t \equiv e^{-i \hat{H}_{\text{PBC}} t}$. For quadratic Hamiltonians, the dynamics of canonical operators are fully captured by symplectic transformation, $\hat{U}_t^{\dagger}\hat{\mathbf{r}}\hat{U}_t = e^{\Omega \mathcal{H}_{\text{PBC}}t} \hat{\mathbf{r}} \equiv S_t \hat{\mathbf{r}}$~\cite{serafini2017quantum}. The time evolution of a characteristic function thus reads,
\begin{equation}
    \chi\left(\xi,t\right) = \Tr \left[ e^{i\xi^T \hat{\mathbf{r}}} \hat{U}_t \rho \hat{U}_t^{\dagger} \right] = \Tr \left[  e^{i\xi^T \hat{U}_t^{\dagger}\hat{\mathbf{r}}\hat{U}_t}  \rho  \right] = \Tr \left[  e^{i\xi^T S_{t}  \hat{\mathbf{r}}}  \rho  \right]   \ \ \ .
\end{equation}
As such, the time evolution is captured by the transformation $\xi \rightarrow S_t^T \xi$. For a 2Pc state, the time evolution thus reads,
\begin{equation}
    \chi_{2Pc}\left(\xi,t\right) = e^{-\frac{1}{4} \xi^T S_t \sigma_{G} S_t^T \xi} \left(1-\frac{1}{4} \xi^T S_t \sigma_{1} S_t^T \xi \right) \left(1-\frac{1}{4} \xi^T S_t \sigma_{2} S_t^T \xi \right) \ \ \ ,
\label{eq:timeevo2Pc}
\end{equation}
which amounts to a linear transformation of $\sigma_G$, $\sigma_1$, and $\sigma_2$ by $S_t$. Together, Eqs.~\eqref{eq:timeevo2Pc} and~\eqref{eq:cmfor2Pc} completely determine the time evolution of the (vacuum subtracted) second moments, which serve as valuable quantities for tracking the dynamics and benchmarking MPS calculations. Notably, Eq.~\eqref{eq:timeevo2Pc} indicates that, throughout the time evolution of counter-propagating wave packets in the lattice free scalar field, the two wave packets remain dynamically independent. This suggests that counter-propagating wave packets do not generate half-space bipartite entanglement of the post-scattering wave packets, which will be further demonstrated in the main text.

\subsection{Closed-from expression of cut Renyi-2 entropy}
\label{app:freefield2}

This section begins by detailing the partial trace formalism for the 2Pc state and calculating the Renyi-2 entropy for the corresponding reduced state, both formulated in terms of characteristic functions. As a result, a closed-from expression for the Renyi-2 entropy is derived, enabling its evaluation with polynomial computational complexity with respect to the lattice volume. Beyond numerically demonstrating that counter-propagating wave packets in the lattice free scalar field do not generate entanglement, this derivation provides an exact benchmark for the Renyi-2 entropy of a subsystem obtained from MPS simulations, establishing a foundation for the analysis of digitization artifacts in~\ref{sec:IV2}. In this section, the vector of phase space operators are rearranged to $\hat{\boldsymbol{r}} = \left( \hat{\phi}_{-\hat{\ell}}, \hat{\pi}_{-\hat{\ell}} \dots \hat{\phi}_{\hat{\ell}},\hat{\pi}_{\hat{\ell}}  \right)^{T}$, such that the symplectic matrix becomes $\Omega_{2\hat{L}} \overset{\mathrm{rearrange}}{=} \bigoplus_{j=1}^{\hat{L}}\Omega_{2}$.

Following Ref.~\cite{serafini2017quantum}, consider an $(n+m)$-site system partitioned into subsystems A and B containing $n$ and $m$ modes,
\begin{equation}
    \xi = \begin{pmatrix}
        \xi_A \\ \xi_B
    \end{pmatrix}, \quad \sigma_G = \begin{pmatrix}
        \sigma_{G,A} & \sigma_{G,AB} \\ \sigma_{G,AB}^T & \sigma_{G,B}
    \end{pmatrix}, \quad \sigma_i = \begin{pmatrix}
        \sigma_{i,A} & \sigma_{i,AB} \\ \sigma_{i,AB}^T & \sigma_{i,B}
    \end{pmatrix}, \quad i\in\{1,2\} \ \ \ .
\end{equation}
As a tensor product operator, multimode displacement is expressed in these variables as $\hat{D}(\xi) = \hat{D}(\xi_A) \otimes \hat{D}(\xi_B)$. By the Fourier-Weyl relation, the density matrix and its partial trace over subsystem B admit phase-space representations,
\begin{equation}
 \hat{\rho} = \frac{1}{\left(2 \pi \right)^{m+n}} \int_{\mathcal{R}^{2(m+n)}} d\xi \, \chi\left(\xi\right)  \hat{D}(-\xi_A) \otimes \hat{D}(-\xi_B), \quad \Tr_B\left[\hat{\rho}\right] = \frac{1}{\left(2 \pi \right)^{n}} \int_{\mathcal{R}^{2n}} d\xi \, \left.\chi\left(\xi\right)\right|_{\xi_B = 0}  \hat{D}(-\xi_A)  \ \ \ ,
\label{eq:partialtrace2Pc}
\end{equation}
where the partial trace applies to $\hat{D}(-\xi_B)$ and yields $\Tr_B\left[\hat{D}(-\xi_B)\right]=\left(2\pi\right)^m \delta^{2m}\left(\xi_B\right)$. As such, the reduced ensemble $\Tr_B\left[\hat{\rho}\right]$ retains a characteristic function in the general 2Pc form [Eq.~\eqref{eq:1Pcha}] with $\xi$, $\sigma_G$, and $\sigma_i$ replaced by their corresponding subsystem quantities, $\xi_A$, $\sigma_{G,A}$, and $\sigma_{i,A}$.

A closed-form expression for the Renyi-2 entropy can be derived for states of the general 2Pc form. Consider an $n$-site system. Following Eq.~\eqref{eq:partialtrace2Pc}, the purity of a general quantum state is calculated as
\begin{equation}
    \mu\left(\hat{\rho}\right) = \Tr\left[\hat{\rho}^2\right] = \frac{1}{\left(2\pi\right)^n} \int_{\mathcal{R}^{2n}} d\xi \, \left| \chi \left(\xi\right)\right|^2 \ \ \ ,
\end{equation}
where $\chi\left( - \xi\right) = \chi\left(\xi\right)^*$ and the trace of the displacement operator yields a delta function, reducing the expression to a single integral. Since the characteristic function of the general 2Pc form is real, the purity may be expressed as
\begin{equation}
\scalemath{0.94}{\mu\left(\hat{\rho}_{2Pc}\right) = \Tr\left[\hat{\rho}^2_{2Pc}\right] = \frac{1}{\left(2\pi\right)^n} \int_{\mathcal{R}^{2n}} d\xi \, e^{-\frac{1}{2} \xi^T \sigma_{G} \xi} \left(1-\frac{1}{4} \xi^T \sigma_{1} \xi \right)^2 \left(1-\frac{1}{4} \xi^T \sigma_{2} \xi \right)^2 =\mathcal{I}_{0} + \mathcal{I}_{2} + \mathcal{I}_{4} + \mathcal{I}_{6} + \mathcal{I}_{8}}  \ \ \  ,
\end{equation}
where $\mathcal{I}_{i}$ denotes the Gaussian integral corresponding to the $i$-th moment of the squared vacuum characteristic function. The Renyi-2 entropy then follows as the negative logarithm of this integral sum,
\begin{equation}
    S_2\left(\hat{\rho}_{2Pc}\right) = -\log\left(\mu\left(\hat{\rho}_{2Pc}\right)\right) \ \ \ .
\end{equation}
From standard Gaussian integration with $\sigma_G^T = \sigma_G$,
\begin{equation}
    \frac{1}{\left(2\pi\right)^n} \int_{\mathcal{R}^{2n}} d\xi \, e^{-\frac{1}{2} \xi^T \sigma_{G} \xi} = \frac{1}{\sqrt{\det (\sigma_{G})}}, \quad \frac{1}{\left(2\pi\right)^n} \int_{\mathcal{R}^{2n}} d\xi \, \xi_{j_1} \xi_{j_2} e^{-\frac{1}{2} \xi^T \sigma_{G} \xi} = \frac{\left(\sigma_{G}^{-1}\right)_{j_{1}j_{2}}}{\sqrt{\det (\sigma_{G})}} \ \ \ ,
\end{equation}
$\mathcal{I}_{0}$ and $\mathcal{I}_{2}$ read,
\begin{equation}
\mathcal{I}_{0} =  \frac{1}{\sqrt{\det (\sigma_{G})}} \ \ \ ,
\end{equation}
\begin{equation}
    \mathcal{I}_{2} = \sum_{j_{1}j_{2}}-\left(\frac{\sigma_1}{2}+\frac{\sigma_2}{2}\right)_{j_{1}j_{2}}\frac{\left(\sigma_{G}^{-1}\right)_{j_{1}j_{2}}}{\sqrt{\det (\sigma_{G})}} = -\frac{1}{2\sqrt{\det (\sigma_{G})}} \Tr\left[\left(\sigma_1+ \sigma_2\right) \sigma_{G}^{-1}\right]  \ \ \ ,
\end{equation}
where $\mathcal{I}_{0}$ is the familiar expression for the purity of Gaussian states. Higher moments can be obtained from Wick's theorem~\cite{PhysRev.80.268}, 
\begin{equation}
    \frac{1}{\left(2\pi\right)^n} \int_{\mathcal{R}^{2n}} d\xi \, \xi_{j_1} \xi_{j_2} \cdots \xi_{j_{2m}} e^{-\frac{1}{2} \xi^T \sigma_{G} \xi} = \frac{1}{\sqrt{\det (\sigma_{G})}} \sum_{\mathcal{P}} \prod_{\left(a\, b\right) \in \mathcal{P}} \left(\sigma_{G}^{-1}\right)_{j_{a}j_{b}}  \ \ \ ,
\end{equation}
where $\mathcal{P}$ runs over all pairings of the index set $\{1,\ldots,2m\}$. The higher order moment integrals are calculated to be,
\begin{multline}
\mathcal{I}_{4} =  \frac{1}{16\sqrt{\det (\sigma_{G})}} \left(  \Tr\left[\sigma_1 \sigma_{G}^{-1}\right]^2  + \Tr\left[\sigma_2 \sigma_{G}^{-1}\right]^2 +2\Tr\left[\sigma_1 \sigma_{G}^{-1}\sigma_1 \sigma_{G}^{-1}\right]+ 2\Tr\left[\sigma_2 \sigma_{G}^{-1} \sigma_2 \sigma_{G}^{-1}\right] \right. \\ \left. + 4\Tr\left[\sigma_1 \sigma_{G}^{-1}\right]\Tr\left[\sigma_2 \sigma_{G}^{-1}\right] + 8\Tr\left[\sigma_1 \sigma_{G}^{-1} \sigma_2 \sigma_{G}^{-1}\right] \right) \ \ \ ,
\end{multline}
\begin{multline}
\mathcal{I}_{6} =  -\frac{1}{32\sqrt{\det (\sigma_{G})}} \left(  \Tr\left[\left(\sigma_1+\sigma_2\right) \sigma_{G}^{-1}\right]\Tr\left[\sigma_1 \sigma_{G}^{-1}\right]\Tr\left[\sigma_2 \sigma_{G}^{-1}\right]    + 2\Tr\left[\sigma_2 \sigma_{G}^{-1}\right]\Tr\left[\sigma_1 \sigma_{G}^{-1} \sigma_1 \sigma_{G}^{-1}\right]   \right. \\ \left. + 2\Tr\left[\sigma_1 \sigma_{G}^{-1}\right]\Tr\left[\sigma_2 \sigma_{G}^{-1} \sigma_2 \sigma_{G}^{-1}\right]  + 4\Tr\left[\left(\sigma_1 +\sigma_2 \right)\sigma_{G}^{-1}\right]\Tr\left[\sigma_1 \sigma_{G}^{-1} \sigma_2 \sigma_{G}^{-1}\right] \right. \\ \left. + 8\Tr\left[\sigma_2 \sigma_{G}^{-1}\sigma_2 \sigma_{G}^{-1}\sigma_1 \sigma_{G}^{-1}\right]+8\Tr\left[\sigma_1 \sigma_{G}^{-1}\sigma_1 \sigma_{G}^{-1}\sigma_2 \sigma_{G}^{-1}\right]  \right) \ \ \  ,
\end{multline}
\begin{multline}
\mathcal{I}_{8} =  \frac{1}{256\sqrt{\det (\sigma_{G})}} \left(  \Tr\left[\sigma_1 \sigma_{G}^{-1}\right]^2\Tr\left[\sigma_2 \sigma_{G}^{-1}\right]^2 + 2\Tr\left[\sigma_1 \sigma_{G}^{-1}\right]^2\Tr\left[\sigma_2 \sigma_{G}^{-1} \sigma_2 \sigma_{G}^{-1}\right] \right. \\ \left. + 2\Tr\left[\sigma_2 \sigma_{G}^{-1}\right]^2\Tr\left[\sigma_1 \sigma_{G}^{-1} \sigma_1 \sigma_{G}^{-1}\right] +4\Tr\left[\sigma_1 \sigma_{G}^{-1} \sigma_1 \sigma_{G}^{-1} \right]\Tr\left[\sigma_2 \sigma_{G}^{-1} \sigma_2 \sigma_{G}^{-1} \right]\right. \\ \left.+ 8\Tr\left[\sigma_1 \sigma_{G}^{-1} \sigma_2 \sigma_{G}^{-1}\right]^2 + 8\Tr\left[\sigma_1 \sigma_{G}^{-1}\right]\Tr\left[\sigma_2 \sigma_{G}^{-1}\right]\Tr\left[\sigma_1 \sigma_{G}^{-1} \sigma_2 \sigma_{G}^{-1}\right]  \right. \\ \left. + 16\Tr\left[\sigma_1 \sigma_{G}^{-1}\right]\Tr\left[\sigma_1 \sigma_{G}^{-1} \sigma_2 \sigma_{G}^{-1}\sigma_2 \sigma_{G}^{-1}\right]   +  16\Tr\left[\sigma_2 \sigma_{G}^{-1}\right]\Tr\left[\sigma_2 \sigma_{G}^{-1} \sigma_1 \sigma_{G}^{-1}\sigma_1 \sigma_{G}^{-1}\right]   \right. \\ \left. + 16\Tr\left[\sigma_1 \sigma_{G}^{-1} \sigma_2 \sigma_{G}^{-1}\sigma_1 \sigma_{G}^{-1} \sigma_2 \sigma_{G}^{-1}\right]  + 32\Tr\left[\sigma_1 \sigma_{G}^{-1} \sigma_1 \sigma_{G}^{-1}\sigma_2 \sigma_{G}^{-1} \sigma_2 \sigma_{G}^{-1}\right]\right) \ \ \ ,
\end{multline}
yielding a closed-form expression for the Renyi-2 entropy of 2Pc ensembles. While $S_2$ must vanish for pure states, replacing $\sigma_{1,2}, \sigma_G$ in the above expressions by their corresponding subsystem quantities allows direct calculation of the purity or Renyi-2 entropy of 2Pc subsystems.

\end{document}